\documentclass[iop,apj]{piner2}
\usepackage[figuresright]{myrotating}
\begin{document}

\shorttitle{Blazar Jet Accelerations}

\title{Relativistic Jets in the Radio Reference Frame Image Database II:\\
Blazar Jet Accelerations from the First 10 Years of Data (1994 -- 2003)}

\author{B.~G.~Piner\altaffilmark{1}, 
A.~B.~Pushkarev\altaffilmark{2,3,4}, Y.~Y.~Kovalev\altaffilmark{5,4},
C.~J.~Marvin\altaffilmark{1}, J.~G.~Arenson\altaffilmark{1},\\
P.~Charlot\altaffilmark{6,7}, A.~L.~Fey\altaffilmark{8}, A.~Collioud\altaffilmark{6,7},
\& P.~A.~Voitsik\altaffilmark{5}}

\altaffiltext{1}{Department of Physics and Astronomy, Whittier College,
13406 E. Philadelphia Street, Whittier, CA 90608; gpiner@whittier.edu}

\altaffiltext{2}{Pulkovo Astronomical Observatory, Pulkovskoe Chaussee 65/1, 196140 St. Petersburg, Russia}

\altaffiltext{3}{Crimean Astrophysical Observatory, 98409 Nauchny, Crimea, Ukraine}

\altaffiltext{4}{Max-Planck-Institut f\"{u}r Radioastronomie, Auf dem H\"{u}gel 69, D-53121 Bonn, Germany}

\altaffiltext{5}{Astro Space Center of Lebedev Physical Institute, Profsoyuznaya Str. 84/32, 117997 Moscow, Russia}

\altaffiltext{6}{Univ. Bordeaux, LAB, UMR 5804, F-33270 Floirac, France}

\altaffiltext{7}{CNRS, LAB, UMR 5804, F-33270 Floirac, France}

\altaffiltext{8}{United States Naval Observatory, 3450 Massachusetts Ave., NW, Washington D.C 20392}

\begin{abstract}
We analyze blazar jet apparent speeds
and accelerations from the RDV series of
astrometric and geodetic VLBI experiments. From these experiments, we have produced
and analyzed 2753 global VLBI images
of 68 sources at 8~GHz with a median beam size of 0.9~milliarcseconds (mas),
and a median of 43 epochs per source.
From this sample, we analyze the motions of 225 jet components in 66 sources.
The distribution of the fastest measured apparent speed in each source has a
median of 8.3$c$ and a maximum of 44$c$. Sources in the 2FGL {\em Fermi}
LAT catalog display higher apparent speeds than those that have not been
detected. On average, components farther from the core in a given source
have significantly higher apparent speeds than components closer to the core;
for example, for a typical source, components at $\sim3$~mas from
the core ($\sim15$~pc projected at $z\sim0.5$)
have apparent speeds about 50\% higher than those of components at $\sim1$~mas from the core
($\sim5$~pc projected at $z\sim0.5$).
We measure accelerations of components in orthogonal directions parallel
and perpendicular to their average velocity vector.
Parallel accelerations have significantly larger magnitudes than
perpendicular accelerations, implying observed accelerations
are predominantly due to changes in the Lorentz factor (bulk or pattern)
rather than projection effects from jet bending. Positive parallel accelerations are significantly more common
than negative ones, so the Lorentz factor (bulk or pattern)
tends to increase on the scales observed here. Observed parallel
accelerations correspond to modest source frame increases in the bulk or pattern Lorentz factor.
\end{abstract}


\section{Introduction}
\label{intro}
The bulk outflow of material at high Lorentz factors
\footnote{The Lorentz factor $\Gamma=1/(1-\beta^2)^{1/2}$, where $\beta$
is the velocity expressed as a fraction of the speed of light.}
in collimated relativistic jets is a well-established property of powerful blazars.
Such high Lorentz factors can be directly observed through the high-speed
apparent motions of jet components in VLBI imaging (e.g., Lister et al. 2009b, hereafter L09),
and they are also required to explain blazar spectral energy distributions
(e.g., Hartman et al. 2001), gamma-ray time variability (e.g., Dondi \& Ghisellini 1995),
and high radio-core brightness temperatures (e.g., Tingay et al. 2001).
These relativistic jets must be accelerated over some length scale between 
about $10^{3}$ gravitational radii from the central 
black hole and the parsec scale where they are
directly observed with VLBI (e.g., Sikora 2005; Vlahakis \& K\"{o}nigl 2004).
Although observations of high Lorentz factor flows are well-established,
the theoretical mechanism by which these outflows are accelerated, and the length
scale over which it operates, is not completely understood.

In the general framework of magnetic jet acceleration in blazars
(e.g., Sikora 2005), the energy of the flow
begins as magnetic energy, or Poynting flux, which is then
converted into bulk kinetic energy during an acceleration phase, and finally
into particle kinetic energy at shocks, which can then be radiated away.
Magnetic acceleration has been investigated through general relativistic magnetohydrodynamic simulations 
(e.g., McKinney 2006); also see Komissarov (2011) and K\"{o}nigl (2010) for summaries
of the theory of magnetic acceleration of relativistic jets.

A number of details within this general framework remain to be addressed:
such as whether the jet is produced in a steady-state or whether the acceleration
is impulsive (Granot et al. 2011; Lyutikov \& Lister 2010), and whether the acceleration is complete
before the scales observed with VLBI, or is still occurring on these parsec scales.
Vlahakis \& K\"{o}nigl (2004) argue that magnetic acceleration can continue to act
out to the parsec scales, and they interpreted two specific
observed acceleration events in NGC~6251 and 3C~345
as evidence for this ``magnetic driving'' on parsec scales, but at the time of that paper
there were insufficient VLBI observations to address the question of
whether acceleration on parsec scales was a common property of blazar jets in large samples.
Even if parsec-scale acceleration events are observed to be commonplace, it does not
necessarily prove the direct observation of conversion of Poynting flux to kinetic energy, since
there may also be hydrodynamic means to produce accelerations in matter-dominated jets
(e.g., Daly \& Marscher 1988; Kadler et al. 2008),
or the observations may be showing an increase in the Lorentz factors of patterns in the underlying flow.

Direct observations of intrinsic acceleration through VLBI imaging are difficult. 
Precise measurements of component positions at many epochs are needed to reliably measure a
second derivative in a position versus time plot. For individual jet components,
the apparent speed is given by the well-known formula: 
\begin{equation}
\label{speedeqn}
\beta_\mathrm{app}=\frac{\beta\sin\theta}{1-\beta\cos\theta},
\end{equation}
where $\beta c$ is the intrinsic speed and $\theta$ is the angle of the motion to the line of sight.
When observed in a single component then, changes in the apparent speed can be produced
either from a change in the intrinsic speed or the viewing angle.
Observations of many apparently accelerating components are needed to statistically distinguish
between these two cases. In practice then, observations of many sources at many epochs, totaling
thousands of VLBI images, are needed to measure variations in intrinsic speeds.
While observations of either individual apparent component accelerations 
(e.g., Unwin et al. 1997; Homan et al. 2003)
or apparent component accelerations in smaller samples of blazars
(e.g., Homan et al. 2001; Jorstad et al. 2005) have been previously noted,
the MOJAVE survey with 2424 total images was the first to investigate 
blazar jet accelerations through a large statistical sample
(Homan et al. 2009, hereafter H09).

In this paper we present a 
continuation of our blazar jet kinematics study from Piner et al. (2007), hereafter Paper I, 
that is designed to enable measurements of blazar jet
accelerations using the RDV (Research \& Development -- VLBA) series of experiments on the VLBA
(Petrov et al. 2009).
The RDV series of experiments is observed primarily for the purposes of astrometry
and geodesy, but because the experiments have occurred roughly every two months 
since the VLBA opened, and produce quality images, they are also useful for blazar astrophysics
(e.g., Kovalev et al. 2008; Pushkarev \& Kovalev 2012).
This is only the second large-scale study of accelerations in the apparent motions of extragalactic jets
(following H09). 

In Paper I, we analyzed jet kinematics
using 19 VLBI experiments observed over a 5 year time baseline from 1994 to 1998
(RDVs 1 through 10 and 12, plus 8 similar VLBI experiments
that were conducted on the VLBA before the RDV series began).
In that paper, we studied all sources that had been observed at 3 or more epochs
over those 19 experiments, yielding a total of 966 images of 87 sources,
which were used to measure apparent jet speeds.

In this paper, we expand on our study from Paper I by extending the analysis to a total of 50
VLBI experiments over a 10 year time baseline from 1994 to 2003 (adding the 31 new experiments
RDVs 11 and 13 through 42), and studying the kinematics of all sources that have been observed at 20 or
more epochs over those 50 experiments. This survey is hereafter referred to as the RDV survey: it
now comprises 2753 VLBI images of 68 sources, with a median of 43 epochs of observation per source.
The number of images is approximately tripled compared to Paper I, and slightly exceeds
the 2424 images in the MOJAVE survey (Lister et al. 2009a).
Note also that the maximum number of epochs per source from Paper I (19) is now less then the minimum
number of epochs per source considered in this paper (20).

The RDV experiments have continued to the present; the most recent available at this
writing is RDV~93 observed on 28 June 2012. Thus, there are already an additional 51 RDV experiments
in the VLBA archive above what is included in this paper.
If these additional experiments are completely imaged and model fit, they have the potential to
approximately double the RDV survey size compared to what is included in this paper:
to approximately 6000 total images and approximately 100 epochs per source.
At the present time, imaging of the RDV experiments is continuing so
that studies such as those presented in this paper could be extended in
the future.

The organization of this paper is as follows: in $\S$~\ref{sample} we describe 
our sample selection. In $\S$~\ref{models} we describe the VLBI imaging and model fitting,
and present a large table of Gaussian model components.
In $\S$~\ref{speeds} we present our measurement of component speeds, and in $\S$~\ref{acc} our
measurements of component accelerations. In $\S$~\ref{discussion}
we discuss the physical implications of these results,
and in $\S$~\ref{conclusions} we present our major conclusions.
Throughout the paper, we assume cosmological parameters of
$H_{0}=71$ km s$^{-1}$ Mpc$^{-1}$, $\Omega_{m}=0.27$, and $\Omega_{\Lambda}=0.73$.

\section{Sample Selection}
\label{sample}
Our sample for this paper is drawn from the RDV series of astrometric and geodetic VLBI experiments.
This series of experiments was fully described in Paper I, here we review 
and summarize some of their important properties.
The RDV experiments are conducted using the 10 antennas of the National Radio Astronomy Observatory's
Very Long Baseline Array (VLBA), along with the addition of up to 10 geodetic VLBI antennas 
in both the northern and southern hemispheres that provide global VLBI coverage.
Observations are made in a simultaneous dual-frequency mode at both S-band (2~GHz) and X-band (8~GHz).
Results of the precise geodesy and astrometry afforded by these observations have been presented elsewhere
(e.g., Petrov \& Ma 2003; Fey et~al. 2004). Observations in this mode
also allow for simultaneous dual-frequency imaging at 8 and 2~GHz, it is the results
from the 8~GHz imaging that form the basis of the work discussed here.

\begin{table*}
\begin{center}
{\small \caption{Observation Log}
\label{obstab}
\begin{tabular}{l c c c c} \tableline \tableline \\[-5pt]
& Decimal & VLBA Observation & & Image \\
\multicolumn{1}{c}{Epoch} & Date & Code & Antennas$^{a}$ & Reference \\ \tableline \\[-5pt]
1994 Jul 8  & 1994.52 & BR005  & VLBA                      & 1,2 \\
1995 Apr 12 & 1995.28 & BR025  & VLBA                      & 1,3 \\
1995 Jul 24 & 1995.56 & RDGEO2 & VLBA                      & 1   \\
1995 Oct 2  & 1995.75 & RDGEO3 & VLBA                      & 1   \\
1995 Oct 12 & 1995.78 & BF012  & VLBA                      & 1,3 \\
1996 Apr 23 & 1996.31 & BE010A & VLBA                      & 1   \\
1997 Jan 10 & 1997.03 & BF025A & VLBA                      & 1,4 \\
1997 Jan 11 & 1997.03 & BF025B & VLBA                      & 1,4 \\
1997 Jan 30 & 1997.08 & RDV01  & VLBA+GcGnKkMcOnWf         & 1   \\
1997 Mar 31 & 1997.25 & RDV02  & VLBA+GcGnKkMcOnWf         & 1   \\
1997 May 19 & 1997.38 & RDV03  & VLBA+GcGnKkMcOnWf         & 1   \\
1997 Jul 24 & 1997.56 & RDV04  & VLBA+GcGnKkMcOnWf         & 1   \\
1997 Sep 8  & 1997.69 & RDV05  & VLBA+GcGnKkOnWf           & 1   \\
1997 Dec 17 & 1997.96 & RDV06  & VLBA+GcGnKkMcOnWf         & 1   \\
1998 Feb 9  & 1998.11 & RDV07  & VLBA+GcGnKkMcNyOnWf       & 1   \\
1998 Apr 15 & 1998.29 & RDV08  & VLBA+GcGnKkMcNyOnWf       & 1   \\
1998 Jun 24 & 1998.48 & RDV09  & VLBA+GcGnKkMcNyOnWf       & 1   \\
1998 Aug 10 & 1998.61 & RDV10  & VLBA+GcGnKkMcNyOn         & 1   \\
1998 Oct 1  & 1998.75 & RDV11  & VLBA+GcGnKkMcNyOnWf       & 5   \\
1998 Dec 21 & 1998.97 & RDV12  & VLBA+GcGnKkMcNyWf         & 1   \\
1999 Mar 8  & 1999.18 & RDV13  & VLBA+GcGnHhKkMcNyOnWfWz   & 5   \\
1999 Apr 15 & 1999.29 & RDV14  & VLBA+GcHhKkMcNyOnTsWfWz   & 1   \\
1999 May 10 & 1999.36 & RDV15  & VLBA+GcHhKkMcNyOnTsWfWz   & 5   \\
1999 Jun 22 & 1999.47 & RDV16  & VLBA+GcHhKkMcNyOnTsWfWz   & 1   \\
1999 Aug 2  & 1999.59 & RDV17  & VLBA+GcHhKkMcNyOnWfWz     & 1   \\
1999 Dec 20 & 1999.97 & RDV18  & VLBA+GcGnHhKkMcNyOnTsWfWz & 5   \\
2000 Jan 31 & 2000.08 & RDV19  & VLBA+GcHhKkMaMcNyOnTsWfWz & 1   \\
2000 Mar 13 & 2000.20 & RDV20  & VLBA+GcHhKkMaMcNyOnTsWfWz & 6   \\
2000 May 22 & 2000.39 & RDV21  & VLBA+GcHhKkMaMcNyTsWfWz   & 5   \\
2000 Jul 6  & 2000.51 & RDV22  & VLBA+GcHhKkMaNyTsWfWz     & 1   \\
2000 Oct 23 & 2000.81 & RDV23  & VLBA+GcHhKkMaMcNyTsWfWz   & 1   \\
2000 Dec 4  & 2000.93 & RDV24  & VLBA+GcHhKkMaMcNyTsWfWz   & 5   \\
2001 Jan 29 & 2001.08 & RDV25  & VLBA+GcHhKkMaMcNyOnTsWfWz & 1   \\
2001 Mar 12 & 2001.19 & RDV26  & VLBA+HhKkMaMcNyOnTsWz     & 6   \\
2001 Apr 9  & 2001.27 & RDV27  & VLBA+GcHhKkMaMcNyTsWfWz   & 5   \\
2001 May 9  & 2001.35 & RDV28  & VLBA+GcHhKkMaMcNyOnTsWfWz & 1   \\
2001 Jul 5  & 2001.51 & RDV29  & VLBA+GcHhKkMaMcNyTsWfWz   & 5   \\
2001 Oct 29 & 2001.83 & RDV30  & VLBA+GcHhKkMaMcNyTsWfWz   & 5   \\
2002 Jan 16 & 2002.04 & RDV31  & VLBA+GcKkMaMcNyOnTsWfWz   & 5   \\
2002 Mar 6  & 2002.18 & RDV32  & VLBA+GcKkMaMcNtOnTsWfWz   & 5   \\
2002 May 8  & 2002.35 & RDV33  & VLBA+ApGcGgHhKkMaMcOnWfWz & 5   \\
2002 Jul 24 & 2002.56 & RDV34  & VLBA+GcKkMaMcNyOnTcWfWz   & 5   \\
2002 Sep 25 & 2002.73 & RDV35  & VLBA+GcKkMaMcOnTcTsWfWz   & 5   \\
2002 Dec 11 & 2002.95 & RDV36  & VLBA+GcKkMaMcNyOnTcWfWz   & 5   \\
2003 Mar 12 & 2003.19 & RDV37  & VLBA+KkMaMcOnTcTsWfWz     & 5   \\
2003 May 7  & 2003.35 & RDV38  & VLBA+KkMaMcOnTcTsWfWz     & 5   \\
2003 Jun 19 & 2003.47 & RDV39  & VLBA+KkMaNyOnTcTsWfWz     & 5   \\
2003 Jul 9  & 2003.52 & RDV40  & VLBA+GcMaNyOnTcTsWfWz     & 1   \\
2003 Sep 17 & 2003.71 & RDV41  & VLBA+GcKkMaMcNyOnTsWfWz   & 5   \\
2003 Dec 17 & 2003.96 & RDV42  & VLBA+GcMaMcNyOnTcTsWfWz   & 6   \\ \tableline \\[-5pt] 
\end{tabular}}
\end{center}
{\bf Notes.}\\
$a$: Non-VLBA antennas are indicated by two-letter codes. 
Sizes and locations of non-VLBA antennas are as follows:
Ap: 46~m, Algonquin Park, Ontario, Canada;
Gc: 26~m, Gilmore Creek, Fairbanks, AK, USA;
Gg: 5~m, Greenbelt, MD, USA;
Gn: 20~m, Green Bank, WV, USA;
Hh: 26~m, Hartebeesthoek, South Africa;
Kk: 20~m, Kokee Park, HI, USA;
Ma: 20~m, Matera, Italy;
Mc: 32~m, Medicina, Italy;
Ny: 20~m, Ny Alesund, Norway;
On: 20~m, Onsala, Sweden,
Tc: 6~m, Concepcion, Chile;
Ts: 32~m, Tsukuba, Japan;
Wf: 18~m, Westford, MA, USA;
Wz: 20~m, Wettzell, Germany\\
{\bf References.---}
(1) http://rorf.usno.navy.mil/RRFID/;
(2) Fey et al. (1996);
(3) Fey \& Charlot (1997);
(4) Fey \& Charlot (2000);
(5) http://astrogeo.org/vlbi\_images/;
(6) http://www.obs.u-bordeaux1.fr/BVID/ \\
\end{table*}
 
\begin{table*}[!t]
\begin{center}
{\scriptsize \caption{Sources in the RDV Sample} 
\label{sources}
\begin{tabular}{l l c c c c c} \tableline \tableline \\[-5pt]
& \multicolumn{1}{c}{Common} & Number of & Optical & & & {\em Fermi}$^{d}$ \\ 
\multicolumn{1}{c}{Source$^{a}$} & \multicolumn{1}{c}{Name} & Epochs & Class$^{b}$ & $z^{b}$ 
& MOJAVE$^{c}$ & 2LAC \\ \tableline \\[-5pt]
0003$-$066       & NRAO~5    & 39 & B     & 0.35      & Y &   \\
0014+813         &           & 43 & Q     & 3.39      &   &   \\
0048$-$097$^{e}$ &           & 42 & B(HP) & 0.63      & Y & Y \\
0059+581         &           & 45 & Q     & 0.64      & Y &   \\
0104$-$408       &           & 37 & Q     & 0.58      &   &   \\
0119+041         &           & 41 & Q(HP) & 0.64      &   &   \\
0119+115         &           & 42 & Q(HP) & 0.57      & Y &   \\
0133+476         & DA~55     & 44 & Q(HP) & 0.86      & Y & Y \\
0201+113         &           & 41 & Q     & 3.61      &   &   \\
0202+149         &           & 43 & G     & 0.41      & Y & Y \\
0229+131         &           & 43 & Q     & 2.07      &   &   \\
0234+285         &           & 43 & Q(HP) & 1.21      & Y & Y \\
0235+164         &           & 25 & Q(HP) & 0.94      & Y & Y \\
0336$-$019       & CTA~26    & 44 & Q(HP) & 0.85      & Y & Y \\
0402$-$362       &           & 39 & Q     & 1.42      &   & Y \\
0430+052         & 3C~120    & 42 & G     & 0.03      & Y &   \\
0454$-$234       &           & 45 & Q(HP) & 1.00      &   & Y \\
0458$-$020       &           & 41 & Q(HP) & 2.29      & Y & Y \\
0528+134         &           & 44 & Q     & 2.07      & Y & Y \\
0537$-$441$^{f}$ &           & 34 & Q(HP) & 0.89      &   & Y \\
0552+398         &           & 49 & Q     & 2.36      & Y &   \\
0642+449         & OH~471    & 43 & Q     & 3.41      & Y &   \\
0727$-$115       &           & 50 & Q     & 1.59      & Y &   \\
0804+499         &           & 44 & Q(HP) & 1.43      & Y &   \\
0823+033         &           & 45 & B(HP) & 0.51      & Y & Y \\
0851+202         & OJ~287    & 45 & B(HP) & 0.31      & Y & Y \\
0919$-$260       &           & 42 & Q     & 2.30      &   &   \\
0920$-$397       &           & 39 & Q     & 0.59      &   &   \\
0923+392         & 4C~+39.25 & 45 & Q     & 0.70      & Y &   \\
0955+476         & OK~492    & 45 & Q     & 1.87      & Y & Y \\
1034$-$293       &           & 36 & Q(HP) & 0.31      &   &   \\
1044+719         &           & 45 & Q     & 1.15      &   & Y \\
1101+384         & Mrk~421   & 43 & B(HP) & 0.03      &   & Y \\
1124$-$186       &           & 42 & Q     & 1.05      & Y & Y \\ 
1128+385         &           & 46 & Q     & 1.73      &   &   \\  
1144$-$379$^{f}$ &           & 34 & Q(HP) & 1.05      &   & Y \\
1145$-$071       &           & 40 & Q     & 1.34      &   & Y \\
1156+295         & 4C~+29.45 & 43 & Q(HP) & 0.73      & Y & Y \\
1228+126         & M87       & 43 & G     & 0.004     & Y & Y \\
1308+326         &           & 43 & Q(HP) & 1.00      & Y & Y \\
1313$-$333$^{f}$ &           & 42 & Q     & 1.21 &   & Y \\
1334$-$127       &           & 40 & Q(HP) & 0.54 & Y & Y \\
1357+769$^{g}$   &           & 45 & Q     & 1.59 &   & Y \\
1424$-$418$^{f}$ &           & 36 & Q(HP) & 1.52 &   & Y \\
1448+762         &           & 24 & G     & 0.90 &   &   \\
1451$-$375       &           & 33 & Q     & 0.31 &   &   \\
1514$-$241       & AP~Lib    & 41 & B(HP) & 0.05 &   & Y \\
1606+106         &           & 45 & Q     & 1.23 & Y & Y \\
1611+343         & DA~406    & 44 & Q     & 1.40 & Y & Y \\
1622$-$253       &           & 39 & Q     & 0.79 &   & Y \\
1638+398         & NRAO~512  & 45 & Q(HP) & 1.67 & Y & Y \\
1642+690         & 4C~+69.21 & 25 & Q(HP) & 0.75 &   &   \\
1657$-$261       &           & 22 & U     & ...  &   &   \\
1726+455         &           & 20 & Q     & 0.71 & Y & Y \\
1739+522         & OT~566    & 45 & Q(HP) & 1.38 & Y & Y \\
1741$-$038       &           & 46 & Q(HP) & 1.06 & Y &   \\
1745+624         & 4C~+62.29 & 43 & Q     & 3.89 &   &   \\
1749+096         & OT~081    & 50 & Q(HP) & 0.32 & Y & Y \\
1803+784         &           & 43 & Q(HP) & 0.68 & Y & Y \\
1908$-$201       &           & 41 & Q     & 1.12 &   & Y \\
1921$-$293       & OV~$-$236 & 43 & Q(HP) & 0.35 &   & Y \\
1954$-$388$^{f}$ &           & 36 & Q(HP) & 0.63 &   & Y \\
2052$-$474$^{f}$ &           & 21 & Q     & 1.49 &   & Y \\
2145+067         &           & 50 & Q     & 1.00 & Y & Y \\
2200+420         & BL Lac    & 43 & B(HP) & 0.07 & Y & Y \\
2223$-$052       & 3C~446    & 26 & Q(HP) & 1.40 & Y & Y \\
2234+282         &           & 45 & Q(HP) & 0.80 &   & Y \\
2243$-$123       &           & 41 & Q(HP) & 0.63 & Y &   \\ \tableline \\[-5pt]
\end{tabular}}
\end{center}
{\bf Notes.}\\
$a$: Epoch 1950 IAU source name.\\
$b$: Unless otherwise noted, optical class and redshift are from V\'{e}ron-Cetty \& V\'{e}ron (2010).
Q=quasar, B=BL Lac object, G=galaxy, HP=high polarization, U=unidentified.\\
$c$: Whether or not source is is MOJAVE survey, using sample 
listed in Table 1 of Lister et al. (2009a). (Y=Yes)\\
$d$: Whether or not source is in the {\em Fermi} LAT 2 year AGN catalog, Ackermann et al. (2011). (Y=Yes)\\
$e$: Tentative redshift from NED. (Redshift not in V\'{e}ron-Cetty \& V\'{e}ron (2010).)\\
$f$: Source is in the TANAMI sample (Ojha et al. 2010).\\
$g$: Optical class and redshift are from Ackermann et al. (2011). (Source not in V\'{e}ron-Cetty \& V\'{e}ron (2010).)\\
\end{table*}

The analysis presented in this paper uses the imaging results at 8~GHz only, because the higher resolution
afforded by the 8~GHz observations is needed for precise measurements of jet kinematics.
The 8~GHz observations presented in this paper that were recorded after 1997 
have similar angular resolution to the 
observations from the MOJAVE survey (Lister et al. 2009a), since the post-1997 RDV experiments use
global VLBI baselines at 8~GHz (see Table~\ref{obstab}), while the MOJAVE survey uses VLBA-only baselines at 15~GHz
\footnote{While the exact comparison is declination dependent,
a typical naturally-weighted beam from an RDV observation of a far northern source is about
10\% larger than a typical naturally-weighted beam from a MOJAVE observation of the same source.}. 
The median RDV survey beam size taken over all beam major and minor axes from all 2753
images is 0.9~mas, which corresponds to a linear size of about 7~pc at $z=1$.

The RDV experiments occur roughly every two months, so that commonly observed sources
have an observing cadence of about six times per year.
Of order 100 sources are observed in a single 24-hour experiment, for an
average time on source per experiment of about 15 minutes.
This time on source is divided into scans of a minute to a few minutes in length
that are spread throughout the 24-hour observing period.
A typical observation consisting of 15 minutes on source with from 10 to 20 antennas yields 
rms noise levels for images of typical mid-latitude sources of about 1~mJy~beam$^{-1}$.
Only right circular polarization is recorded, so linear polarization intensity
and the electric vector position angle are not available from these observations.

For this paper, we have used the complete series of 42 RDV experiments conducted through the end of 2003
(RDVs 1 to 42), plus 8 similar geodetic VLBI experiments that were conducted on the VLBA before the RDV series began.
This yields a total of 50 VLBI experiments observed over a ten-year time baseline from 1994 to 2003.
These 50 VLBI experiments are summarized in Table~\ref{obstab}.
The results from 19 of these 50 VLBI experiments formed the sample used in
Paper I; the present paper then adds an additional 31 VLBI experiments to the 19 considered in Paper I.
Most of these 31 new VLBI experiments had not been previously
imaged, and were imaged by the authors for the purposes of this and other projects
(e.g., Pushkarev \& Kovalev 2012).
The RDV experiment series has continued to observe every two
months through the present (and is currently up to RDV~93), but the epochs
after RDV~42 are not fully imaged and model fit.

For the analysis in this paper we selected all sources that were observed at 20 or more epochs 
over the series of 50 VLBI experiments listed in Table~\ref{obstab}.
This yielded a sample of 72 sources, from which we excluded two that were 
below $-50\arcdeg$ declination (0208$-$512 and 1815$-$553) and so too far south to be adequately imaged with
the available antennas. 
Two other sources (0238$-$084 and 1404+286) had 8 GHz structures that were two-sided
(hindering identification of the core), and/or
so smooth and complex at 8 GHz that we were not able to reliably follow components from epoch to epoch.
The remaining 68 sources in the final RDV sample are listed in Table~\ref{sources}.
The total number of observations of all 68 sources is 2753, and there is
a median of 43 epochs of observation per source.

In terms of optical identifications, the sources in the RDV sample are predominantly quasars. 
From the optical class identifications by V\'{e}ron-Cetty \& V\'{e}ron (2010),
56 sources are quasars, 7 are BL~Lac objects, 4 are galaxies, and 1 is unidentified.
Approximately half of the sources are also members of the MOJAVE survey: comparing
Table~\ref{sources} with the MOJAVE source list from Lister et al. (2009a), we
find that 37 of the 68 sources are in the MOJAVE survey, while 31 are not.
Particularly notable is the inclusion in the RDV sample of a substantial number of southern
sources that are not present in MOJAVE, due to the inclusion of southern
hemisphere telescopes in the RDV experiments (see Table~\ref{obstab}). 
Of these southern sources, six are also being observed by the TANAMI project (Ojha et al. 2010)
of southern hemisphere VLBI observations.
About 60\% of the sources in the RDV sample (43 of 68) are detected by the
{\em Fermi} LAT gamma-ray telescope after the first 24 months of scientific operation
(Ackermann et al. 2011), these LAT sources are
noted in Table~\ref{sources}.

The source list in Table~\ref{sources} is somewhat different than the corresponding source
list from Paper I, because of the different selection criteria. Compared with the source
list from Paper I, 24 sources have been dropped for not meeting the selection criteria
of the current study, and 5 sources (0235+164, 1448+762, 1642+690, 1657$-$261, and 2223$-$052)
have been added for this paper.
A few of the sources in Table~\ref{sources} did not have measurable jet kinematics.
Of the 68 sources in Table~\ref{sources}, two (0235+164 and 2052$-$474)
were very compact and were modeled as only a single core component
at almost all epochs, and so had no measurable proper motions.
In addition, one source (1657$-$261) did not have a measured redshift, so that its proper motion
could not be converted to an apparent speed. This yields a total of 66 sources with
measured proper motions, and 65 sources with measured apparent speeds.

\begin{figure*}[!t]
\begin{center}
\includegraphics[scale=0.50]{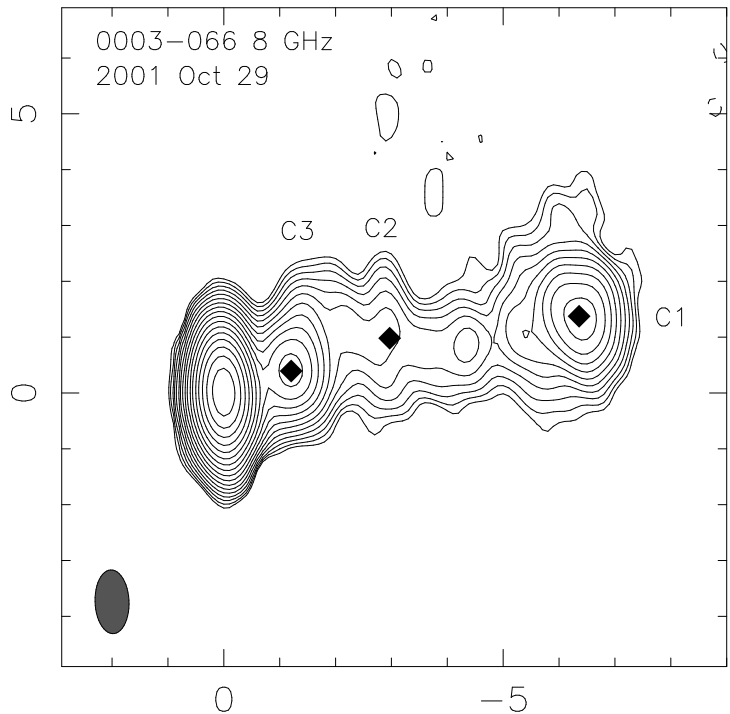}
\end{center}
\caption{
Image of 0003$-$066 from RDV30 on 2001 Oct 29. The axes are labeled in milliarcseconds (mas).
The lowest contour is set to three times the rms noise level of 0.9~mJy~beam$^{-1}$, and each
successive contour is a factor of square root of two higher.
The peak flux density is 0.96~Jy~beam$^{-1}$.
The beam size is 1.14 by 0.61 mas at a position angle of 2.3$\arcdeg$, and is shown at the bottom left of the image.
The three filled diamonds indicate features in the jet that are followed
by Gaussian model fitting. Parameters of the Gaussian models are given in Table~\ref{mfit}.}
\end{figure*}

\section{Imaging and Model Fitting}
\label{models}
All VLBI experiments were calibrated and fringe-fit using 
standard routines from the NRAO AIPS software package;
and self-calibration, imaging, and model fitting were done in the Caltech DIFMAP
software package. Calibration and imaging procedures 
for these RDV experiments have been described in detail in Paper I
and in Pushkarev \& Kovalev (2012).
We show an example image of the source 0003$-$066 from a middle epoch in Figure~1.
Paper I also displayed a set of sample images from these experiments for cases of
good, adequate, and poor $(u,v)$-plane coverage. 
A subset of images is also displayed in the article references given in the final column of Table~\ref{obstab},
and in Pushkarev \& Kovalev (2012). 
All of the images used for this paper are publicly available online, and the online image reference
for the various RDV experiments is given in the final column of Table~\ref{obstab}
\footnote{The amount of associated material available for each image and the specific file formats
depend on the specific archive where it is located. As given in Table~\ref{obstab},
these archives are: the Radio Reference Frame Image Database (http://rorf.usno.navy.mil/RRFID/),
the Bordeaux VLBI Image Database (http://www.obs.u-bordeaux1.fr/BVID/), and
the Astrogeo Center (http://astrogeo.org/vlbi\_images/).}.
We note that the total number of new VLBI images produced 
from the 31 new RDV experiments in Table~\ref{obstab} by the
authors was much higher (approximately 6000), than the total number of images used
in this paper (2753), because the 2~GHz images and the images of
sources observed at less than 20 epochs over these 50 experiments are not used in the present paper.
However, these additional presently unused images are also available 
in the online archives listed in Table~\ref{obstab}, and so they are now 
available to the community for any subsequent studies (e.g., of the frequency-dependent
shift of the core position).

After self-calibration and imaging,
Gaussian models were fit to the calibrated visibilities associated with each image
using the {\em modelfit} task in DIFMAP.
Such Gaussian models provide a concise mathematical description of the
location and properties of the various jet components in each image.
Model fitting in the visibility plane versus the image plane allows sub-beam resolution
to be attained in cases of high signal-to-noise; see also the discussion of visibility-plane
versus image-plane model fitting by L09 and the visibility-plane resolution limit
by Kovalev et al. (2005). 

Our model-fitting procedure was described in detail in Paper I, but we review
and summarize it here. The exact number of Gaussian components used, and the choice between
elliptical Gaussians or circular Gaussians, can be subjective, but
in our case was motivated by consideration of the simplicity of the resulting model,
and consistency of the model fits for a given source from epoch to epoch.
Elliptical components were used sparingly, and only to represent the core
or a bright jet component when the residuals remaining from a circular Gaussian
fit were so large as to hinder further model fitting using the residual map.
To maintain consistency from epoch to epoch, the final model from a previous or later
epoch was often used as the starting guess for the epoch under consideration.
Occasionally, some model fits from Paper I were redone for consistency with later epochs.
A subset of sources was modeled independently by multiple authors to check
the consistency of the results, and consistent kinematic results 
were obtained by the different modelers in the vast majority ($\sim 95\%$) of cases.
However, despite all precautions, VLBI model fits are not unique, and represent 
only one mathematically possible deconvolution of complex source structure
(see, e.g., the comparison of RDV and 2~cm Survey results from Paper I).

The complete results from the Gaussian model fitting are presented in machine-readable form in Table~\ref{mfit}.
Table~\ref{mfit} contains a total of 8571 Gaussian components fit to the 2753 images,
or an average of about 3 components per image (a core and two jet components).
Columns (2)--(8) of Table~\ref{mfit} correspond directly to the DIFMAP {\em modelfit} results,
and are suitable for reading directly into DIFMAP with the {\em rmodel} command.
Positions of components in Table~\ref{mfit} have not been shifted to place the core at the origin, so that the positions
in Table~\ref{mfit} correspond directly to positions on the publicly available images.
Note that the flux density measurements in column (2) are not highly accurate in the
case of relatively closely spaced components, where the division of flux density
between the components can be ambiguous.

\begin{table*}[!t]
\begin{center}
\caption{Gaussian Models}
\label{mfit}
{\small \begin{tabular}{l r r r r r r c c c r r r} \tableline \tableline \\[-5pt]
& \multicolumn{1}{c}{$S$} & \multicolumn{1}{c}{$r$} & \multicolumn{1}{c}{PA} & \multicolumn{1}{c}{$a$} & & \multicolumn{1}{c}{PA$_{maj}$}
& & & & \multicolumn{1}{c}{$a_{beam}$} & \multicolumn{1}{c}{$b_{beam}$} & \multicolumn{1}{c}{$\theta_{beam}$} \\
\multicolumn{1}{c}{Source} & \multicolumn{1}{c}{(Jy)} & \multicolumn{1}{c}{(mas)} & \multicolumn{1}{c}{(deg)}
& \multicolumn{1}{c}{(mas)} & \multicolumn{1}{c}{$(b/a)$} & \multicolumn{1}{c}{(deg)} & Type & Epoch &
\multicolumn{1}{c}{Comp.} & \multicolumn{1}{c}{(mas)} & \multicolumn{1}{c}{(mas)} & \multicolumn{1}{c}{(deg)} \\
\multicolumn{1}{c}{(1)} & \multicolumn{1}{c}{(2)} & \multicolumn{1}{c}{(3)} & \multicolumn{1}{c}{(4)} &
\multicolumn{1}{c}{(5)} & \multicolumn{1}{c}{(6)} & \multicolumn{1}{c}{(7)} & \multicolumn{1}{c}{(8)} &
\multicolumn{1}{c}{(9)} & \multicolumn{1}{c}{(10)} & \multicolumn{1}{c}{(11)} & \multicolumn{1}{c}{(12)} &
\multicolumn{1}{c}{(13)} \\ \tableline \\[-5pt]
0003$-$066 & 1.599 & 0.079 &   148.3 & 0.633 & 0.387 & $-$16.3 & 1 & 1995.78 &  0 & 2.29 & 0.95 & $-$1.1 \\
           & 0.645 & 1.040 & $-$60.5 & 1.384 & 1.000 &     0.0 & 1 &         & 99 &      &      &        \\
           & 0.156 & 5.145 & $-$74.5 & 3.222 & 1.000 &     0.0 & 1 &         &  1 &      &      &        \\
           & 1.209 & 0.032 &   114.2 & 0.529 & 0.000 &    21.2 & 1 & 1997.08 &  0 & 2.03 & 0.75 & $-$5.8 \\
           & 0.225 & 0.786 & $-$48.9 & 0.520 & 1.000 &     0.0 & 1 &         &  3 &      &      &        \\
           & 0.194 & 2.131 & $-$71.1 & 1.416 & 1.000 &     0.0 & 1 &         &  2 &      &      &        \\
           & 0.083 & 5.586 & $-$75.2 & 2.455 & 1.000 &     0.0 & 1 &         &  1 &      &      &        \\ \tableline \\[-5pt]
\end{tabular}}
\end{center}
{\bf Notes.} Table~\ref{mfit} is published in its entirety in machine-readable form
in the electronic edition of the journal. A portion is shown here for guidance regarding
its form and content.
Columns are as follows: Column 1: B1950 source name. Column 2: flux density in Janskys.
Columns 3 and 4: $r$ and PA (Position Angle) are the polar coordinates of the
Gaussian center. Position Angle is measured from north through east.
Columns 5--7: $a$ and $b$ are the FWHM of the major and minor axes of the Gaussian,
and PA$_{maj}$ is the position angle of the major axis.
$(b/a)$ and PA$_{maj}$ are set to 1.0 and 0.0 for circular components, respectively.
Column 8: component type for the DIFMAP `modelfit' command. Type 1 indicates a Gaussian component.
Type 0 indicates a delta function. Column 9: epoch of observation.
Column 10: component identification. Component `0' indicates the presumed core. Other components are numbered from 1 to 11,
from the outermost component inward. A component ID of `99' indicates a flagged component not used in the analysis.
Columns 11--13: $a_{beam}$, $b_{beam}$, and $\theta_{beam}$ are the major axis FWHM,
minor axis FWHM, and position angle of the major axis of the
naturally weighted restoring beam (uvweight 0,$-$1~in DIFMAP).
\end{table*}

\begin{figure*}
\begin{center}
\includegraphics[angle=90,scale=0.45]{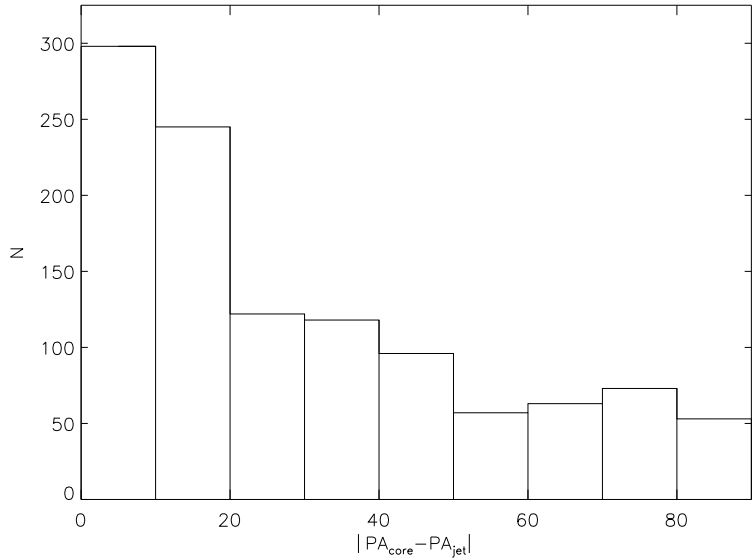}
\end{center}
\caption{
Histogram of the misalignment between the major axis position angle of elliptical core components and
the position angle of the closest downstream jet component.}
\end{figure*}

After model fitting all epochs for a given source, jet components needed to be
cross-identified from epoch to epoch in order to study their kinematics. 
This component identification is given in column (10)
of Table~\ref{mfit}. A component identification of `0' indicates the presumed core. 
Other components are numbered from 1 to 11,
from the outermost component inward. A component ID of `99' indicates an unidentified component not used in the analysis.
We identify the core in each source as
the compact component at the end of the one-sided jet structure --- often,
but not always, it is also the brightest component.
As noted above, we excluded sources known to show two-sided VLBI structures at these scales.
Jet component identifications were made based on consistency in flux, radial position, position angle, and size
from epoch to epoch. With the large number of epochs per source used here, and their close time spacing, such
identifications are expected to be robust.
In cases where a model component could not be directly identified with model components seen at other
epochs, it was given an identification of `99' in Table~\ref{mfit}
to flag it as a model component not used in the analysis.
This typically happened when a somewhat lower resolution image blended together what was seen as two
separate components in other model fits (a `merger'), or when a
low-dynamic-range component was detected in only a few images with a poorly constrained position.

Some overall statistics for the fits in Table~\ref{mfit} are given below.
Of the total of 8571 components, 2753 are core components
while 5818 are jet components. About 84\% of the components (7205) are circular Gaussians, 
while about 16\% (1366) are elliptical Gaussians. Of the 1366 elliptical Gaussians,
1277 (about 93\%) have been used to represent core components, only 89 (about 7\%)
are used to represent jet components. 
This means that about 46\% of the core components are represented by ellipses,
while only about 2\% of the jet components are represented by ellipses.
When the core is modeled by an elliptical component, then the position angle of the major axis
tends to align with the jet position angle, as shown in Figure~2.
This figure shows a histogram of the difference between the major axis position
angle of elliptical core components and the position angle of the closest downstream jet
component, for the 1125 elliptical core components with a downstream jet component modeled at
the same epoch. The excess at small misaglignments is clear, suggesting that the elliptical core
components are modeling the beginning of the elongated jet structure. A similar result was found
for elliptical core components by Kovalev et al. (2005) for the 2~cm Survey.

Of the 5818 jet components in Table~\ref{mfit}, 5069 (about 87\%) have been identified with a specific
component identification in column (10) of Table~\ref{mfit}, while
749 (about 13\%) are unidentified (an ID of `99' in column (10)).
There are a total of 225 unique jet components  with at least four epochs
of observation (which we require for the kinematic analysis) identified in 66 sources in Table~\ref{mfit}.
With 5069 total observations of identified jet components, this 
gives a mean of about 23 observations of each of the 225 unique components.

Table~\ref{comptab} shows a comparison of some of the important properties of the
RDV survey compared with the MOJAVE survey as published by Lister et al. (2009a, 2009b).
While the total image count is similar between the two surveys, the MOJAVE survey
has studied about twice the number of sources, and has about twice the total number
of components as the RDV survey. However, the RDV survey has a higher
mean number of images per source, and a lower mean time gap between the images, both
of which are useful for studying the jet kinematics including acceleration analysis.

\begin{table*}
\begin{center}
\caption{MOJAVE and RDV survey comparison}
\label{comptab}
{\small \begin{tabular}{c c c} \tableline \tableline \\[-5pt]
Property & MOJAVE$^{a}$ & RDV \\ \tableline \\[-5pt]
Total number of sources                          & 135  & 68   \\
Number of sources with proper motion             & 127  & 66   \\
Total number of images                           & 2424 & 2753 \\
%Median number of images per source              & 15   & 43   \\
%Minimum number of images per source             & 5    & 20   \\
Total time span (years)                          & 13   & 10   \\
Mean time gap between images of a source (years) & 0.7  & 0.2  \\
Number of jet components                         & 526  & 225  \\ \tableline \\[-5pt]
\end{tabular}}
\end{center}
{\bf Notes.}\\
$a$: Using published images and kinematics from the MOJAVE survey from Lister et al. (2009a; 2009b).
\end{table*}

\begin{figure*}
\begin{center}
\includegraphics[angle=90,scale=0.45]{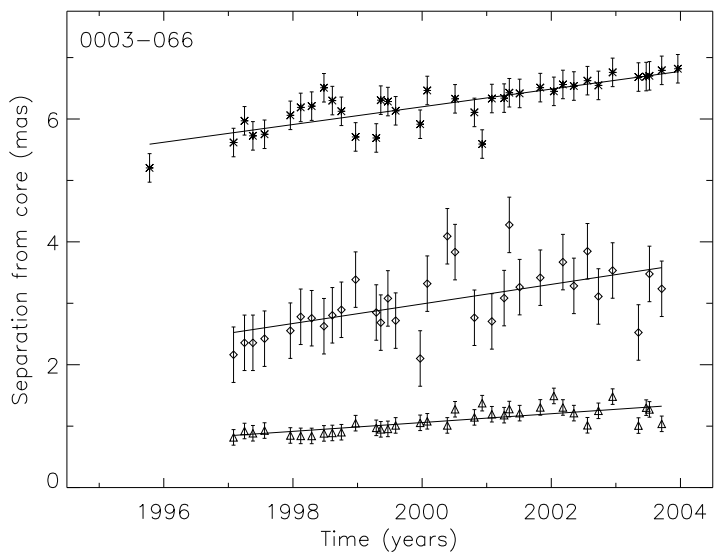}
\end{center}
\caption{Angular separations from the core of Gaussian model component centers
as a function of time. The straight lines are the least-squares fits to radial motion at constant speed
for components detected at four or more epochs.
For each source, asterisks are used to represent component 1, diamonds component 2, triangles component 3,
squares component 4, x's component 5, and circles component 6. 
This cycle of plotting symbols then repeats starting at component 7.
Unidentified components from Table~\ref{mfit} are not plotted.
Some error bars are smaller than the plotting symbols.
The B1950 source name is given at the top left of each panel.
The complete figure set (66 images, one for each source in Table~\ref{speedtab}) is available in the online journal.}
\end{figure*}

\section{Apparent Speeds}
\label{speeds}
\subsection{Fitting Methods} 
\label{fit}
We performed two types of fits to each of the 225 components' position versus time data in order to
study the jet kinematics. The first fit was a linear fit to $r$ versus $t$, with two free parameters.
These fits yield a proper motion as the rate of change in $r$, equal to the slope
of the best fit line on a separation versus time
plot, and they allow for a direct comparison with the apparent speed measurements from Paper I.

The second type of fit was a second-order polynomial fit to $x(t)$ and $y(t)$ separately for each component,
and provides information on the apparent acceleration of each component.
The nonlinear fitting method used here is identical to that described by Homan et al. (2001), and
subsequently used in the acceleration measurements for the MOJAVE survey by L09 and H09.
We use the same parametrization for the fits as Homan et al. (2001) and H09, allowing for
direct comparison of our results with the MOJAVE acceleration results. We summarize this nonlinear fitting
method below.

In these nonlinear fits there are three
fit parameters for both $x(t)$ and $y(t)$, for a total of six free parameters for each component.
The vector proper motion for each fit is defined
from the average proper motions in the $x$ and $y$ directions ($\mu_{x}$ and $\mu_{y}$),
or equivalently the proper motion vector at the mid-time
$t_{mid}=(t_{i}+t_{f})/2$, where $t_{i}$ and $t_{f}$ are the times of the initial and final
observation of the component, respectively.
The magnitude of this average proper motion vector is denoted $\mu$, and the direction of this
average proper motion vector is denoted $\phi$. The magnitude $\mu$ is then converted to the
observed apparent speed, $\beta_\mathrm{app}$. The quantity $|PA-\phi|$ gives the difference between the
weighted mean position angle $PA$ of the component and the direction of its average apparent velocity vector.

The apparent angular acceleration is computed from $\ddot{x}$ and $\ddot{y}$  ($\dot{\mu}_{x}$ and $\dot{\mu}_{y}$).
This apparent angular acceleration is resolved into two components parallel and perpendicular
to the nominal velocity direction $\phi$. These two components are denoted by
$\dot{\mu}_{\parallel}$ and $\dot{\mu}_{\perp}$, and they represent the apparent angular acceleration
due to changes in apparent speed and due to changes in direction, respectively.
To more easily compare apparent accelerations among high and low speed jets,
the relative parallel acceleration is defined as
\begin{equation}
\label{relpar}
\dot{\eta}_{\parallel}\equiv\dot{\beta}_\mathrm{\parallel app}/\beta_\mathrm{app}=(1+z)\dot{\mu}_{\parallel}/\mu, 
\end{equation}
and the relative perpendicular acceleration is defined as
\begin{equation}
\label{relperp}
\dot{\eta}_{\perp}\equiv\dot{\beta}_\mathrm{\perp app}/\beta_\mathrm{app}=(1+z)\dot{\mu}_{\perp}/\mu.
\end{equation}
Thus a component with a relative parallel acceleration of 0.1, for example, has an apparent speed that
is increasing at a rate of 10\% per year; note also that relative acceleration defined in this
fashion has dimensions of inverse time. These fits assume a constant acceleration over
the time span of observation of the component, although we cannot exclude the possibility that more
complicated and time variable accelerations may also act.

\begin{table*}
\begin{center}
\caption{Apparent Component Speeds from Linear Fits}
\label{speedtab}
{\scriptsize \begin{tabular}{l c r r r r c} \tableline \tableline \\[-5pt]
& & \multicolumn{1}{c}{$\langle{S}\rangle$} & \multicolumn{1}{c}{$\langle{r}\rangle$} & \multicolumn{1}{c}{$\mu$} 
& & \multicolumn{1}{c}{$t_{0}^{a}$} \\
\multicolumn{1}{c}{Source} & Comp. & \multicolumn{1}{c}{(Jy)} & \multicolumn{1}{c}{(mas)} 
& \multicolumn{1}{c}{($\mu as$ yr$^{-1}$)} & \multicolumn{1}{c}{$\beta_\mathrm{app}$} & \multicolumn{1}{c}{(yr)} \\
\multicolumn{1}{c}{(1)} & \multicolumn{1}{c}{(2)} & \multicolumn{1}{c}{(3)} & \multicolumn{1}{c}{(4)} & \multicolumn{1}{c}{(5)} &
\multicolumn{1}{c}{(6)} & \multicolumn{1}{c}{(7)} \\ \tableline \\[-5pt]
0003$-$066 &  1 &  0.138 & $ 6.25\pm 0.04$ & $ 144\pm  18$ & $  3.1\pm 0.4$ & $1957.02\pm   5.36$ \\
         &  2 &  0.158 & $ 3.03\pm 0.08$ & $ 159\pm  40$ & $  3.5\pm 0.9$ & $1981.23\pm   5.10$ \\
         &  3 &  0.199 & $ 1.09\pm 0.02$ & $  72\pm  11$ & $  1.6\pm 0.2$ & $1985.24\pm   2.31$ \\
0014+813 &  1 &  0.006 & $ 9.39\pm 0.07$ & $  45\pm  34$ & $  4.9\pm 3.6$ & ... \\  
         &  2 &  0.010 & $ 5.36\pm 0.06$ & $  28\pm  30$ & $  3.0\pm 3.2$ & ... \\
         &  3 &  0.113 & $ 0.67\pm 0.01$ & $  -5\pm   4$ & $ -0.6\pm 0.4$ & ... \\
0048$-$097 &  1 &  0.008 & $ 3.09\pm 0.36$ & $ -79\pm 183$ & $ -2.9\pm 6.7$ & ... \\
         &  3 &  0.022 & $ 0.60\pm 0.04$ & $  32\pm  53$ & $  1.2\pm 1.9$ & ... \\
         &  4 &  0.063 & $ 0.86\pm 0.09$ & $ 339\pm 191$ & $ 12.4\pm 7.0$ & ... \\
         &  5 &  0.137 & $ 0.67\pm 0.06$ & $ 178\pm  70$ & $  6.5\pm 2.6$ & ... \\
0059+581 &  1 &  0.026 & $ 2.51\pm 0.04$ & $ -24\pm  20$ & $ -0.9\pm 0.7$ & ... \\
         &  2 &  0.036 & $ 1.35\pm 0.03$ & $  43\pm  16$ & $  1.6\pm 0.6$ & ... \\
         &  3 &  0.091 & $ 0.65\pm 0.02$ & $  -2\pm  16$ & $ -0.1\pm 0.6$ & ... \\
         &  4 &  0.102 & $ 0.57\pm 0.01$ & $ 103\pm   8$ & $  3.8\pm 0.3$ & $1995.13\pm   0.42$ \\
         &  5 &  1.666 & $ 0.16\pm 0.01$ & $  67\pm  59$ & $  2.5\pm 2.2$ & ... \\
0104$-$408 &  1 &  0.076 & $ 2.36\pm 0.15$ & $  37\pm  69$ & $  1.3\pm 2.3$ & ... \\
         &  2 &  0.218 & $ 0.53\pm 0.05$ & $  78\pm  26$ & $  2.6\pm 0.9$ & $1994.13\pm   2.56$ \\
0119+041 &  1 &  0.314 & $ 0.73\pm 0.02$ & $  39\pm   9$ & $  1.5\pm 0.3$ & $1981.16\pm   4.47$ \\
0119+115 &  1 &  0.023 & $20.33\pm 0.11$ & $ 284\pm 143$ & $  9.5\pm 4.8$ & ... \\
         &  2 &  0.009 & $14.24\pm 0.08$ & $  61\pm  93$ & $  2.0\pm 3.1$ & ... \\
         &  3 &  0.105 & $ 1.64\pm 0.05$ & $ 242\pm  23$ & $  8.1\pm 0.8$ & $1993.53\pm   0.65$ \\
         &  4 &  0.247 & $ 1.39\pm 0.04$ & $ 317\pm  43$ & $ 10.6\pm 1.4$ & $1998.15\pm   0.61$ \\
0133+476 &  1 &  0.052 & $ 2.61\pm 0.04$ & $  -9\pm  16$ & $ -0.4\pm 0.8$ & ... \\
         &  2 &  0.087 & $ 1.07\pm 0.03$ & $  58\pm  18$ & $  2.7\pm 0.9$ & $1982.36\pm   6.50$ \\
         &  3 &  0.176 & $ 0.53\pm 0.02$ & $  31\pm   9$ & $  1.5\pm 0.4$ & $1983.58\pm   5.38$ \\
0201+113 &  1 &  0.034 & $ 1.46\pm 0.04$ & $  19\pm  19$ & $  2.1\pm 2.1$ & ... \\
         &  2 &  0.114 & $ 1.21\pm 0.01$ & $  31\pm   6$ & $  3.4\pm 0.7$ & $1960.65\pm   8.65$ \\
0202+149 &  1 &  0.215 & $ 4.81\pm 0.02$ & $-113\pm   9$ & $ -2.8\pm 0.2$ & ... \\
         &  2 &  0.462 & $ 0.60\pm 0.02$ & $  98\pm  17$ & $  2.5\pm 0.4$ & $1993.01\pm   1.07$ \\
         &  3 &  1.234 & $ 0.52\pm 0.05$ & $ 101\pm  71$ & $  2.5\pm 1.8$ & ... \\
0229+131 &  1 &  0.010 & $ 7.46\pm 0.24$ & $ 132\pm 116$ & $ 11.2\pm 9.8$ & ... \\
         &  2 &  0.010 & $ 3.18\pm 0.08$ & $-193\pm 109$ & $-16.4\pm 9.2$ & ... \\
         &  3 &  0.043 & $ 2.78\pm 0.11$ & $ 327\pm  35$ & $ 27.8\pm 3.0$ & $1990.23\pm   0.93$ \\
         &  4 &  0.065 & $ 1.77\pm 0.04$ & $  55\pm  26$ & $  4.6\pm 2.2$ & ... \\
         &  5 &  0.151 & $ 0.51\pm 0.02$ & $  11\pm  13$ & $  1.0\pm 1.1$ & ... \\
         &  6 &  0.515 & $ 0.24\pm 0.02$ & $-100\pm  47$ & $ -8.5\pm 4.0$ & ... \\
0234+285 &  1 &  0.140 & $ 6.03\pm 0.08$ & $ 445\pm  50$ & $ 26.9\pm 3.1$ & $1985.34\pm   1.56$ \\
         &  2 &  0.211 & $ 4.01\pm 0.03$ & $ 298\pm  12$ & $ 18.0\pm 0.8$ & $1986.69\pm   0.57$ \\
         &  3 &  0.093 & $ 1.03\pm 0.04$ & $  48\pm  25$ & $  2.9\pm 1.5$ & ... \\
         &  4 &  0.196 & $ 0.45\pm 0.01$ & $  31\pm  14$ & $  1.9\pm 0.8$ & ... \\
         &  5 &  0.816 & $ 0.33\pm 0.02$ & $ 117\pm  39$ & $  7.1\pm 2.4$ & $2000.55\pm   1.06$ \\
0336$-$019 &  1 &  0.024 & $ 5.97\pm 0.22$ & $  14\pm 136$ & $  0.7\pm 6.3$ & ... \\
         &  2 &  0.025 & $ 3.63\pm 0.09$ & $  85\pm  95$ & $  3.9\pm 4.4$ & ... \\
         &  3 &  0.082 & $ 2.94\pm 0.05$ & $ 187\pm  22$ & $  8.7\pm 1.0$ & $1984.39\pm   1.91$ \\
         &  4 &  0.270 & $ 1.50\pm 0.03$ & $ 123\pm  15$ & $  5.7\pm 0.7$ & $1987.30\pm   1.52$ \\
         &  5 &  0.444 & $ 0.93\pm 0.02$ & $ 277\pm  14$ & $ 12.9\pm 0.7$ & $1998.40\pm   0.18$ \\
         &  6 &  0.762 & $ 0.35\pm 0.04$ & $  74\pm  41$ & $  3.4\pm 1.9$ & ... \\
0402$-$362 &  1 &  0.044 & $ 2.76\pm 0.11$ & $ 174\pm  69$ & $ 11.8\pm 4.6$ & ... \\
         &  2 &  0.304 & $ 0.79\pm 0.04$ & $ 109\pm  20$ & $  7.3\pm 1.4$ & $1993.05\pm   1.41$ \\
0430+052 &  1 &  0.272 & $ 5.57\pm 0.06$ & $1455\pm  83$ & $  2.9\pm 0.2$ & $1994.40\pm   0.22$ \\
         &  2 &  0.801 & $ 2.48\pm 0.04$ & $1728\pm  52$ & $  3.5\pm 0.1$ & $1996.80\pm   0.04$ \\
         &  3 &  0.426 & $ 2.27\pm 0.08$ & $1835\pm 181$ & $  3.7\pm 0.4$ & $1997.40\pm   0.12$ \\
         &  4 &  0.131 & $ 8.76\pm 0.12$ & $2142\pm 113$ & $  4.3\pm 0.2$ & $1998.07\pm   0.22$ \\
         &  5 &  0.373 & $ 5.31\pm 0.05$ & $1903\pm  62$ & $  3.8\pm 0.1$ & $1999.48\pm   0.09$ \\
0454$-$234 &  1 &  0.151 & $ 0.85\pm 0.05$ & $ -12\pm  21$ & $ -0.7\pm 1.1$ & ... \\
0458$-$020 &  1 &  0.070 & $ 4.57\pm 0.07$ & $ 296\pm  35$ & $ 26.6\pm 3.1$ & $1984.97\pm   1.85$ \\
         &  2 &  0.121 & $ 1.78\pm 0.05$ & $ 198\pm  23$ & $ 17.8\pm 2.1$ & $1991.21\pm   1.06$ \\
0528+134 &  1 &  0.090 & $ 3.66\pm 0.04$ & $  75\pm  19$ & $  6.4\pm 1.6$ & $1951.52\pm  13.05$ \\
         &  2 &  0.266 & $ 1.43\pm 0.02$ & $ 125\pm  11$ & $ 10.7\pm 0.9$ & $1989.09\pm   0.99$ \\
         &  3 &  0.564 & $ 0.46\pm 0.01$ & $   5\pm   7$ & $  0.4\pm 0.6$ & ... \\
         &  4 &  1.166 & $ 0.23\pm 0.04$ & $ 201\pm 133$ & $ 17.1\pm11.3$ & ... \\ 
0537$-$441 &  1 &  0.186 & $ 2.50\pm 0.07$ & $  -3\pm  27$ & $ -0.2\pm 1.3$ & ... \\
         &  2 &  0.621 & $ 0.97\pm 0.14$ & $  34\pm 262$ & $  1.6\pm12.7$ & ... \\
0552+398 &  1 &  1.076 & $ 0.65\pm 0.00$ & $  -5\pm   1$ & $ -0.4\pm 0.1$ & ... \\
0642+449 &  1 &  0.012 & $ 3.41\pm 0.06$ & $  -5\pm  27$ & $ -0.5\pm 2.9$ & ... \\
         &  3 &  0.708 & $ 0.28\pm 0.01$ & $  -8\pm   3$ & $ -0.9\pm 0.3$ & ... \\
0727$-$115 &  1 &  0.181 & $ 2.21\pm 0.05$ & $  37\pm  20$ & $  2.7\pm 1.5$ & ... \\
         &  2 &  0.253 & $ 0.70\pm 0.08$ & $  66\pm  69$ & $  4.8\pm 5.0$ & ... \\
         &  3 &  1.028 & $ 0.27\pm 0.02$ & $   7\pm  10$ & $  0.5\pm 0.7$ & ... \\
0804+499 &  1 &  0.011 & $ 2.59\pm 0.08$ & $  69\pm  60$ & $  4.7\pm 4.1$ & ... \\
         &  2 &  0.080 & $ 1.12\pm 0.02$ & $  55\pm   9$ & $  3.7\pm 0.6$ & $1979.94\pm   3.27$ \\
         &  3 &  0.070 & $ 0.30\pm 0.02$ & $ -19\pm  12$ & $ -1.3\pm 0.8$ & ... \\
0823+033 &  1 &  0.015 & $ 9.80\pm 0.21$ & $ 655\pm 235$ & $ 19.9\pm 7.1$ & ... \\
         &  2 &  0.038 & $ 4.05\pm 0.08$ & $ 122\pm  63$ & $  3.7\pm 1.9$ & ... \\
         &  3 &  0.090 & $ 2.62\pm 0.03$ & $ -13\pm  14$ & $ -0.4\pm 0.4$ & ... \\
         &  4 &  0.164 & $ 1.02\pm 0.02$ & $  59\pm  11$ & $  1.8\pm 0.3$ & $1982.21\pm   3.43$ \\
         &  5 &  0.106 & $ 0.59\pm 0.03$ & $ 131\pm  38$ & $  4.0\pm 1.1$ & $1997.89\pm   1.41$ \\
         &  6 &  0.304 & $ 0.33\pm 0.02$ & $ -58\pm  65$ & $ -1.8\pm 2.0$ & ... \\
\end{tabular}}
\end{center}
\end{table*}

\begin{table*}
\begin{center}
Table 5 (Continued) \\
{\scriptsize \begin{tabular}{l c r r r r c} \tableline \tableline \\[-5pt]
& & \multicolumn{1}{c}{$\langle{S}\rangle$} & \multicolumn{1}{c}{$\langle{r}\rangle$} & \multicolumn{1}{c}{$\mu$} 
& & \multicolumn{1}{c}{$t_{0}^{a}$} \\
\multicolumn{1}{c}{Source} & Comp. & \multicolumn{1}{c}{(Jy)} & \multicolumn{1}{c}{(mas)} 
& \multicolumn{1}{c}{($\mu as$ yr$^{-1}$)} & \multicolumn{1}{c}{$\beta_\mathrm{app}$} & \multicolumn{1}{c}{(yr)} \\ 
\multicolumn{1}{c}{(1)} & \multicolumn{1}{c}{(2)} & \multicolumn{1}{c}{(3)} & \multicolumn{1}{c}{(4)} & \multicolumn{1}{c}{(5)} &
\multicolumn{1}{c}{(6)} & \multicolumn{1}{c}{(7)} \\ \tableline \\[-5pt]
0851+202 &  1 &  0.023 & $ 3.61\pm 0.12$ & $  40\pm  84$ & $  0.8\pm 1.6$ & ... \\
         &  2 &  0.104 & $ 2.55\pm 0.07$ & $ 344\pm  27$ & $  6.7\pm 0.5$ & $1992.84\pm   0.58$ \\
         &  3 &  0.125 & $ 1.21\pm 0.03$ & $ 230\pm  32$ & $  4.5\pm 0.6$ & $1994.10\pm   0.75$ \\
         &  4 &  0.364 & $ 0.92\pm 0.03$ & $ 199\pm  16$ & $  3.9\pm 0.3$ & $1996.66\pm   0.38$ \\
         &  5 &  0.420 & $ 0.67\pm 0.03$ & $ 258\pm  25$ & $  5.0\pm 0.5$ & $1999.85\pm   0.25$ \\
0919$-$260 &  1 &  0.020 & $ 6.14\pm 0.09$ & $ 146\pm  43$ & $ 13.2\pm 3.9$ & $1957.71\pm  13.52$ \\
         &  2 &  0.038 & $ 2.02\pm 0.21$ & $-137\pm 172$ & $-12.4\pm15.5$ & ... \\
         &  3 &  0.118 & $ 1.39\pm 0.04$ & $ 121\pm  16$ & $ 10.9\pm 1.4$ & $1988.74\pm   1.57$ \\
0920$-$397 &  1 &  0.058 & $ 6.45\pm 0.32$ & $ 390\pm 199$ & $ 13.4\pm 6.9$ & ... \\
         &  2 &  0.094 & $ 4.08\pm 0.13$ & $ 256\pm  50$ & $  8.8\pm 1.7$ & $1985.15\pm   3.26$ \\
0923+392 &  1 &  1.103 & $ 2.52\pm 0.05$ & $   6\pm  44$ & $  0.2\pm 1.7$ & ... \\
         &  2 &  7.089 & $ 2.11\pm 0.02$ & $  52\pm   9$ & $  2.1\pm 0.3$ & $1959.29\pm   7.07$ \\
         &  3 &  1.851 & $ 1.46\pm 0.03$ & $ 164\pm  15$ & $  6.5\pm 0.6$ & $1991.62\pm   0.81$ \\
0955+476 &  1 &  0.014 & $ 1.10\pm 0.13$ & $ 212\pm 116$ & $ 17.0\pm 9.3$ & ... \\
         &  2 &  0.031 & $ 0.57\pm 0.05$ & $  84\pm  36$ & $  6.8\pm 2.9$ & ... \\
         &  3 &  0.110 & $ 0.21\pm 0.02$ & $  43\pm  13$ & $  3.5\pm 1.1$ & $1997.17\pm   1.61$ \\
1034$-$293 &  1 &  0.029 & $ 3.05\pm 0.23$ & $ 130\pm 151$ & $  2.5\pm 2.9$ & ... \\
         &  2 &  0.061 & $ 2.03\pm 0.04$ & $ 212\pm  20$ & $  4.1\pm 0.4$ & $1990.78\pm   0.92$ \\
         &  3 &  0.101 & $ 1.31\pm 0.04$ & $ 158\pm  18$ & $  3.1\pm 0.3$ & $1992.20\pm   0.96$ \\
         &  4 &  0.299 & $ 0.53\pm 0.02$ & $  45\pm  14$ & $  0.9\pm 0.3$ & $1989.67\pm   4.20$ \\
1044+719 &  1 &  0.034 & $ 0.81\pm 0.06$ & $ 176\pm  56$ & $ 10.2\pm 3.3$ & $1992.41\pm   1.63$ \\
         &  2 &  0.546 & $ 0.53\pm 0.01$ & $  79\pm   4$ & $  4.6\pm 0.3$ & $1994.60\pm   0.38$ \\
1101+384 &  1 &  0.016 & $ 5.38\pm 0.09$ & $ -79\pm  41$ & $ -0.2\pm 0.1$ & ... \\
         &  2 &  0.012 & $ 2.84\pm 0.06$ & $ -58\pm  27$ & $ -0.1\pm 0.1$ & ... \\
         &  3 &  0.027 & $ 1.46\pm 0.03$ & $  23\pm  15$ & $  0.05\pm 0.03$ & ... \\
         &  4 &  0.061 & $ 0.55\pm 0.02$ & $   2\pm   9$ & $  0.004\pm 0.019$ & ... \\
1124$-$186 &  1 &  0.011 & $ 2.70\pm 0.20$ & $ 500\pm 122$ & $ 27.3\pm 6.7$ & $1993.11\pm   1.40$ \\
         &  2 &  0.072 & $ 0.87\pm 0.05$ & $  41\pm  23$ & $  2.3\pm 1.3$ & ... \\
1128+385 &  1 &  0.078 & $ 0.84\pm 0.02$ & $ -18\pm   9$ & $ -1.4\pm 0.7$ & ... \\
         &  2 &  0.154 & $ 0.37\pm 0.01$ & $  11\pm   4$ & $  0.9\pm 0.3$ & ... \\
1144$-$379 &  1 &  0.046 & $ 3.75\pm 0.20$ & $ -80\pm 151$ & $ -4.4\pm 8.3$ & ... \\
         &  2 &  0.150 & $ 1.12\pm 0.10$ & $ 218\pm  51$ & $ 11.9\pm 2.8$ & $1994.14\pm   1.27$ \\
1145$-$071 &  1 &  0.103 & $ 2.20\pm 0.01$ & $  52\pm   6$ & $  3.4\pm 0.4$ & $1958.15\pm   4.55$ \\
1156+295 &  1 &  0.049 & $ 7.22\pm 0.27$ & $ -75\pm 207$ & $ -3.1\pm 8.5$ & ... \\
         &  2 &  0.080 & $ 6.29\pm 0.08$ & $ 636\pm  37$ & $ 26.2\pm 1.5$ & $1990.38\pm   0.58$ \\
         &  3 &  0.117 & $ 2.24\pm 0.11$ & $ 494\pm  49$ & $ 20.3\pm 2.0$ & $1995.61\pm   0.46$ \\
         &  5 &  0.209 & $ 0.56\pm 0.03$ & $  50\pm  19$ & $  2.1\pm 0.8$ & ... \\
1228+126 &  1 &  0.125 & $21.37\pm 0.29$ & $-114\pm 142$ & $ -0.03\pm 0.04$ &  ... \\
         &  2 &  0.096 & $11.54\pm 0.24$ & $  63\pm 120$ & $  0.02\pm 0.03$ & ... \\
         &  3 &  0.108 & $ 6.57\pm 0.07$ & $  88\pm  36$ & $  0.02\pm 0.01$ & ... \\
         &  4 &  0.140 & $ 2.97\pm 0.08$ & $  90\pm  39$ & $  0.02\pm 0.01$ & ... \\
         &  5 &  0.285 & $ 1.48\pm 0.04$ & $  -4\pm  22$ & $ -0.001\pm 0.006$ &  ... \\
         &  6 &  0.336 & $ 0.57\pm 0.02$ & $   6\pm  11$ & $  0.001\pm 0.003$ &  ... \\
1308+326 &  1 &  0.534 & $ 1.74\pm 0.02$ & $ 398\pm  12$ & $ 21.0\pm 0.6$ & $1995.96\pm   0.13$ \\
         &  2 &  0.319 & $ 1.49\pm 0.01$ & $ 486\pm  13$ & $ 25.6\pm 0.7$ & $1999.23\pm   0.08$ \\
1313$-$333 &  1 &  0.028 & $ 7.45\pm 0.07$ & $ 704\pm 115$ & $ 42.6\pm 7.0$ & $1989.58\pm   1.78$ \\
         &  2 &  0.049 & $ 2.02\pm 0.05$ & $ 383\pm  68$ & $ 23.2\pm 4.1$ & $1992.65\pm   0.97$ \\
         &  3 &  0.121 & $ 2.08\pm 0.07$ & $ 497\pm  34$ & $ 30.1\pm 2.1$ & $1996.27\pm   0.29$ \\
1334$-$127 &  1 &  0.175 & $ 2.78\pm 0.04$ & $  89\pm  28$ & $  2.9\pm 0.9$ & $1967.95\pm  10.71$ \\
         &  2 &  0.122 & $ 1.71\pm 0.04$ & $ 225\pm  14$ & $  7.2\pm 0.4$ & $1993.04\pm   0.46$ \\
         &  3 &  0.395 & $ 0.98\pm 0.04$ & $ 297\pm  44$ & $  9.5\pm 1.4$ & $1998.82\pm   0.50$ \\
1357+769 &  1 &  0.008 & $ 2.44\pm 0.20$ & $  79\pm  71$ & $  5.8\pm 5.2$ &  ... \\
         &  2 &  0.015 & $ 1.39\pm 0.05$ & $ 104\pm  33$ & $  7.6\pm 2.4$ & $1987.76\pm   4.74$ \\
         &  3 &  0.034 & $ 0.54\pm 0.02$ & $  93\pm  12$ & $  6.8\pm 0.9$ & $1992.07\pm   0.75$ \\
         &  4 &  0.106 & $ 0.21\pm 0.02$ & $   5\pm  16$ & $  0.4\pm 1.2$ & ... \\
1424$-$418 &  1 &  0.075 & $ 2.77\pm 0.09$ & $  15\pm  49$ & $  1.1\pm 3.5$ & ... \\
1448+762 &  1 &  0.014 & $ 1.57\pm 0.05$ & $  -4\pm  39$ & $ -0.2\pm 1.9$ & ... \\
         &  2 &  0.077 & $ 0.96\pm 0.04$ & $ -60\pm  33$ & $ -2.9\pm 1.6$ & ... \\
         &  3 &  0.144 & $ 0.54\pm 0.02$ & $ -87\pm  17$ & $ -4.2\pm 0.8$ & ... \\
1451$-$375 &  1 &  0.025 & $ 7.73\pm 0.44$ & $ 337\pm 212$ & $  6.5\pm 4.1$ & ... \\
         &  2 &  0.090 & $ 2.05\pm 0.16$ & $ 293\pm  72$ & $  5.7\pm 1.4$ & $1993.66\pm   1.82$ \\
1514$-$241 &  1 &  0.061 & $11.54\pm 0.49$ & $2964\pm1075$ & $  9.8\pm 3.6$ & ... \\
         &  2 &  0.058 & $ 7.58\pm 0.21$ & $1240\pm 285$ & $  4.1\pm 0.9$ & $1992.08\pm   1.48$ \\
         &  3 &  0.124 & $ 9.96\pm 0.19$ & $1651\pm 163$ & $  5.5\pm 0.5$ & $1995.66\pm   0.60$ \\
         &  4 &  0.115 & $ 5.07\pm 0.20$ & $1605\pm 391$ & $  5.3\pm 1.3$ & $1998.71\pm   0.82$ \\
         &  5 &  0.405 & $ 1.77\pm 0.08$ & $ 556\pm 160$ & $  1.8\pm 0.5$ & $1995.53\pm   1.00$ \\
1606+106 &  1 &  0.015 & $ 7.63\pm 0.06$ & $   4\pm  30$ & $  0.2\pm 1.8$ & ... \\
         &  2 &  0.035 & $ 2.48\pm 0.05$ & $  88\pm  22$ & $  5.4\pm 1.3$ & $1971.60\pm   7.41$ \\
         &  3 &  0.109 & $ 1.53\pm 0.02$ & $  15\pm  10$ & $  0.9\pm 0.6$ & ... \\
         &  4 &  0.250 & $ 0.53\pm 0.02$ & $ -30\pm   7$ & $ -1.9\pm 0.5$ & ... \\
1611+343 &  1 &  0.573 & $ 3.59\pm 0.03$ & $  60\pm  15$ & $  4.0\pm 1.0$ & $1940.33\pm  16.37$ \\
         &  2 &  0.204 & $ 4.03\pm 0.02$ & $ 108\pm  18$ & $  7.2\pm 1.2$ & $1963.53\pm   6.33$ \\
         &  3 &  0.350 & $ 2.84\pm 0.01$ & $  25\pm   4$ & $  1.7\pm 0.3$ & $1886.21\pm  19.59$ \\
         &  4 &  0.099 & $ 1.38\pm 0.03$ & $ 183\pm  18$ & $ 12.3\pm 1.2$ & $1992.17\pm   0.73$ \\
         &  5 &  0.386 & $ 0.73\pm 0.01$ & $ 214\pm   8$ & $ 14.3\pm 0.6$ & $1997.78\pm   0.13$ \\
         &  6 &  0.730 & $ 0.49\pm 0.03$ & $ 333\pm  55$ & $ 22.3\pm 3.7$ & $2001.49\pm   0.25$ \\
\end{tabular}}
\end{center}
\end{table*}

\begin{table*}
\begin{center}
Table 5 (Continued) \\
{\scriptsize \begin{tabular}{l c r r r r c} \tableline \tableline \\[-5pt]
& & \multicolumn{1}{c}{$\langle{S}\rangle$} & \multicolumn{1}{c}{$\langle{r}\rangle$} & \multicolumn{1}{c}{$\mu$} 
& & \multicolumn{1}{c}{$t_{0}^{a}$} \\
\multicolumn{1}{c}{Source} & Comp. & \multicolumn{1}{c}{(Jy)} & \multicolumn{1}{c}{(mas)} 
& \multicolumn{1}{c}{($\mu as$ yr$^{-1}$)} & \multicolumn{1}{c}{$\beta_\mathrm{app}$} & \multicolumn{1}{c}{(yr)} \\ 
\multicolumn{1}{c}{(1)} & \multicolumn{1}{c}{(2)} & \multicolumn{1}{c}{(3)} & \multicolumn{1}{c}{(4)} & \multicolumn{1}{c}{(5)} &
\multicolumn{1}{c}{(6)} & \multicolumn{1}{c}{(7)} \\ \tableline \\[-5pt]
1622$-$253 &  1 &  0.039 & $ 2.69\pm 0.09$ & $  43\pm  56$ & $  1.9\pm 2.4$ & ... \\
         &  2 &  0.131 & $ 1.08\pm 0.07$ & $ 184\pm  38$ & $  8.1\pm 1.7$ & $1994.30\pm   1.25$ \\
1638+398 &  1 &  0.083 & $ 0.56\pm 0.02$ & $  27\pm   9$ & $  2.0\pm 0.7$ & ... \\
         &  2 &  0.118 & $ 0.40\pm 0.01$ & $  30\pm   7$ & $  2.3\pm 0.5$ & $1987.95\pm   3.30$ \\
         &  3 &  0.238 & $ 0.17\pm 0.01$ & $  13\pm  10$ & $  1.0\pm 0.7$ & ... \\
1642+690 &  1 &  0.070 & $ 9.63\pm 0.03$ & $  57\pm  23$ & $  2.4\pm 0.9$ & ... \\
         &  2 &  0.016 & $ 4.84\pm 0.04$ & $ 577\pm  58$ & $ 24.3\pm 2.5$ & $1993.80\pm   0.86$ \\
         &  3 &  0.041 & $ 3.82\pm 0.04$ & $ 340\pm  30$ & $ 14.3\pm 1.3$ & $1990.64\pm   0.99$ \\
         &  4 &  0.020 & $ 2.81\pm 0.02$ & $ 355\pm  21$ & $ 14.9\pm 0.9$ & $1994.21\pm   0.47$ \\
         &  5 &  0.022 & $ 1.67\pm 0.04$ & $ 208\pm  39$ & $  8.7\pm 1.6$ & $1993.61\pm   1.57$ \\
         &  6 &  0.042 & $ 1.20\pm 0.02$ & $ 164\pm  12$ & $  6.9\pm 0.5$ & $1994.57\pm   0.56$ \\
         &  7 &  0.074 & $ 0.43\pm 0.02$ & $  56\pm  19$ & $  2.3\pm 0.8$ & ... \\
1657$-$261 &  1 &  0.036 & $ 0.85\pm 0.12$ & $ 158\pm  95$ &    ...         & ... \\
1726+455 &  1 &  0.057 & $ 1.81\pm 0.06$ & $ 181\pm  34$ & $  7.3\pm 1.4$ & $1988.65\pm   1.92$ \\
         &  2 &  0.095 & $ 0.93\pm 0.07$ & $ 293\pm  43$ & $ 11.8\pm 1.7$ & $1996.35\pm   0.47$ \\
1739+522 &  1 &  0.101 & $ 1.16\pm 0.11$ & $  54\pm  97$ & $  3.6\pm 6.4$ & ... \\
         &  2 &  0.154 & $ 0.37\pm 0.02$ & $  58\pm  12$ & $  3.9\pm 0.8$ & $1995.08\pm   1.34$ \\
1741$-$038 &  1 &  0.019 & $ 1.82\pm 0.12$ & $ -14\pm  75$ & $ -0.8\pm 4.1$ & ... \\ 
         &  2 &  0.065 & $ 0.98\pm 0.05$ & $ -35\pm  54$ & $ -2.0\pm 3.0$ &  ... \\
         &  3 &  1.365 & $ 0.43\pm 0.02$ & $  30\pm   6$ & $  1.7\pm 0.4$ & $1985.88\pm   3.21$ \\
1745+624 &  1 &  0.007 & $ 2.58\pm 0.07$ & $  76\pm  30$ & $  8.7\pm 3.5$ & ... \\
         &  2 &  0.015 & $ 1.46\pm 0.03$ & $  64\pm  17$ & $  7.4\pm 1.9$ & $1976.72\pm   6.44$ \\
         &  3 &  0.018 & $ 1.10\pm 0.05$ & $ 132\pm  62$ & $ 15.1\pm 7.1$ & ... \\
         &  4 &  0.035 & $ 0.55\pm 0.05$ & $  70\pm  54$ & $  8.1\pm 6.3$ & ... \\
         &  5 &  0.287 & $ 0.24\pm 0.01$ & $  10\pm   3$ & $  1.2\pm 0.4$ & $1977.39\pm   7.75$ \\
1749+096 &  1 &  0.027 & $ 3.89\pm 0.18$ & $ 789\pm 143$ & $ 15.8\pm 2.9$ & $1991.66\pm   0.93$ \\
         &  2 &  0.030 & $ 2.46\pm 0.12$ & $ 711\pm  92$ & $ 14.2\pm 1.8$ & $1993.64\pm   0.46$ \\
         &  3 &  0.101 & $ 1.06\pm 0.08$ & $ 558\pm  95$ & $ 11.2\pm 1.9$ & $1996.42\pm   0.33$ \\
         &  4 &  0.073 & $ 1.92\pm 0.06$ & $ 449\pm  38$ & $  9.0\pm 0.8$ & $1996.89\pm   0.37$ \\
         &  5 &  0.169 & $ 0.86\pm 0.03$ & $ 187\pm  24$ & $  3.7\pm 0.5$ & $1997.39\pm   0.59$ \\
1803+784 &  1 &  0.040 & $ 7.15\pm 0.09$ & $ -56\pm  42$ & $ -2.2\pm 1.6$ & ... \\
         &  2 &  0.051 & $ 3.45\pm 0.08$ & $  80\pm  38$ & $  3.1\pm 1.5$ & ... \\
         &  3 &  0.083 & $ 1.83\pm 0.02$ & $ -38\pm   9$ & $ -1.5\pm 0.4$ & ... \\
         &  4 &  0.217 & $ 1.44\pm 0.01$ & $ -22\pm   5$ & $ -0.8\pm 0.2$ & ... \\
         &  5 &  0.118 & $ 1.03\pm 0.02$ & $   1\pm  14$ & $  0.04\pm 0.54$ & ... \\
         &  6 &  0.257 & $ 0.47\pm 0.01$ & $  22\pm   7$ & $  0.9\pm 0.3$ & $1979.21\pm   7.09$ \\
1908$-$201 &  1 &  0.036 & $ 5.41\pm 0.21$ & $ 342\pm  99$ & $ 19.6\pm 5.7$ & $1984.36\pm   5.03$ \\
         &  2 &  0.229 & $ 2.85\pm 0.04$ & $ 193\pm  17$ & $ 11.0\pm 1.0$ & $1985.55\pm   1.34$ \\
         &  3 &  0.206 & $ 1.27\pm 0.07$ & $ 249\pm  70$ & $ 14.3\pm 4.0$ & $1995.07\pm   1.56$ \\
         &  4 &  0.623 & $ 0.82\pm 0.04$ & $ 187\pm  27$ & $ 10.7\pm 1.5$ & $1997.21\pm   0.64$ \\
1921$-$293 &  1 &  1.457 & $ 6.20\pm 0.07$ & $ 176\pm  31$ & $  3.8\pm 0.7$ & $1964.78\pm   6.32$ \\
         &  2 &  0.730 & $ 2.99\pm 0.19$ & $ 142\pm  76$ & $  3.1\pm 1.6$ &  ... \\
         &  3 &  0.716 & $ 1.28\pm 0.08$ & $ 229\pm  82$ & $  5.0\pm 1.8$ &  ... \\
1954$-$388 &  1 &  0.050 & $ 2.38\pm 0.11$ & $ 102\pm  50$ & $  3.7\pm 1.8$ &  ... \\
         &  2 &  0.341 & $ 0.82\pm 0.06$ & $ 102\pm  28$ & $  3.7\pm 1.0$ & $1992.33\pm   2.35$ \\
2145+067 &  1 &  0.041 & $ 5.39\pm 0.12$ & $  10\pm  49$ & $  0.5\pm 2.6$ & ... \\
         &  2 &  0.035 & $ 2.51\pm 0.06$ & $ -56\pm  30$ & $ -3.0\pm 1.6$ & ... \\
         &  3 &  0.426 & $ 1.15\pm 0.04$ & $ 125\pm  21$ & $  6.6\pm 1.1$ & $1988.91\pm   1.55$ \\
         &  4 &  1.633 & $ 0.81\pm 0.01$ & $  84\pm   7$ & $  4.4\pm 0.4$ & $1991.13\pm   0.85$ \\
         &  5 &  1.735 & $ 0.53\pm 0.02$ & $  25\pm  21$ & $  1.3\pm 1.1$ & ... \\
2200+420 &  1 &  0.098 & $ 7.71\pm 0.17$ & $ 993\pm 130$ & $  4.6\pm 0.6$ & $1990.81\pm   1.03$ \\
         &  2 &  0.118 & $ 7.47\pm 0.11$ & $ 572\pm  98$ & $  2.6\pm 0.5$ & $1988.62\pm   2.31$ \\
         &  3 &  0.281 & $ 3.93\pm 0.08$ & $ 666\pm 123$ & $  3.1\pm 0.6$ & $1992.39\pm   1.13$ \\
         &  4 &  0.359 & $ 3.06\pm 0.08$ & $ 807\pm  65$ & $  3.7\pm 0.3$ & $1994.60\pm   0.31$ \\
         &  5 &  0.215 & $ 3.98\pm 0.06$ & $ 822\pm  34$ & $  3.8\pm 0.2$ & $1995.60\pm   0.20$ \\
         &  6 &  0.376 & $ 2.82\pm 0.04$ & $ 562\pm  20$ & $  2.6\pm 0.1$ & $1995.19\pm   0.18$ \\
         &  7 &  0.156 & $ 2.11\pm 0.03$ & $ 599\pm  37$ & $  2.8\pm 0.2$ & $1996.66\pm   0.22$ \\
         &  8 &  0.166 & $ 2.23\pm 0.06$ & $ 611\pm  40$ & $  2.8\pm 0.2$ & $1997.86\pm   0.24$ \\
         &  9 &  0.186 & $ 1.97\pm 0.05$ & $ 662\pm  54$ & $  3.1\pm 0.2$ & $1999.45\pm   0.24$ \\
         & 10 &  0.129 & $ 1.28\pm 0.04$ & $ 200\pm  61$ & $  0.9\pm 0.3$ & $1996.39\pm   2.13$ \\
         & 11 &  0.491 & $ 0.34\pm 0.01$ & $ -17\pm   6$ & $ -0.08\pm 0.03$ & ... \\
2223$-$052 &  1 &  0.072 & $ 5.76\pm 0.11$ & $ 260\pm  61$ & $ 17.4\pm 4.1$ & $1979.08\pm   5.49$ \\
         &  2 &  0.151 & $ 3.15\pm 0.04$ & $  99\pm  27$ & $  6.6\pm 1.8$ & $1969.64\pm   9.28$ \\
         &  3 &  0.073 & $ 1.31\pm 0.06$ & $  57\pm  34$ & $  3.8\pm 2.3$ &  ... \\
         &  4 &  1.058 & $ 0.47\pm 0.02$ & $ 105\pm  15$ & $  7.0\pm 1.0$ & $1997.16\pm   0.67$ \\
2234+282 &  1 &  0.081 & $ 0.84\pm 0.04$ & $  14\pm  22$ & $  0.6\pm 1.0$ & ... \\
         &  2 &  0.412 & $ 0.51\pm 0.02$ & $  37\pm  16$ & $  1.6\pm 0.7$ & ... \\
         &  3 &  0.388 & $ 0.51\pm 0.01$ & $  73\pm   6$ & $  3.2\pm 0.3$ & $1994.72\pm   0.56$ \\
2243$-$123 &  1 &  0.056 & $10.82\pm 0.08$ & $ -58\pm  41$ & $ -2.1\pm 1.5$ & ... \\
         &  2 &  0.178 & $ 3.33\pm 0.03$ & $  99\pm  13$ & $  3.6\pm 0.5$ & $1966.80\pm   4.40$ \\
         &  3 &  0.480 & $ 1.40\pm 0.01$ & $  87\pm   7$ & $  3.2\pm 0.3$ & $1984.25\pm   1.34$ \\
\tableline \\[-5pt]
\end{tabular}}
\end{center}
{\bf Notes.} (1) Source name; (2) Component ID; (3) Mean flux density; 
(4) Weighted mean radial separation
from core; (5) Proper
motion; (6) Apparent speed in units of the speed of light; 
(7) Ejection time is given for proper motions with significance above 3$\sigma$.
\end{table*}

For both fitting methods (the linear and the nonlinear),
the errors on the position measurements were assigned according to the
method described by Homan et al. (2001) and subsequently used by L09 and H09. This method uses the scatter
of the data points about the fit to assign a constant error to each point, such that the
reduced $\chi^{2}$ of the fit has a value of 1.0.
This differs from the beam-based method for assigning positional errors that we used in Paper I.
The new method has the disadvantage that the reduced $\chi^{2}$ value cannot be used to determine the
suitability of the chosen model, but under the assumption that the model is appropriate it does
yield good uncertainties for the model parameters. It also
addresses in a straightforward manner the well-known problem in VLBI model fitting that
with a large number of position measurements there is no method for obtaining robust and statistically
accurate uncertainties that will work for all components in a dataset, and that will take
into account both errors caused by measurement uncertainties in the visibilities and systematic
effects due to changes in the number and shape of model components, or effects introduced during calibration
and imaging.

The results of the linear fits are presented in Table~\ref{speedtab}, and shown in the form of separation versus
time plots in Figure~3. Table~\ref{speedtab} of this paper is similar to Table~4 of Paper I, but
there are on average about four times as many data points per component in this paper compared to Paper I.
Thus, we have dropped the `Quality Code' that we associated with each component in Paper I, as now essentially
all of the fits are of Good or Excellent quality, as defined in that paper.
Table~\ref{speedtab} gives the mean flux density and weighted mean separation of each component, the radial proper motion
and apparent speed, and the fitted epoch when the component separated from the core, for components that have
a proper motion significance above 3$\sigma$.
The complete set of 66 plots in Figure~3 is available in the online journal.
The linear fits in Table~\ref{speedtab} are presented mainly for consistency with Paper I; however, most of the
subsequent analysis in this paper uses the results from the nonlinear fits from Table~\ref{acctab}.

\begin{sidewaysfigure*}
\begin{center}
\includegraphics[angle=90,scale=0.80]{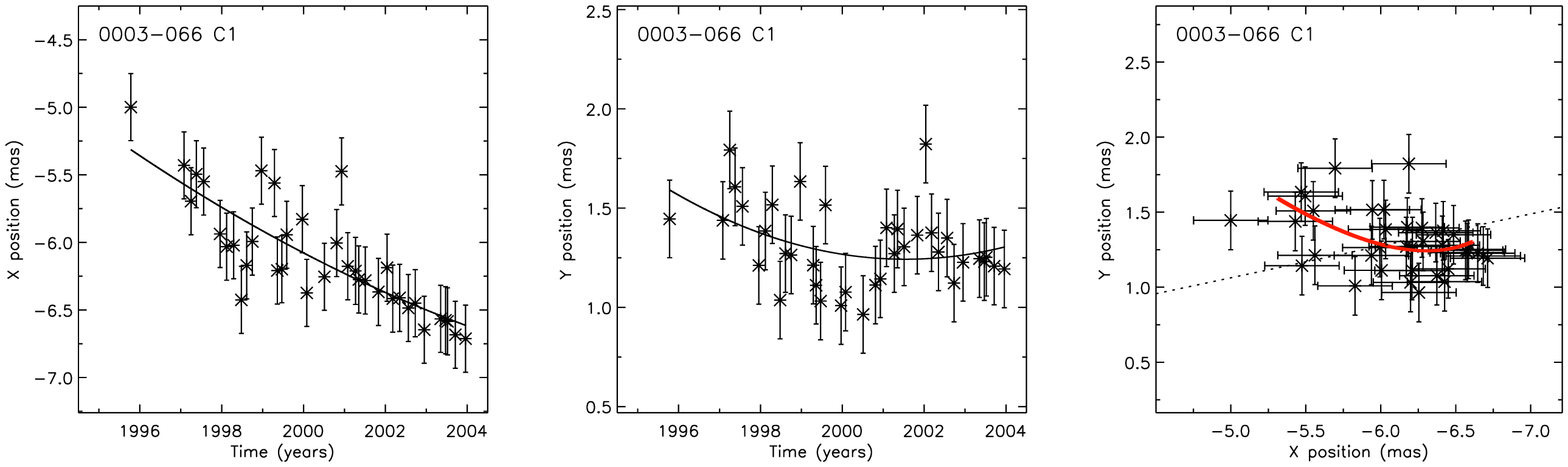}
\end{center}
\caption{The left and center panels show the $x$ and $y$ positions respectively
of Gaussian model component centers relative to the core
as a function of time. The curves are the least-squares fits to motion at constant 
acceleration for $x(t)$ and $y(t)$, for components detected at four or more epochs.
The right panel shows the $(x,y)$ positions of the Gaussian model component centers
from the left and center panels. The dotted line shows the
radial direction toward and away from the core at the component's weighted mean position
angle from Table~\ref{acctab}. The fitted $(x,y)$ trajectory 
is plotted as a red curve for components that
are members of the 48 or 64-component subsamples used in studying apparent accelerations
(see $\S$~\ref{acc}). Plotting symbols are the same as those in Figure~3.
The B1950 source name and component number are given at the top left of each panel.
The complete figure set (225 images, one for each component in 
Table~\ref{acctab}) is available in the online journal.}
\end{sidewaysfigure*}

The results of the nonlinear fits to each of the 225 components are presented in Table~\ref{acctab}, and are shown
graphically in Figure~4, in the form of second-order polynomial fits to both $x(t)$ and $y(t)$ for each component
in Table~\ref{acctab}. The complete set of 225 plots in Figure~4 is available
in the online journal. Note that no entries are given in columns (9) through (14)
of Table~\ref{acctab} if the error in the fitted direction of motion is $>90\arcdeg$.
This generally results from components that are nearly stationary, so that the direction of motion,
and thus all subsequent columns that depend on the direction of motion, are undefined.
The column headings in Table~\ref{acctab} are identical to those in the corresponding table from H09 (Table~1),
to aid in comparisons between the two acceleration analyses.
One major difference between Table~\ref{acctab} and the corresponding table from H09 is that
Table~1 of H09 shows the acceleration analyses only for a high-quality subsample
of 203 components that satisfied specific selection criteria,
out of the 526 total components in the MOJAVE survey. In this paper, Table~\ref{acctab} presents the second-order fits
for all 225 components in the RDV survey, and we then introduce quality cuts on the data similar to those introduced by H09
before undertaking the acceleration analysis in $\S$~\ref{acc}.

\begin{sidewaystable*}
\begin{center}
\caption{Results of Acceleration Analysis}
\label{acctab}
{\scriptsize \begin{tabular}{l c c r r r r r r r r r r r} \tableline \tableline \\[5pt]
\multicolumn{1}{c}{Source} & ID & N & \multicolumn{1}{c}{$\langle{r}\rangle$} & \multicolumn{1}{c}{PA} & 
\multicolumn{1}{c}{$d_{proj}$} & \multicolumn{1}{c}{$\mu$} & \multicolumn{1}{c}{$\beta_\mathrm{app}$} & 
\multicolumn{1}{c}{$\phi$} & \multicolumn{1}{c}{$|PA-\phi|$} & \multicolumn{1}{c}{$\dot{\mu}_{\parallel}$} &
\multicolumn{1}{c}{$\dot{\mu}_{\perp}$} & \multicolumn{1}{c}{$\dot{\eta}_{\parallel}$} &
\multicolumn{1}{c}{$\dot{\eta}_{\perp}$} \\
 & & & \multicolumn{1}{c}{(mas)} & \multicolumn{1}{c}{(deg)} & \multicolumn{1}{c}{(pc)} & 
\multicolumn{1}{c}{($\mu as$ yr$^{-1}$)} & & \multicolumn{1}{c}{(deg)} & \multicolumn{1}{c}{(deg)} & 
\multicolumn{1}{c}{($\mu as$ yr$^{-2}$)} & \multicolumn{1}{c}{($\mu as$ yr$^{-2}$)} & \multicolumn{1}{c}{(yr$^{-1}$)} & \multicolumn{1}{c}{(yr$^{-1}$)} \\
\multicolumn{1}{c}{(1)} & \multicolumn{1}{c}{(2)} & \multicolumn{1}{c}{(3)} & \multicolumn{1}{c}{(4)} & \multicolumn{1}{c}{(5)} &
\multicolumn{1}{c}{(6)} & \multicolumn{1}{c}{(7)} & \multicolumn{1}{c}{(8)} & \multicolumn{1}{c}{(9)} & \multicolumn{1}{c}{(10)} &
\multicolumn{1}{c}{(11)} & \multicolumn{1}{c}{(12)} & \multicolumn{1}{c}{(13)} & \multicolumn{1}{c}{(14)} \\ \tableline \\[5pt]
0003$-$066 &  1$^{b}$ & 37 &  6.25 &  $-$78.0 &  30.82 & $  162\pm 19$ & $ 3.5\pm 0.4$ & $-102.5\pm 5.6$ & $ 24.4\pm  5.7$ & $  -15\pm  16$ & $   18\pm  15$ & $ -0.13\pm  0.14$ & $  0.15\pm  0.13$ \\
           &  2       & 33 &  3.03 &  $-$74.8 &  14.92 & $  153\pm 36$ & $ 3.3\pm 0.8$ & $ -77.8\pm 9.3$ & $  3.0\pm  9.3$ & $  -97\pm  40$ & $   24\pm  32$ & $ -0.86\pm  0.41$ & $  0.21\pm  0.29$ \\
           &  3       & 36 &  1.10 &  $-$70.3 &   5.41 & $   94\pm  8$ & $ 2.1\pm 0.2$ & $-108.2\pm 7.6$ & $ 37.9\pm  7.7$ & $  -38\pm   8$ & $   -3\pm  15$ & $ -0.55\pm  0.13$ & $ -0.05\pm  0.22$ \\
0014+813   &  1       & 26 &  9.39 & $-$171.6 &  71.06 & $   52\pm 32$ & $ 5.7\pm 3.5$ & $-138.2\pm33.6$ & $ 33.4\pm 33.6$ & $  -43\pm  39$ & $  -22\pm  49$ & $ -3.62\pm  3.93$ & $ -1.88\pm  4.24$ \\
           &  2       & 34 &  5.36 & $-$166.1 &  40.60 & $   40\pm 31$ & $ 4.4\pm 3.4$ & $ 160.4\pm33.4$ & $ 33.6\pm 33.4$ & $   20\pm  34$ & $   27\pm  34$ & $  2.20\pm  4.17$ & $  2.98\pm  4.40$ \\
           &  3       & 42 &  0.67 & $-$179.4 &   5.08 & $   13\pm  1$ & $ 1.4\pm 0.2$ & $  72.4\pm13.9$ & $108.2\pm 13.9$ & $   -7\pm   3$ & $    9\pm   3$ & $ -2.54\pm  1.21$ & $  3.07\pm  1.22$ \\
0048$-$097 &  1       &  4 &  3.07 &      9.8 &  21.03 & $  195\pm 88$ & $ 7.1\pm 3.2$ & $ 121.9\pm37.2$ & $112.0\pm 37.3$ & $  250\pm 294$ & $  365\pm 284$ & $  2.09\pm  2.63$ & $  3.05\pm  2.74$ \\
           &  3       & 10 &  0.59 &  $-$81.3 &   4.07 & $  112\pm 45$ & $ 4.1\pm 1.7$ & $-144.6\pm26.1$ & $ 63.3\pm 26.2$ & $  -87\pm 110$ & $  -85\pm 109$ & $ -1.27\pm  1.67$ & $ -1.23\pm  1.66$ \\
           &  4       &  7 &  0.86 &  $-$12.8 &   5.91 & $  379\pm166$ & $13.8\pm 6.1$ & $ -13.2\pm16.4$ & $  0.4\pm 16.7$ & $-1113\pm 632$ & $  423\pm 632$ & $ -4.79\pm  3.43$ & $  1.82\pm  2.83$ \\
           &  5       & 12 &  0.67 &  $-$11.8 &   4.57 & $  179\pm 69$ & $ 6.5\pm 2.5$ & $ -18.7\pm 9.2$ & $  6.9\pm  9.3$ & $   26\pm 129$ & $   26\pm 131$ & $  0.24\pm  1.18$ & $  0.24\pm  1.20$ \\
0059+581   &  1$^{c}$ & 35 &  2.51 & $-$115.1 &  17.31 & $   25\pm 19$ & $ 0.9\pm 0.7$ & $  70.5\pm32.9$ & $174.4\pm 32.9$ & $  -14\pm  16$ & $    6\pm  18$ & $ -0.92\pm  1.30$ & $  0.42\pm  1.22$ \\
           &  2$^{c}$ & 34 &  1.35 & $-$133.5 &   9.32 & $   42\pm 13$ & $ 1.6\pm 0.5$ & $-136.0\pm19.3$ & $  2.6\pm 19.3$ & $    3\pm  15$ & $   23\pm  15$ & $  0.16\pm  0.61$ & $  0.94\pm  0.67$ \\
           &  3       &  9 &  0.65 & $-$153.8 &   4.48 & $   60\pm 43$ & $ 2.2\pm 1.6$ & $ -67.6\pm22.2$ & $ 86.3\pm 22.5$ & $  -24\pm  67$ & $   -4\pm  48$ & $ -0.68\pm  1.90$ & $ -0.11\pm  1.33$ \\
           &  4$^{b}$ & 36 &  0.58 & $-$126.1 &   4.03 & $  112\pm  7$ & $ 4.2\pm 0.3$ & $-103.6\pm 5.8$ & $ 22.5\pm  6.0$ & $    1\pm   9$ & $   15\pm  13$ & $  0.02\pm  0.14$ & $  0.23\pm  0.20$ \\
           &  5       &  5 &  0.16 &    168.5 &   1.11 & $   67\pm 50$ & $ 2.5\pm 1.9$ & $-164.3\pm26.5$ & $ 27.2\pm 26.6$ & $ 1001\pm 584$ & $ -127\pm 573$ & $ 24.52\pm 23.45$ & $ -3.12\pm 14.24$ \\
0104$-$408 &  1       & 17 &  2.33 &     18.9 &  15.38 & $   60\pm 56$ & $ 2.1\pm 1.9$ & $  48.2\pm58.1$ & $ 29.2\pm 58.1$ & $   84\pm  69$ & $   39\pm 119$ & $  2.21\pm  2.75$ & $  1.04\pm  3.27$ \\
           &  2       &  5 &  0.53 &     33.6 &   3.50 & $   73\pm 21$ & $ 2.5\pm 0.7$ & $  22.8\pm15.9$ & $ 10.8\pm 16.5$ & $   95\pm  53$ & $ -134\pm  46$ & $  2.05\pm  1.29$ & $ -2.88\pm  1.30$ \\
0119+041   &  1       & 38 &  0.73 &    109.9 &   5.06 & $   64\pm  4$ & $ 2.4\pm 0.2$ & $ 164.5\pm 6.1$ & $ 54.6\pm  6.1$ & $    5\pm   5$ & $  -37\pm   6$ & $  0.15\pm  0.14$ & $ -0.94\pm  0.17$ \\
0119+115   &  1       & 11 & 20.33 &      3.6 & 132.93 & $  342\pm137$ & $11.5\pm 4.6$ & $ -22.6\pm18.5$ & $ 26.2\pm 18.5$ & $ -697\pm 470$ & $  326\pm 518$ & $ -3.19\pm  2.50$ & $  1.50\pm  2.45$ \\
           &  2       &  8 & 14.24 &      2.5 &  93.13 & $   83\pm 75$ & $ 2.8\pm 2.5$ & $ -46.5\pm52.9$ & $ 49.1\pm 52.9$ & $ -366\pm 191$ & $  -49\pm 379$ & $ -6.92\pm  7.24$ & $ -0.94\pm  7.23$ \\
           &  3$^{a}$ & 41 &  1.65 &      8.3 &  10.77 & $  197\pm 15$ & $ 6.6\pm 0.5$ & $  12.4\pm 2.3$ & $  4.0\pm  2.4$ & $  102\pm  13$ & $    1\pm   8$ & $  0.82\pm  0.13$ & $  0.01\pm  0.07$ \\
           &  4$^{a}$ & 17 &  1.39 &   $-$0.1 &   9.08 & $  317\pm 44$ & $10.6\pm 1.5$ & $  -0.6\pm 1.7$ & $  0.5\pm  1.8$ & $    0\pm 116$ & $   94\pm  25$ & $  0.00\pm  0.57$ & $  0.47\pm  0.14$ \\
0133+476   &  1       & 39 &  2.61 &  $-$35.4 &  20.20 & $   63\pm 12$ & $ 3.0\pm 0.6$ & $  68.6\pm11.4$ & $104.0\pm 11.4$ & $   15\pm  15$ & $  -37\pm  13$ & $  0.44\pm  0.45$ & $ -1.10\pm  0.44$ \\
           &  2       & 31 &  1.08 &  $-$29.3 &   8.34 & $   61\pm 12$ & $ 2.9\pm 0.6$ & $ -51.6\pm13.0$ & $ 22.3\pm 13.1$ & $  -15\pm  18$ & $  -13\pm  17$ & $ -0.47\pm  0.57$ & $ -0.40\pm  0.54$ \\
           &  3$^{c}$ & 39 &  0.53 &  $-$27.9 &   4.07 & $   31\pm  7$ & $ 1.5\pm 0.3$ & $ -33.9\pm11.5$ & $  6.0\pm 11.5$ & $    6\pm   7$ & $    5\pm   7$ & $  0.38\pm  0.47$ & $  0.30\pm  0.43$ \\
0201+113   &  1       & 22 &  1.46 &  $-$56.0 &  10.79 & $   49\pm 22$ & $ 5.5\pm 2.5$ & $-115.4\pm23.5$ & $ 59.4\pm 23.6$ & $  -17\pm  21$ & $  -24\pm  20$ & $ -1.62\pm  2.10$ & $ -2.26\pm  2.16$ \\
           &  2       & 40 &  1.21 &  $-$30.4 &   8.97 & $   40\pm  4$ & $ 4.6\pm 0.5$ & $ -51.8\pm 6.9$ & $ 21.4\pm  6.9$ & $  -17\pm   4$ & $  -16\pm   4$ & $ -1.95\pm  0.56$ & $ -1.82\pm  0.55$ \\
0202+149   &  1$^{a}$ & 41 &  4.81 &  $-$52.0 &  26.19 & $  112\pm  9$ & $ 2.8\pm 0.2$ & $ 123.5\pm 4.9$ & $175.5\pm  4.9$ & $    6\pm   9$ & $   25\pm   9$ & $  0.09\pm  0.12$ & $  0.32\pm  0.12$ \\
           &  2$^{a}$ & 29 &  0.58 &  $-$58.9 &   3.16 & $   95\pm 14$ & $ 2.4\pm 0.4$ & $ -71.5\pm 4.7$ & $ 12.6\pm  4.8$ & $   72\pm  19$ & $  -31\pm  13$ & $  1.06\pm  0.33$ & $ -0.46\pm  0.20$ \\
           &  3       & 10 &  0.52 &  $-$22.1 &   2.84 & $  120\pm 67$ & $ 3.0\pm 1.7$ & $ -17.2\pm25.2$ & $  4.9\pm 25.5$ & $ -169\pm 173$ & $   -9\pm 154$ & $ -1.98\pm  2.32$ & $ -0.12\pm  1.80$ \\
0229+131   &  1       &  8 &  7.46 &     55.2 &  63.30 & $  150\pm114$ & $12.8\pm 9.7$ & $  81.8\pm39.2$ & $ 26.6\pm 39.2$ & $  -31\pm  84$ & $  -50\pm  79$ & $ -0.65\pm  1.80$ & $ -1.02\pm  1.81$ \\
           &  2       &  4 &  3.17 &     34.3 &  26.89 & $  228\pm279$ & $19.4\pm23.8$ & $-171.6\pm27.2$ & $154.1\pm 27.3$ & $ -219\pm1297$ & $  777\pm 784$ & $ -2.95\pm 17.83$ & $ 10.46\pm 16.62$ \\
           &  3$^{b}$ & 11 &  2.76 &     42.1 &  23.39 & $  323\pm 30$ & $27.5\pm 2.6$ & $  51.5\pm 5.2$ & $  9.4\pm  5.5$ & $   53\pm  28$ & $    8\pm  26$ & $  0.51\pm  0.28$ & $  0.08\pm  0.25$ \\
           &  4       & 26 &  1.77 &     40.7 &  15.05 & $   62\pm 22$ & $ 5.3\pm 1.9$ & $  66.6\pm20.6$ & $ 25.9\pm 20.6$ & $  -34\pm  31$ & $  -36\pm  31$ & $ -1.69\pm  1.67$ & $ -1.79\pm  1.67$ \\
           &  5$^{c}$ & 21 &  0.51 &     70.5 &   4.33 & $   47\pm 10$ & $ 4.0\pm 0.9$ & $ 142.6\pm13.8$ & $ 72.1\pm 13.9$ & $    5\pm  11$ & $  -14\pm  12$ & $  0.38\pm  0.77$ & $ -0.92\pm  0.85$ \\
           &  6       &  5 &  0.24 &     95.6 &   2.04 & $  106\pm 47$ & $ 9.1\pm 4.1$ & $ -78.5\pm14.3$ & $174.1\pm 14.5$ & $ -387\pm 299$ & $ -374\pm 302$ & $-11.15\pm  9.98$ & $-10.78\pm  9.97$ \\
0234+285   &  1$^{a}$ & 28 &  6.03 &   $-$9.0 &  50.63 & $  353\pm 38$ & $21.4\pm 2.3$ & $ -11.8\pm 2.4$ & $  2.8\pm  2.4$ & $  175\pm  31$ & $   58\pm  16$ & $  1.09\pm  0.23$ & $  0.37\pm  0.11$ \\
           &  2$^{a}$ & 43 &  4.01 &  $-$11.4 &  33.65 & $  289\pm 13$ & $17.5\pm 0.8$ & $  -2.7\pm 1.1$ & $  8.7\pm  1.2$ & $   18\pm   8$ & $    6\pm   5$ & $  0.14\pm  0.06$ & $  0.05\pm  0.04$ \\
           &  3$^{c}$ & 14 &  1.03 &  $-$15.9 &   8.64 & $   47\pm 23$ & $ 2.9\pm 1.4$ & $ -29.0\pm17.8$ & $ 13.0\pm 17.8$ & $    5\pm  19$ & $    3\pm  13$ & $  0.24\pm  0.92$ & $  0.16\pm  0.64$ \\
           &  4$^{c}$ & 23 &  0.45 &  $-$27.4 &   3.81 & $   48\pm  6$ & $ 2.9\pm 0.4$ & $ -77.1\pm16.1$ & $ 49.7\pm 16.2$ & $  -12\pm  22$ & $   19\pm  22$ & $ -0.59\pm  1.06$ & $  0.91\pm  1.03$ \\
           &  5       &  7 &  0.32 &  $-$27.4 &   2.70 & $  122\pm 45$ & $ 7.4\pm 2.7$ & $ -16.3\pm 8.0$ & $ 11.1\pm  8.2$ & $ -110\pm 192$ & $ -225\pm 180$ & $ -1.98\pm  3.54$ & $ -4.06\pm  3.58$ \\
0336$-$019 &  1       &  6 &  5.88 &     60.5 &  45.33 & $   76\pm190$ & $ 3.6\pm 8.9$ & $ -23.7\pm75.5$ & $ 84.3\pm 75.5$ & $   67\pm 286$ & $  165\pm 208$ & $  1.63\pm  8.02$ & $  4.01\pm 11.18$ \\
           &  2       & 11 &  3.62 &     58.1 &  27.94 & $  146\pm120$ & $ 6.8\pm 5.6$ & $ 113.3\pm42.2$ & $ 55.2\pm 42.2$ & $  856\pm 416$ & $ -166\pm 719$ & $ 10.85\pm 10.38$ & $ -2.10\pm  9.28$ \\
           &  3$^{b}$ & 34 &  2.94 &     57.5 &  22.67 & $  198\pm 20$ & $ 9.2\pm 1.0$ & $  61.5\pm 5.2$ & $  4.0\pm  5.2$ & $  -41\pm  17$ & $    2\pm  15$ & $ -0.39\pm  0.17$ & $  0.02\pm  0.15$ \\
           &  4       & 34 &  1.50 &     71.2 &  11.55 & $  118\pm 12$ & $ 5.5\pm 0.6$ & $  71.4\pm 7.0$ & $  0.3\pm  7.1$ & $  -17\pm  12$ & $  -18\pm  12$ & $ -0.28\pm  0.19$ & $ -0.28\pm  0.19$ \\
           &  5$^{a}$ & 23 &  0.95 &     50.2 &   7.33 & $  290\pm 17$ & $13.5\pm 0.8$ & $  68.5\pm 4.2$ & $ 18.3\pm  4.4$ & $   37\pm  35$ & $   47\pm  34$ & $  0.24\pm  0.23$ & $  0.30\pm  0.22$ \\
           &  6       &  8 &  0.34 &     39.8 &   2.60 & $   80\pm 32$ & $ 3.8\pm 1.5$ & $  21.0\pm12.0$ & $ 18.8\pm 12.5$ & $  118\pm  59$ & $  123\pm  58$ & $  2.71\pm  1.74$ & $  2.81\pm  1.75$ \\
\end{tabular}}
\end{center}
\end{sidewaystable*}

\begin{sidewaystable*}
\begin{center}
Table 6 (Continued) \\
{\scriptsize \begin{tabular}{l c c r r r r r r r r r r r} \tableline \tableline \\[5pt]
\multicolumn{1}{c}{Source} & Comp. & N & \multicolumn{1}{c}{$\langle{r}\rangle$} & \multicolumn{1}{c}{PA} & 
\multicolumn{1}{c}{$d_{proj}$} & \multicolumn{1}{c}{$\mu$} & \multicolumn{1}{c}{$\beta_\mathrm{app}$} & 
\multicolumn{1}{c}{$\phi$} & \multicolumn{1}{c}{$|PA-\phi|$} & \multicolumn{1}{c}{$\dot{\mu}_{\parallel}$} &
\multicolumn{1}{c}{$\dot{\mu}_{\perp}$} & \multicolumn{1}{c}{$\dot{\eta}_{\parallel}$} &
\multicolumn{1}{c}{$\dot{\eta}_{\perp}$} \\
 & & & \multicolumn{1}{c}{(mas)} & \multicolumn{1}{c}{(deg)} & \multicolumn{1}{c}{(pc)} & 
\multicolumn{1}{c}{($\mu as$ yr$^{-1}$)} & & \multicolumn{1}{c}{(deg)} & \multicolumn{1}{c}{(deg)} & 
\multicolumn{1}{c}{($\mu as$ yr$^{-2}$)} & \multicolumn{1}{c}{($\mu as$ yr$^{-2}$)} & \multicolumn{1}{c}{(yr$^{-1}$)} & \multicolumn{1}{c}{(yr$^{-1}$)} \\
\multicolumn{1}{c}{(1)} & \multicolumn{1}{c}{(2)} & \multicolumn{1}{c}{(3)} & \multicolumn{1}{c}{(4)} & \multicolumn{1}{c}{(5)} &
\multicolumn{1}{c}{(6)} & \multicolumn{1}{c}{(7)} & \multicolumn{1}{c}{(8)} & \multicolumn{1}{c}{(9)} & \multicolumn{1}{c}{(10)} &
\multicolumn{1}{c}{(11)} & \multicolumn{1}{c}{(12)} & \multicolumn{1}{c}{(13)} & \multicolumn{1}{c}{(14)} \\ \tableline \\[5pt]
0402$-$362 &  1       & 11 &  2.75 &     27.5 &  23.54 & $  183\pm 53$ & $12.4\pm 3.6$ & $  -0.5\pm14.7$ & $ 28.0\pm 14.8$ & $ -247\pm  82$ & $  -21\pm  96$ & $ -3.27\pm  1.45$ & $ -0.28\pm  1.28$ \\
           &  2$^{b}$ & 29 &  0.79 &     18.7 &   6.75 & $  111\pm 18$ & $ 7.5\pm 1.2$ & $  20.1\pm 5.6$ & $  1.4\pm  5.7$ & $   26\pm  15$ & $   20\pm  16$ & $  0.58\pm  0.34$ & $  0.46\pm  0.36$ \\
0430+052   &  1$^{a}$ & 12 &  5.57 & $-$110.6 &   3.31 & $ 1458\pm 91$ & $ 2.9\pm 0.2$ & $-110.3\pm 3.3$ & $  0.3\pm  3.4$ & $   94\pm 300$ & $  501\pm 285$ & $  0.07\pm  0.21$ & $  0.35\pm  0.20$ \\
           &  2$^{a}$ & 13 &  2.55 & $-$116.3 &   1.52 & $ 1736\pm 39$ & $ 3.5\pm 0.1$ & $-112.1\pm 1.8$ & $  4.2\pm  2.0$ & $   85\pm 146$ & $  246\pm 153$ & $  0.05\pm  0.09$ & $  0.15\pm  0.09$ \\
           &  3       &  9 &  2.24 & $-$118.5 &   1.34 & $ 1869\pm160$ & $ 3.7\pm 0.3$ & $-106.7\pm 3.8$ & $ 11.8\pm  4.0$ & $ 1253\pm 709$ & $ 1981\pm 714$ & $  0.69\pm  0.40$ & $  1.09\pm  0.40$ \\
           &  4$^{a}$ & 18 &  8.66 & $-$110.5 &   5.16 & $ 2198\pm117$ & $ 4.4\pm 0.2$ & $ -98.0\pm 1.1$ & $ 12.6\pm  1.1$ & $  278\pm 222$ & $  628\pm 173$ & $  0.13\pm  0.10$ & $  0.29\pm  0.08$ \\
           &  5$^{a}$ & 14 &  5.30 & $-$120.8 &   3.15 & $ 1910\pm 47$ & $ 3.8\pm 0.1$ & $-115.8\pm 1.3$ & $  5.0\pm  1.3$ & $  332\pm 146$ & $   18\pm 127$ & $  0.18\pm  0.08$ & $  0.01\pm  0.07$ \\
0454$-$234 &  1       & 32 &  0.85 &    177.7 &   6.85 & $   54\pm 19$ & $ 2.9\pm 1.0$ & $ -71.7\pm22.1$ & $110.6\pm 22.3$ & $  -17\pm  14$ & $   -2\pm  16$ & $ -0.65\pm  0.56$ & $ -0.08\pm  0.60$ \\
0458$-$020 &  1$^{b}$ & 38 &  4.57 &  $-$52.7 &  38.21 & $  287\pm 28$ & $25.8\pm 2.6$ & $ -44.4\pm 5.7$ & $  8.3\pm  5.8$ & $  -74\pm  29$ & $   14\pm  29$ & $ -0.85\pm  0.35$ & $  0.17\pm  0.34$ \\
           &  2$^{b}$ & 41 &  1.77 &  $-$47.9 &  14.82 & $  196\pm 21$ & $17.7\pm 1.9$ & $ -60.4\pm 5.7$ & $ 12.5\pm  5.9$ & $   13\pm  17$ & $   12\pm  18$ & $  0.23\pm  0.30$ & $  0.21\pm  0.31$ \\
0528+134   &  1       & 44 &  3.66 &     23.8 &  31.06 & $   78\pm 11$ & $ 6.7\pm 1.0$ & $  56.1\pm11.1$ & $ 32.4\pm 11.1$ & $   17\pm  12$ & $  -47\pm  14$ & $  0.69\pm  0.50$ & $ -1.86\pm  0.62$ \\
           &  2$^{a}$ & 39 &  1.42 &     38.6 &  12.06 & $  127\pm  9$ & $10.8\pm 0.8$ & $  27.0\pm 3.0$ & $ 11.6\pm  3.1$ & $  -10\pm   9$ & $  -15\pm   7$ & $ -0.25\pm  0.23$ & $ -0.36\pm  0.19$ \\
           &  3$^{c}$ & 35 &  0.46 &     50.9 &   3.87 & $   20\pm  5$ & $ 1.7\pm 0.5$ & $ -17.9\pm17.1$ & $ 68.8\pm 17.1$ & $    9\pm   6$ & $  -12\pm   6$ & $  1.43\pm  1.06$ & $ -1.97\pm  1.07$ \\
           &  4       &  5 &  0.22 &     34.6 &   1.90 & $  205\pm 80$ & $17.5\pm 6.8$ & $  29.1\pm24.7$ & $  5.5\pm 25.3$ & $-1378\pm 703$ & $ -135\pm 956$ & $-20.56\pm 13.19$ & $ -2.02\pm 14.28$ \\
0537$-$441 &  1       & 12 &  2.50 &     55.6 &  19.59 & $  191\pm102$ & $ 9.2\pm 4.9$ & $ 149.4\pm26.8$ & $ 93.8\pm 27.3$ & $   28\pm  81$ & $  -36\pm  76$ & $  0.28\pm  0.82$ & $ -0.36\pm  0.78$ \\
           &  2       &  4 &  0.98 &     65.7 &   7.70 & $  693\pm 55$ & $33.5\pm 2.7$ & $ -15.5\pm 1.8$ & $ 81.2\pm  2.2$ & $-6474\pm 357$ & $ 2753\pm 363$ & $-17.64\pm  1.71$ & $  7.50\pm  1.16$ \\
0552+398   &  1       & 48 &  0.65 &  $-$71.8 &   5.41 & $    6\pm  0$ & $ 0.6\pm 0.1$ & $ 131.6\pm 8.8$ & $156.6\pm  8.8$ & $    0\pm   0$ & $    3\pm   0$ & $ -0.38\pm  0.48$ & $  1.86\pm  0.51$ \\
0642+449   &  1       & 24 &  3.41 &     90.8 &  25.79 & $   17\pm 18$ & $ 1.9\pm 2.0$ & ...             & ...             & ...            & ...            & ...               & ...               \\
           &  3       & 38 &  0.28 &     95.2 &   2.15 & $    9\pm  2$ & $ 1.0\pm 0.3$ & $ -57.2\pm12.1$ & $152.4\pm 12.1$ & $   -3\pm   2$ & $    7\pm   2$ & $ -1.60\pm  1.40$ & $  3.48\pm  1.69$ \\
0727$-$115 &  1       & 43 &  2.21 &  $-$45.3 &  18.98 & $   67\pm 16$ & $ 4.9\pm 1.2$ & $  19.8\pm14.7$ & $ 65.1\pm 14.8$ & $   19\pm  12$ & $  -13\pm  13$ & $  0.75\pm  0.53$ & $ -0.51\pm  0.54$ \\
           &  2       &  4 &  0.75 &  $-$87.7 &   6.41 & $   55\pm137$ & $ 4.0\pm10.0$ & ...             & ...             & ...            & ...            & ...               & ...               \\
           &  3$^{c}$ & 29 &  0.27 & $-$122.5 &   2.30 & $   35\pm 10$ & $ 2.6\pm 0.8$ & $ -32.4\pm13.7$ & $ 90.0\pm 14.2$ & $   10\pm  15$ & $  -24\pm  10$ & $  0.80\pm  1.12$ & $ -1.80\pm  0.92$ \\
0804+499   &  1       & 10 &  2.59 &    136.5 &  22.13 & $   98\pm 96$ & $ 6.7\pm 6.5$ & $ 104.3\pm50.6$ & $ 32.2\pm 50.7$ & $  -41\pm 151$ & $   38\pm 151$ & $ -1.02\pm  3.86$ & $  0.96\pm  3.85$ \\
           &  2       & 44 &  1.12 &    136.6 &   9.54 & $   61\pm  7$ & $ 4.2\pm 0.5$ & $ 144.9\pm 6.2$ & $  8.3\pm  6.3$ & $  -14\pm   6$ & $   -9\pm   5$ & $ -0.59\pm  0.27$ & $ -0.39\pm  0.24$ \\
           &  3$^{c}$ & 36 &  0.30 &     52.6 &   2.59 & $   14\pm 10$ & $ 1.0\pm 0.7$ & $ -87.3\pm40.1$ & $139.9\pm 40.2$ & $  -18\pm  22$ & $   28\pm  17$ & $ -3.19\pm  4.49$ & $  4.76\pm  4.55$ \\
0823+033   &  1       & 11 &  9.80 &     27.4 &  60.49 & $  711\pm259$ & $21.6\pm 7.9$ & $   1.2\pm20.4$ & $ 26.1\pm 20.5$ & $ 1076\pm 917$ & $  508\pm 971$ & $  2.29\pm  2.12$ & $  1.08\pm  2.10$ \\
           &  2       & 18 &  4.05 &     12.7 &  24.99 & $  131\pm 67$ & $ 4.0\pm 2.0$ & $ -11.7\pm18.2$ & $ 24.4\pm 18.2$ & $   53\pm 100$ & $   43\pm  95$ & $  0.61\pm  1.19$ & $  0.50\pm  1.13$ \\
           &  3       & 37 &  2.61 &     18.8 &  16.10 & $   71\pm 11$ & $ 2.2\pm 0.3$ & $ 121.3\pm12.0$ & $102.5\pm 12.0$ & $   45\pm   9$ & $    4\pm  14$ & $  0.96\pm  0.25$ & $  0.10\pm  0.31$ \\
           &  4       & 36 &  1.02 &     31.1 &   6.32 & $   72\pm  7$ & $ 2.2\pm 0.2$ & $  61.9\pm 7.2$ & $ 30.8\pm  7.2$ & $    5\pm   6$ & $  -23\pm   8$ & $  0.11\pm  0.14$ & $ -0.48\pm  0.18$ \\
           &  5       & 12 &  0.59 &     33.3 &   3.66 & $  134\pm 29$ & $ 4.1\pm 0.9$ & $  22.1\pm11.4$ & $ 11.3\pm 11.6$ & $  114\pm  73$ & $   45\pm  76$ & $  1.29\pm  0.88$ & $  0.51\pm  0.87$ \\
           &  6       &  6 &  0.33 &     17.3 &   2.05 & $   75\pm 47$ & $ 2.3\pm 1.5$ & $-121.1\pm48.5$ & $138.4\pm 48.5$ & $  -65\pm 441$ & $  437\pm 371$ & $ -1.31\pm  8.90$ & $  8.79\pm  9.33$ \\
0851+202   &  1       & 12 &  3.60 & $-$111.6 &  16.37 & $  109\pm 53$ & $ 2.1\pm 1.0$ & $-160.4\pm24.0$ & $ 48.8\pm 24.1$ & $ -216\pm 132$ & $ -251\pm 120$ & $ -2.60\pm  2.03$ & $ -3.01\pm  2.06$ \\
           &  2$^{a}$ & 30 &  2.57 & $-$106.9 &  11.68 & $  358\pm 24$ & $ 7.0\pm 0.5$ & $-117.9\pm 3.2$ & $ 11.0\pm  3.3$ & $   -2\pm  14$ & $   -7\pm  15$ & $ -0.01\pm  0.05$ & $ -0.03\pm  0.06$ \\
           &  3       & 17 &  1.21 & $-$103.8 &   5.51 & $  228\pm 33$ & $ 4.4\pm 0.7$ & $ -99.6\pm 6.6$ & $  4.1\pm  6.7$ & $  -16\pm  74$ & $ -159\pm  58$ & $ -0.10\pm  0.43$ & $ -0.92\pm  0.36$ \\
           &  4$^{a}$ & 26 &  0.93 & $-$113.9 &   4.21 & $  203\pm 13$ & $ 3.9\pm 0.3$ & $-119.1\pm 2.3$ & $  5.2\pm  2.4$ & $   56\pm  21$ & $   23\pm  10$ & $  0.36\pm  0.14$ & $  0.15\pm  0.07$ \\
           &  5$^{a}$ & 10 &  0.67 & $-$121.4 &   3.03 & $  256\pm 17$ & $ 5.0\pm 0.3$ & $-122.8\pm 3.1$ & $  1.3\pm  3.3$ & $  206\pm  90$ & $   88\pm  63$ & $  1.06\pm  0.47$ & $  0.45\pm  0.32$ \\
0919$-$260 &  1       & 26 &  6.13 &  $-$57.9 &  51.26 & $  149\pm 52$ & $13.5\pm 4.7$ & $ -87.2\pm36.4$ & $ 29.3\pm 36.4$ & $   20\pm  36$ & $   -2\pm  64$ & $  0.44\pm  0.81$ & $ -0.06\pm  1.42$ \\
           &  2       &  4 &  2.02 &  $-$87.7 &  16.90 & $  244\pm127$ & $22.0\pm11.4$ & $  72.5\pm32.6$ & $160.2\pm 32.9$ & $ 1371\pm 714$ & $ -535\pm1020$ & $ 18.52\pm 13.64$ & $ -7.23\pm 14.27$ \\
           &  3       & 42 &  1.39 &  $-$77.1 &  11.62 & $  116\pm 19$ & $10.5\pm 1.8$ & $ -52.7\pm10.2$ & $ 24.3\pm 10.3$ & $    4\pm  13$ & $  -21\pm  13$ & $  0.11\pm  0.39$ & $ -0.61\pm  0.38$ \\
0920$-$397 &  1       &  7 &  6.46 &    176.4 &  42.93 & $  418\pm197$ & $14.4\pm 6.8$ & $ 150.9\pm18.7$ & $ 25.5\pm 18.8$ & $   34\pm 230$ & $ -279\pm 128$ & $  0.13\pm  0.88$ & $ -1.06\pm  0.70$ \\
           &  2       & 16 &  4.09 &    176.4 &  27.18 & $  274\pm 47$ & $ 9.5\pm 1.7$ & $ 157.5\pm 8.1$ & $ 18.8\pm  8.2$ & $  100\pm  56$ & $  -53\pm  59$ & $  0.58\pm  0.34$ & $ -0.31\pm  0.35$ \\
0923+392   &  1       & 20 &  2.52 &     97.5 &  18.08 & $   34\pm 28$ & $ 1.4\pm 1.1$ & $-128.2\pm41.5$ & $134.3\pm 41.5$ & $  197\pm  89$ & $  105\pm 147$ & $  9.69\pm  9.03$ & $  5.18\pm  8.38$ \\
           &  2$^{ac}$ & 44 &  2.11 &    102.4 &  15.16 & $   48\pm  8$ & $ 1.9\pm 0.3$ & $ 106.7\pm 5.0$ & $  4.2\pm  5.0$ & $    9\pm   5$ & $   -9\pm   5$ & $  0.34\pm  0.20$ & $ -0.32\pm  0.20$ \\
           &  3$^{a}$ & 41 &  1.45 &    107.0 &  10.41 & $  165\pm 14$ & $ 6.6\pm 0.6$ & $ 101.7\pm 1.8$ & $  5.3\pm  1.8$ & $  -15\pm  12$ & $   -8\pm  11$ & $ -0.16\pm  0.13$ & $ -0.09\pm  0.12$ \\
0955+476   &  1       &  6 &  1.10 &    130.1 &   9.41 & $  257\pm 71$ & $20.6\pm 5.8$ & $ 114.2\pm17.5$ & $ 15.9\pm 17.9$ & $  336\pm 178$ & $  150\pm 199$ & $  3.75\pm  2.26$ & $  1.68\pm  2.28$ \\
           &  2       & 15 &  0.57 &    137.7 &   4.87 & $  103\pm 24$ & $ 8.3\pm 2.0$ & $ 142.8\pm14.4$ & $  5.1\pm 14.7$ & $ -154\pm  51$ & $   76\pm  59$ & $ -4.31\pm  1.77$ & $  2.13\pm  1.74$ \\
           &  3       & 22 &  0.21 &    138.4 &   1.76 & $   68\pm 12$ & $ 5.5\pm 1.0$ & $ 103.8\pm 8.1$ & $ 34.5\pm  8.9$ & $   -9\pm  21$ & $   -4\pm  20$ & $ -0.40\pm  0.90$ & $ -0.19\pm  0.87$ \\
\end{tabular}}
\end{center}
\end{sidewaystable*}

\begin{sidewaystable*}
\begin{center}
Table 6 (Continued) \\
{\scriptsize \begin{tabular}{l c c r r r r r r r r r r r} \tableline \tableline \\[5pt] 
\multicolumn{1}{c}{Source} & Comp. & N & \multicolumn{1}{c}{$\langle{r}\rangle$} & \multicolumn{1}{c}{PA} & 
\multicolumn{1}{c}{$d_{proj}$} & \multicolumn{1}{c}{$\mu$} & \multicolumn{1}{c}{$\beta_\mathrm{app}$} & 
\multicolumn{1}{c}{$\phi$} & \multicolumn{1}{c}{$|PA-\phi|$} & \multicolumn{1}{c}{$\dot{\mu}_{\parallel}$} &
\multicolumn{1}{c}{$\dot{\mu}_{\perp}$} & \multicolumn{1}{c}{$\dot{\eta}_{\parallel}$} &
\multicolumn{1}{c}{$\dot{\eta}_{\perp}$} \\
 & & & \multicolumn{1}{c}{(mas)} & \multicolumn{1}{c}{(deg)} & \multicolumn{1}{c}{(pc)} & 
\multicolumn{1}{c}{($\mu as$ yr$^{-1}$)} & & \multicolumn{1}{c}{(deg)} & \multicolumn{1}{c}{(deg)} & 
\multicolumn{1}{c}{($\mu as$ yr$^{-2}$)} & \multicolumn{1}{c}{($\mu as$ yr$^{-2}$)} & \multicolumn{1}{c}{(yr$^{-1}$)} & \multicolumn{1}{c}{(yr$^{-1}$)} \\ 
\multicolumn{1}{c}{(1)} & \multicolumn{1}{c}{(2)} & \multicolumn{1}{c}{(3)} & \multicolumn{1}{c}{(4)} & \multicolumn{1}{c}{(5)} &
\multicolumn{1}{c}{(6)} & \multicolumn{1}{c}{(7)} & \multicolumn{1}{c}{(8)} & \multicolumn{1}{c}{(9)} & \multicolumn{1}{c}{(10)} &
\multicolumn{1}{c}{(11)} & \multicolumn{1}{c}{(12)} & \multicolumn{1}{c}{(13)} & \multicolumn{1}{c}{(14)} \\ \tableline \\[5pt]
1034$-$293 &  1       &  4 &  2.64 &    138.7 &  11.99 & $  428\pm510$ & $ 8.3\pm 9.9$ & $-168.2\pm18.1$ & $ 53.1\pm 20.5$ & $ -722\pm1175$ & $ -146\pm 629$ & $ -2.21\pm  4.45$ & $ -0.45\pm  2.00$ \\
           &  2$^{a}$ & 17 &  2.01 &    140.9 &   9.15 & $  224\pm 21$ & $ 4.4\pm 0.4$ & $ 156.8\pm 4.7$ & $ 15.9\pm  4.9$ & $   -9\pm  24$ & $  -32\pm  23$ & $ -0.06\pm  0.14$ & $ -0.19\pm  0.14$ \\
           &  3       & 13 &  1.31 &    131.2 &   5.95 & $  162\pm 20$ & $ 3.2\pm 0.4$ & $ 140.5\pm 7.5$ & $  9.3\pm  7.8$ & $  -38\pm  21$ & $  -14\pm  23$ & $ -0.31\pm  0.18$ & $ -0.11\pm  0.19$ \\
           &  4$^{c}$ & 16 &  0.52 &    125.9 &   2.36 & $   48\pm 24$ & $ 0.9\pm 0.5$ & $  88.2\pm15.3$ & $ 37.7\pm 15.6$ & $    7\pm  29$ & $   13\pm  25$ & $  0.19\pm  0.81$ & $  0.36\pm  0.70$ \\
1044+719   &  1       &  7 &  0.76 &     62.0 &   6.37 & $  230\pm 63$ & $13.4\pm 3.7$ & $  31.1\pm12.5$ & $ 30.9\pm 13.3$ & $  -85\pm 103$ & $   38\pm  76$ & $ -0.79\pm  0.99$ & $  0.36\pm  0.72$ \\
           &  2$^{a}$ & 32 &  0.52 &    147.0 &   4.36 & $   78\pm  4$ & $ 4.6\pm 0.3$ & $ 156.5\pm 2.3$ & $  9.5\pm  2.3$ & $   17\pm   4$ & $  -16\pm   4$ & $  0.48\pm  0.14$ & $ -0.44\pm  0.14$ \\
1101+384   &  1       & 29 &  5.38 &  $-$41.1 &   3.20 & $   81\pm 41$ & $ 0.2\pm 0.1$ & $ 153.1\pm23.3$ & $165.8\pm 23.3$ & $  -31\pm  48$ & $    0\pm  40$ & $ -0.40\pm  0.65$ & $  0.00\pm  0.52$ \\
           &  2       & 20 &  2.84 &  $-$39.9 &   1.69 & $   47\pm 39$ & $ 0.1\pm 0.1$ & $ 137.7\pm47.7$ & $177.6\pm 47.8$ & $   44\pm  66$ & $   47\pm  64$ & $  0.95\pm  1.63$ & $  1.02\pm  1.62$ \\
           &  3       & 40 &  1.45 &  $-$35.7 &   0.87 & $   78\pm 12$ & $ 0.2\pm 0.0$ & $  38.1\pm10.1$ & $ 73.9\pm 10.2$ & $   34\pm  13$ & $  -20\pm  16$ & $  0.46\pm  0.19$ & $ -0.28\pm  0.22$ \\
           &  4       & 39 &  0.55 &  $-$15.0 &   0.33 & $   30\pm  7$ & $ 0.1\pm 0.0$ & $  65.7\pm14.8$ & $ 80.8\pm 14.9$ & $  -31\pm  10$ & $  -23\pm  11$ & $ -1.09\pm  0.43$ & $ -0.82\pm  0.44$ \\
1124$-$186 &  1       &  9 &  2.67 &    173.9 &  21.79 & $  567\pm108$ & $31.0\pm 5.9$ & $-152.9\pm 7.4$ & $ 33.2\pm  7.7$ & $  -62\pm 254$ & $  431\pm 202$ & $ -0.23\pm  0.92$ & $  1.56\pm  0.79$ \\
           &  2       & 33 &  0.87 &    178.8 &   7.08 & $  173\pm 22$ & $ 9.5\pm 1.2$ & $ 119.5\pm 7.8$ & $ 59.3\pm  8.3$ & $  -60\pm  27$ & $   65\pm  27$ & $ -0.72\pm  0.34$ & $  0.78\pm  0.34$ \\
1128+385   &  1       & 33 &  0.84 & $-$153.9 &   7.20 & $   20\pm  6$ & $ 1.6\pm 0.5$ & $  -4.7\pm21.0$ & $149.3\pm 21.1$ & $    1\pm   6$ & $    8\pm   6$ & $  0.19\pm  0.92$ & $  1.17\pm  0.97$ \\
           &  2       & 35 &  0.37 & $-$165.0 &   3.22 & $   16\pm  2$ & $ 1.2\pm 0.2$ & $-120.1\pm13.2$ & $ 44.9\pm 13.2$ & $    3\pm   4$ & $   -3\pm   3$ & $  0.63\pm  0.68$ & $ -0.61\pm  0.67$ \\
1144$-$379 &  1       &  5 &  3.75 &    157.5 &  30.64 & $   86\pm194$ & $ 4.7\pm10.6$ & ...             & ...             & ...            & ...            & ...               & ...               \\
           &  2       & 14 &  1.12 &    138.1 &   9.18 & $  235\pm 51$ & $12.9\pm 2.8$ & $ 115.6\pm15.4$ & $ 22.5\pm 16.1$ & $   98\pm  59$ & $  -85\pm  62$ & $  0.86\pm  0.55$ & $ -0.74\pm  0.57$ \\
1145$-$071 &  1$^{a}$ & 40 &  2.19 &  $-$66.9 &  18.64 & $   87\pm  5$ & $ 5.7\pm 0.4$ & $ -20.1\pm 2.8$ & $ 46.8\pm  2.9$ & $  -22\pm   4$ & $    4\pm   4$ & $ -0.61\pm  0.14$ & $  0.12\pm  0.12$ \\
1156+295   &  1       &  7 &  7.21 &     22.4 &  52.60 & $  457\pm131$ & $18.9\pm 5.4$ & $ -67.7\pm13.4$ & $ 90.1\pm 13.5$ & $  -77\pm 166$ & $  399\pm 125$ & $ -0.29\pm  0.64$ & $  1.51\pm  0.64$ \\
           &  2$^{a}$ & 39 &  6.21 &     20.3 &  45.31 & $  688\pm 44$ & $28.4\pm 1.9$ & $  41.6\pm 3.9$ & $ 21.3\pm  4.0$ & $   28\pm  29$ & $   30\pm  30$ & $  0.07\pm  0.07$ & $  0.08\pm  0.08$ \\
           &  3$^{a}$ & 34 &  2.25 &   $-$5.8 &  16.41 & $  503\pm 57$ & $20.7\pm 2.4$ & $  -7.2\pm 3.0$ & $  1.4\pm  3.2$ & $  -16\pm  36$ & $    2\pm  19$ & $ -0.06\pm  0.13$ & $  0.01\pm  0.07$ \\
           &  5       & 31 &  0.56 &     11.7 &   4.12 & $   57\pm 12$ & $ 2.4\pm 0.5$ & $  45.6\pm12.3$ & $ 34.0\pm 12.3$ & $  -76\pm  23$ & $   49\pm  22$ & $ -2.29\pm  0.85$ & $  1.47\pm  0.75$ \\
1228+126   &  1       & 32 & 21.38 &  $-$69.2 &   1.75 & $  123\pm126$ & $ 0.0\pm 0.0$ & $  98.3\pm43.5$ & $167.4\pm 43.5$ & $  122\pm 148$ & $  120\pm 149$ & $  0.99\pm  1.57$ & $  0.98\pm  1.57$ \\
           &  2       & 35 & 11.54 &  $-$72.4 &   0.94 & $   53\pm113$ & $ 0.0\pm 0.0$ & $ -84.3\pm80.4$ & $ 11.9\pm 80.4$ & $ -158\pm 142$ & $  -65\pm 240$ & $ -2.95\pm  6.78$ & $ -1.21\pm  5.17$ \\
           &  3       & 36 &  6.57 &  $-$76.3 &   0.54 & $   88\pm 34$ & $ 0.0\pm 0.0$ & $ -92.8\pm32.0$ & $ 16.5\pm 32.0$ & $   -1\pm  37$ & $  -12\pm  51$ & $ -0.01\pm  0.42$ & $ -0.15\pm  0.59$ \\
           &  4       & 36 &  2.97 &  $-$80.4 &   0.24 & $  114\pm 33$ & $ 0.0\pm 0.0$ & $-115.7\pm10.2$ & $ 35.3\pm 10.2$ & $   70\pm  32$ & $   56\pm  29$ & $  0.62\pm  0.34$ & $  0.50\pm  0.30$ \\
           &  5       & 37 &  1.48 &  $-$77.7 &   0.12 & $   41\pm 10$ & $ 0.0\pm 0.0$ & $  17.8\pm28.9$ & $ 95.5\pm 28.9$ & $   22\pm  26$ & $  -40\pm  20$ & $  0.54\pm  0.67$ & $ -1.00\pm  0.55$ \\
           &  6       & 39 &  0.57 &  $-$79.6 &   0.05 & $    6\pm  9$ & $ 0.0\pm 0.0$ & $ -40.3\pm86.6$ & $ 39.3\pm 86.6$ & $    6\pm  12$ & $   -4\pm  13$ & $  0.94\pm  2.29$ & $ -0.64\pm  2.21$ \\
1308+326   &  1$^{a}$ & 40 &  1.74 &  $-$73.3 &  14.05 & $  398\pm 10$ & $21.0\pm 0.5$ & $ -71.7\pm 1.2$ & $  1.6\pm  1.2$ & $   52\pm  11$ & $  -24\pm   9$ & $  0.27\pm  0.06$ & $ -0.12\pm  0.05$ \\
           &  2$^{a}$ & 20 &  1.44 &  $-$44.6 &  11.62 & $  488\pm 13$ & $25.8\pm 0.7$ & $ -53.0\pm 1.3$ & $  8.3\pm  1.3$ & $  -77\pm  27$ & $  -23\pm  33$ & $ -0.32\pm  0.11$ & $ -0.10\pm  0.14$ \\
1313$-$333 &  1       &  8 &  7.43 &  $-$81.2 &  62.41 & $  730\pm230$ & $44.2\pm14.0$ & $ -66.4\pm33.5$ & $ 14.8\pm 33.5$ & $ -162\pm 662$ & $ -486\pm1266$ & $ -0.49\pm  2.01$ & $ -1.47\pm  3.86$ \\
           &  2       &  4 &  2.07 & $-$117.5 &  17.37 & $  504\pm365$ & $30.6\pm22.1$ & $ -47.1\pm43.8$ & $ 70.4\pm 44.6$ & $ 2550\pm1770$ & $ 1143\pm2226$ & $ 11.16\pm 11.20$ & $  5.00\pm 10.40$ \\
           &  3$^{a}$ & 38 &  2.08 &  $-$87.1 &  17.49 & $  487\pm 34$ & $29.5\pm 2.1$ & $ -89.2\pm 3.9$ & $  2.2\pm  4.2$ & $   21\pm  37$ & $   12\pm  37$ & $  0.10\pm  0.17$ & $  0.06\pm  0.17$ \\
1334$-$127 &  1       & 22 &  2.78 &    152.7 &  17.68 & $  103\pm 18$ & $ 3.3\pm 0.6$ & $ 122.1\pm14.2$ & $ 30.6\pm 14.2$ & $   -8\pm  25$ & $    3\pm  44$ & $ -0.13\pm  0.39$ & $  0.05\pm  0.67$ \\
           &  2$^{a}$ & 22 &  1.66 &    145.6 &  10.59 & $  228\pm 12$ & $ 7.3\pm 0.4$ & $ 151.8\pm 2.2$ & $  6.2\pm  2.3$ & $   11\pm  18$ & $   -5\pm  20$ & $  0.08\pm  0.12$ & $ -0.04\pm  0.14$ \\
           &  3       & 17 &  0.98 &    139.9 &   6.23 & $  293\pm 36$ & $ 9.4\pm 1.2$ & $ 141.0\pm 7.3$ & $  1.1\pm  7.6$ & $   17\pm  77$ & $   45\pm  75$ & $  0.09\pm  0.41$ & $  0.24\pm  0.40$ \\
1357+769   &  1       &  6 &  2.44 & $-$121.8 &  20.97 & $  124\pm 83$ & $ 9.0\pm 6.1$ & $-165.3\pm35.5$ & $ 43.5\pm 35.9$ & $  114\pm  91$ & $  -33\pm 108$ & $  2.37\pm  2.48$ & $ -0.70\pm  2.30$ \\
           &  2       & 11 &  1.39 & $-$119.0 &  11.94 & $  115\pm 38$ & $ 8.4\pm 2.8$ & $-135.1\pm18.9$ & $ 16.2\pm 19.1$ & $  -77\pm  63$ & $  -53\pm  63$ & $ -1.73\pm  1.53$ & $ -1.20\pm  1.49$ \\
           &  3$^{b}$ & 14 &  0.52 & $-$133.3 &   4.45 & $  115\pm 15$ & $ 8.4\pm 1.1$ & $-106.5\pm 5.5$ & $ 26.8\pm  5.8$ & $  -12\pm  14$ & $  -18\pm  14$ & $ -0.28\pm  0.32$ & $ -0.41\pm  0.33$ \\
           &  4$^{c}$ & 26 &  0.21 &  $-$88.2 &   1.81 & $   12\pm  8$ & $ 0.9\pm 0.6$ & $-161.0\pm65.7$ & $ 72.8\pm 65.7$ & $   14\pm  72$ & $   62\pm  29$ & $  2.99\pm 15.49$ & $ 13.16\pm 10.87$ \\
1424$-$418 &  1       & 19 &  2.64 &     76.9 &  22.66 & $   88\pm 50$ & $ 6.3\pm 3.6$ & ...             & ...             & ...            & ...            & ...               & ...               \\
1448+762   &  1$^{c}$ & 16 &  1.57 &     77.2 &  12.29 & $   24\pm 21$ & $ 1.2\pm 1.1$ & $  11.4\pm87.1$ & $ 65.9\pm 87.1$ & $  -83\pm 188$ & $ -118\pm 142$ & $ -6.38\pm 15.49$ & $ -9.07\pm 13.46$ \\
           &  2       & 18 &  0.96 &     80.5 &   7.52 & $   57\pm 27$ & $ 2.8\pm 1.3$ & $-108.1\pm11.4$ & $171.4\pm 11.4$ & $  132\pm  60$ & $   37\pm  35$ & $  4.40\pm  2.94$ & $  1.26\pm  1.31$ \\
           &  3$^{a}$ & 20 &  0.54 &     81.9 &   4.26 & $   86\pm 17$ & $ 4.2\pm 0.8$ & $-100.3\pm 3.2$ & $177.8\pm  3.3$ & $  -16\pm  34$ & $   -2\pm   8$ & $ -0.37\pm  0.75$ & $ -0.05\pm  0.19$ \\
1451$-$375 &  1       &  8 &  7.65 & $-$149.7 &  34.74 & $  532\pm211$ & $10.3\pm 4.1$ & $-108.2\pm32.0$ & $ 41.5\pm 32.2$ & $ -148\pm 594$ & $ -730\pm 531$ & $ -0.36\pm  1.47$ & $ -1.80\pm  1.49$ \\
           &  2       & 20 &  2.04 & $-$132.7 &   9.27 & $  279\pm 59$ & $ 5.4\pm 1.2$ & $-134.6\pm12.3$ & $  1.9\pm 12.7$ & $   58\pm  68$ & $  100\pm  60$ & $  0.28\pm  0.33$ & $  0.47\pm  0.30$ \\
1514$-$241 &  1       &  9 & 11.54 &    158.9 &  11.18 & $ 2977\pm758$ & $ 9.9\pm 2.5$ & $ 157.0\pm 7.5$ & $  1.9\pm  7.6$ & $10136\pm4015$ & $  436\pm2345$ & $  3.57\pm  1.68$ & $  0.15\pm  0.83$ \\
           &  2       &  8 &  7.58 &    156.1 &   7.35 & $ 1341\pm175$ & $ 4.5\pm 0.6$ & $ 146.4\pm 5.8$ & $  9.7\pm  5.8$ & $ 1824\pm 626$ & $ 1212\pm 508$ & $  1.43\pm  0.52$ & $  0.95\pm  0.42$ \\
           &  3$^{a}$ & 20 &  9.96 &    155.2 &   9.65 & $ 1567\pm144$ & $ 5.2\pm 0.5$ & $ 154.1\pm 3.4$ & $  1.2\pm  3.4$ & $  384\pm 225$ & $  163\pm 137$ & $  0.26\pm  0.15$ & $  0.11\pm  0.09$ \\
           &  4       &  8 &  5.07 &    156.0 &   4.91 & $ 1780\pm341$ & $ 5.9\pm 1.1$ & $ 142.2\pm11.1$ & $ 13.8\pm 11.3$ & $ 1672\pm1491$ & $  332\pm1549$ & $  0.99\pm  0.90$ & $  0.20\pm  0.91$ \\
           &  5       & 10 &  1.76 &    160.2 &   1.70 & $  579\pm171$ & $ 1.9\pm 0.6$ & $ 169.6\pm 7.3$ & $  9.4\pm  7.4$ & $ -834\pm 638$ & $  889\pm 625$ & $ -1.51\pm  1.24$ & $  1.61\pm  1.23$ \\
\end{tabular}}
\end{center}
\end{sidewaystable*}

\begin{sidewaystable*}
\begin{center}
Table 6 (Continued) \\
{\scriptsize \begin{tabular}{l c c r r r r r r r r r r r} \tableline \tableline \\[5pt]
\multicolumn{1}{c}{Source} & Comp. & N & \multicolumn{1}{c}{$\langle{r}\rangle$} & \multicolumn{1}{c}{PA} & 
\multicolumn{1}{c}{$d_{proj}$} & \multicolumn{1}{c}{$\mu$} & \multicolumn{1}{c}{$\beta_\mathrm{app}$} & 
\multicolumn{1}{c}{$\phi$} & \multicolumn{1}{c}{$|PA-\phi|$} & \multicolumn{1}{c}{$\dot{\mu}_{\parallel}$} &
\multicolumn{1}{c}{$\dot{\mu}_{\perp}$} & \multicolumn{1}{c}{$\dot{\eta}_{\parallel}$} &
\multicolumn{1}{c}{$\dot{\eta}_{\perp}$} \\
 & & & \multicolumn{1}{c}{(mas)} & \multicolumn{1}{c}{(deg)} & \multicolumn{1}{c}{(pc)} & 
\multicolumn{1}{c}{($\mu as$ yr$^{-1}$)} & & \multicolumn{1}{c}{(deg)} & \multicolumn{1}{c}{(deg)} & 
\multicolumn{1}{c}{($\mu as$ yr$^{-2}$)} & \multicolumn{1}{c}{($\mu as$ yr$^{-2}$)} & \multicolumn{1}{c}{(yr$^{-1}$)} & \multicolumn{1}{c}{(yr$^{-1}$)} \\
\multicolumn{1}{c}{(1)} & \multicolumn{1}{c}{(2)} & \multicolumn{1}{c}{(3)} & \multicolumn{1}{c}{(4)} & \multicolumn{1}{c}{(5)} &
\multicolumn{1}{c}{(6)} & \multicolumn{1}{c}{(7)} & \multicolumn{1}{c}{(8)} & \multicolumn{1}{c}{(9)} & \multicolumn{1}{c}{(10)} &
\multicolumn{1}{c}{(11)} & \multicolumn{1}{c}{(12)} & \multicolumn{1}{c}{(13)} & \multicolumn{1}{c}{(14)} \\ \tableline \\[5pt]
1606+106   &  1       & 19 &  7.63 &  $-$37.7 &  64.20 & $   14\pm 48$ & $ 0.9\pm 3.0$ & ...             & ...             & ...            & ...            & ...               & ...               \\
           &  2       & 31 &  2.47 &  $-$38.2 &  20.78 & $   79\pm 18$ & $ 4.9\pm 1.2$ & $ -43.3\pm13.2$ & $  5.1\pm 13.2$ & $   17\pm  17$ & $   12\pm  14$ & $  0.49\pm  0.49$ & $  0.36\pm  0.42$ \\
           &  3$^{c}$ & 40 &  1.52 &  $-$54.2 &  12.82 & $   14\pm  8$ & $ 0.9\pm 0.5$ & $ -47.9\pm34.2$ & $  6.3\pm 34.2$ & $    0\pm   9$ & $    1\pm   8$ & $ -0.08\pm  1.40$ & $  0.23\pm  1.24$ \\
           &  4       & 39 &  0.53 &  $-$66.1 &   4.43 & $   31\pm  6$ & $ 1.9\pm 0.4$ & $ 128.8\pm12.2$ & $165.1\pm 12.2$ & $  -12\pm   7$ & $   -3\pm   7$ & $ -0.92\pm  0.54$ & $ -0.23\pm  0.57$ \\
1611+343   &  1       & 43 &  3.59 &    167.7 &  30.68 & $  118\pm 17$ & $ 7.9\pm 1.2$ & $-141.1\pm 8.9$ & $ 51.2\pm  9.0$ & $  -47\pm  12$ & $  -18\pm  13$ & $ -0.96\pm  0.29$ & $ -0.37\pm  0.28$ \\
           &  2       & 28 &  4.01 &    148.1 &  34.26 & $  140\pm 26$ & $ 9.4\pm 1.8$ & $ 116.0\pm 6.3$ & $ 32.1\pm  6.3$ & $   90\pm  38$ & $  -18\pm  19$ & $  1.54\pm  0.72$ & $ -0.31\pm  0.34$ \\
           &  3$^{c}$ & 43 &  2.84 &    174.2 &  24.28 & $   36\pm  5$ & $ 2.5\pm 0.3$ & $ 128.4\pm 7.8$ & $ 45.8\pm  7.8$ & $    3\pm   3$ & $   -4\pm   3$ & $  0.22\pm  0.22$ & $ -0.28\pm  0.22$ \\
           &  4$^{a}$ & 30 &  1.38 &    168.5 &  11.80 & $  183\pm 18$ & $12.3\pm 1.2$ & $ 161.9\pm 3.2$ & $  6.6\pm  3.3$ & $    0\pm  21$ & $    0\pm  16$ & $ -0.01\pm  0.28$ & $ -0.01\pm  0.21$ \\
           &  5$^{a}$ & 28 &  0.72 &    157.5 &   6.18 & $  221\pm  6$ & $14.8\pm 0.4$ & $ 167.8\pm 1.6$ & $ 10.3\pm  1.7$ & $   24\pm   8$ & $    7\pm   8$ & $  0.26\pm  0.10$ & $  0.09\pm  0.09$ \\
           &  6$^{b}$ & 11 &  0.49 &    163.6 &   4.20 & $  340\pm 49$ & $22.8\pm 3.3$ & $ 157.2\pm 5.1$ & $  6.5\pm  5.4$ & $  123\pm 191$ & $   19\pm 101$ & $  0.87\pm  1.36$ & $  0.14\pm  0.72$ \\
1622$-$253 &  1       & 18 &  2.68 &  $-$26.1 &  20.17 & $  316\pm 57$ & $13.9\pm 2.5$ & $ -96.2\pm11.8$ & $ 70.1\pm 12.0$ & $  -93\pm  64$ & $ -133\pm  61$ & $ -0.53\pm  0.38$ & $ -0.76\pm  0.37$ \\
           &  2$^{b}$ & 20 &  1.07 &  $-$16.2 &   8.07 & $  220\pm 30$ & $ 9.7\pm 1.3$ & $  -8.9\pm 5.1$ & $  7.3\pm  5.5$ & $   63\pm  35$ & $ -148\pm  26$ & $  0.52\pm  0.30$ & $ -1.21\pm  0.27$ \\
1638+398   &  1       & 10 &  0.55 & $-$170.9 &   4.73 & $   52\pm 18$ & $ 3.9\pm 1.4$ & $ -96.0\pm 9.2$ & $ 74.9\pm 10.0$ & $   34\pm  11$ & $   15\pm   9$ & $  1.77\pm  0.86$ & $  0.79\pm  0.56$ \\
           &  2$^{c}$ & 15 &  0.40 & $-$162.5 &   3.41 & $   37\pm  8$ & $ 2.8\pm 0.6$ & $-126.7\pm13.9$ & $ 35.8\pm 14.0$ & $    7\pm  10$ & $    5\pm  11$ & $  0.57\pm  0.77$ & $  0.36\pm  0.84$ \\
           &  3       & 14 &  0.16 & $-$104.0 &   1.37 & $   72\pm 13$ & $ 5.4\pm 1.0$ & $ 171.0\pm 5.3$ & $ 85.0\pm  6.3$ & $  176\pm  33$ & $   14\pm  22$ & $  6.51\pm  1.73$ & $  0.54\pm  0.85$ \\
1642+690   &  1       & 25 &  9.63 & $-$165.9 &  71.03 & $   60\pm 22$ & $ 2.6\pm 0.9$ & $-179.1\pm 8.4$ & $ 13.2\pm  8.4$ & $   76\pm  38$ & $   25\pm  26$ & $  2.21\pm  1.38$ & $  0.73\pm  0.80$ \\
           &  2$^{a}$ & 13 &  4.84 & $-$164.3 &  35.72 & $  571\pm 63$ & $24.1\pm 2.7$ & $-163.5\pm 3.1$ & $  0.9\pm  3.1$ & $ -123\pm 183$ & $  -36\pm 141$ & $ -0.38\pm  0.56$ & $ -0.11\pm  0.43$ \\
           &  3$^{a}$ & 25 &  3.82 & $-$161.6 &  28.20 & $  342\pm 25$ & $14.4\pm 1.1$ & $-158.7\pm 3.0$ & $  2.8\pm  3.0$ & $   76\pm  49$ & $  -28\pm  33$ & $  0.39\pm  0.25$ & $ -0.15\pm  0.17$ \\
           &  4$^{a}$ & 21 &  2.81 & $-$170.1 &  20.73 & $  358\pm 20$ & $15.1\pm 0.9$ & $-164.4\pm 2.4$ & $  5.7\pm  2.5$ & $   -8\pm  36$ & $  -29\pm  27$ & $ -0.04\pm  0.18$ & $ -0.14\pm  0.13$ \\
           &  5$^{a}$ & 22 &  1.68 & $-$173.2 &  12.37 & $  226\pm 28$ & $ 9.6\pm 1.2$ & $-165.3\pm 4.9$ & $  8.0\pm  4.9$ & $  197\pm  47$ & $   40\pm  40$ & $  1.52\pm  0.41$ & $  0.31\pm  0.31$ \\
           &  6$^{a}$ & 25 &  1.20 & $-$179.1 &   8.84 & $  174\pm 11$ & $ 7.3\pm 0.5$ & $-159.6\pm 1.9$ & $ 19.4\pm  1.9$ & $   25\pm  18$ & $   30\pm  16$ & $  0.26\pm  0.19$ & $  0.31\pm  0.17$ \\
           &  7       & 25 &  0.43 &    175.2 &   3.15 & $   56\pm 18$ & $ 2.4\pm 0.8$ & $-170.2\pm 7.3$ & $ 14.6\pm  7.3$ & $    8\pm  32$ & $   19\pm  20$ & $  0.28\pm  1.02$ & $  0.61\pm  0.66$ \\
1657$-$261 &  1       & 11 &  0.80 &     28.5 &   ...  & $  236\pm 85$ & ...           & $  -8.7\pm 6.6$ & $ 37.2\pm  7.5$ & $  125\pm 127$ & $  231\pm 110$ & ...               & ...               \\
1726+455   &  1$^{b}$ & 19 &  1.81 &  $-$87.8 &  13.07 & $  204\pm 30$ & $ 8.2\pm 1.2$ & $ -94.4\pm 5.6$ & $  6.6\pm  5.7$ & $   65\pm  24$ & $   -2\pm  17$ & $  0.55\pm  0.22$ & $ -0.02\pm  0.15$ \\
           &  2$^{a}$ & 15 &  0.92 & $-$103.3 &   6.63 & $  322\pm 37$ & $13.0\pm 1.5$ & $ -99.0\pm 3.1$ & $  4.3\pm  3.4$ & $   96\pm  37$ & $   22\pm  19$ & $  0.51\pm  0.21$ & $  0.12\pm  0.10$ \\
1739+522   &  1       &  9 &  1.15 &     16.5 &   9.84 & $  104\pm 61$ & $ 6.9\pm 4.1$ & $ -36.8\pm28.9$ & $ 53.3\pm 28.9$ & $ -201\pm 202$ & $ -334\pm 169$ & $ -4.59\pm  5.35$ & $ -7.62\pm  5.93$ \\
           &  2       & 26 &  0.37 &     32.2 &   3.13 & $   73\pm 10$ & $ 4.9\pm 0.7$ & $  50.3\pm 9.0$ & $ 18.1\pm  9.4$ & $    3\pm  15$ & $  -52\pm  13$ & $  0.10\pm  0.50$ & $ -1.72\pm  0.52$ \\
1741$-$038 &  1       &  6 &  1.83 & $-$174.1 &  15.01 & $  143\pm172$ & $ 7.9\pm 9.5$ & $  93.1\pm28.7$ & $ 92.7\pm 29.9$ & $   70\pm 235$ & $ -143\pm 182$ & $  1.01\pm  3.59$ & $ -2.05\pm  3.59$ \\
           &  2       & 10 &  0.97 & $-$149.0 &   7.91 & $   76\pm 61$ & $ 4.2\pm 3.4$ & $ -26.0\pm28.1$ & $123.1\pm 28.1$ & $  -26\pm  95$ & $  -62\pm  97$ & $ -0.71\pm  2.63$ & $ -1.69\pm  2.95$ \\
           &  3$^{c}$ & 34 &  0.43 & $-$178.4 &   3.53 & $   33\pm  6$ & $ 1.9\pm 0.4$ & $ 164.6\pm 7.1$ & $ 17.0\pm  7.2$ & $   -4\pm   4$ & $    3\pm   4$ & $ -0.25\pm  0.25$ & $  0.20\pm  0.26$ \\
1745+624   &  1       & 15 &  2.57 & $-$141.2 &  18.48 & $   76\pm 29$ & $ 8.8\pm 3.3$ & $-140.9\pm18.9$ & $  0.3\pm 18.9$ & $  -10\pm  29$ & $   -2\pm  35$ & $ -0.66\pm  1.91$ & $ -0.17\pm  2.28$ \\
           &  2       & 28 &  1.46 & $-$145.1 &  10.53 & $   60\pm 12$ & $ 6.9\pm 1.4$ & $-131.8\pm11.7$ & $ 13.2\pm 11.8$ & $   52\pm  15$ & $  -14\pm  18$ & $  4.25\pm  1.51$ & $ -1.19\pm  1.52$ \\
           &  3       &  6 &  1.10 & $-$144.9 &   7.94 & $  134\pm 52$ & $15.4\pm 6.0$ & $-142.3\pm23.6$ & $  2.5\pm 23.7$ & $  -59\pm 284$ & $  275\pm 253$ & $ -2.17\pm 10.40$ & $ 10.05\pm 10.02$ \\
           &  4       &  6 &  0.54 & $-$146.7 &   3.91 & $   62\pm 55$ & $ 7.2\pm 6.4$ & $-149.9\pm43.2$ & $  3.2\pm 43.4$ & $ -108\pm 128$ & $   49\pm 127$ & $ -8.48\pm 12.65$ & $  3.90\pm 10.63$ \\
           &  5       & 40 &  0.24 & $-$135.8 &   1.71 & $   11\pm  2$ & $ 1.3\pm 0.3$ & $-134.9\pm11.6$ & $  0.9\pm 11.7$ & $   12\pm   2$ & $    3\pm   3$ & $  5.23\pm  1.50$ & $  1.55\pm  1.59$ \\
1749+096   &  1       &  9 &  3.92 &     28.3 &  18.19 & $  812\pm133$ & $16.3\pm 2.7$ & $  43.3\pm 9.3$ & $ 15.0\pm  9.5$ & $  110\pm 265$ & $  255\pm 232$ & $  0.18\pm  0.43$ & $  0.42\pm  0.38$ \\
           &  2$^{b}$ & 13 &  2.45 &     25.2 &  11.37 & $  706\pm 83$ & $14.1\pm 1.7$ & $  22.5\pm 5.0$ & $  2.6\pm  5.3$ & $   60\pm 111$ & $  -36\pm  91$ & $  0.11\pm  0.21$ & $ -0.07\pm  0.17$ \\
           &  3$^{b}$ & 11 &  1.08 &     37.8 &   5.02 & $  547\pm 49$ & $10.9\pm 1.0$ & $  47.2\pm 5.4$ & $  9.4\pm  5.7$ & $  566\pm 158$ & $ -148\pm 140$ & $  1.37\pm  0.40$ & $ -0.36\pm  0.34$ \\
           &  4$^{b}$ & 22 &  1.94 &     40.1 &   9.01 & $  445\pm 37$ & $ 8.9\pm 0.8$ & $  35.8\pm 5.4$ & $  4.3\pm  5.7$ & $ -141\pm  50$ & $  -64\pm  46$ & $ -0.42\pm  0.15$ & $ -0.19\pm  0.14$ \\
           &  5$^{b}$ & 21 &  0.86 &      7.3 &   3.99 & $  188\pm 24$ & $ 3.8\pm 0.5$ & $   8.6\pm 5.2$ & $  1.3\pm  5.3$ & $   21\pm  32$ & $   81\pm  23$ & $  0.15\pm  0.23$ & $  0.57\pm  0.18$ \\
1803+784   &  1       & 41 &  7.15 &  $-$96.0 &  50.68 & $   19\pm 37$ & $ 0.7\pm 1.5$ & $ -54.4\pm87.1$ & $ 41.6\pm 87.1$ & $  -64\pm  90$ & $   57\pm  99$ & $ -5.62\pm 13.52$ & $  5.06\pm 13.17$ \\
           &  2       & 41 &  3.45 &  $-$91.8 &  24.44 & $   73\pm 36$ & $ 2.9\pm 1.4$ & $-125.2\pm20.8$ & $ 33.4\pm 20.8$ & $   -7\pm  33$ & $   59\pm  15$ & $ -0.16\pm  0.77$ & $  1.36\pm  0.76$ \\
           &  3       & 31 &  1.83 &  $-$93.4 &  12.96 & $   39\pm  9$ & $ 1.5\pm 0.4$ & $  49.3\pm13.3$ & $142.8\pm 13.3$ & $   17\pm   6$ & $   -7\pm   7$ & $  0.74\pm  0.33$ & $ -0.32\pm  0.35$ \\
           &  4       & 43 &  1.44 &  $-$92.8 &  10.21 & $   33\pm  5$ & $ 1.3\pm 0.2$ & $  23.6\pm 9.1$ & $116.4\pm  9.1$ & $    4\pm   3$ & $    6\pm   3$ & $  0.21\pm  0.18$ & $  0.32\pm  0.18$ \\
           &  5       & 31 &  1.03 &  $-$81.1 &   7.27 & $   59\pm  6$ & $ 2.3\pm 0.3$ & $   9.3\pm12.2$ & $ 90.4\pm 12.2$ & $  -17\pm  17$ & $   55\pm  17$ & $ -0.49\pm  0.49$ & $  1.55\pm  0.52$ \\
           &  6$^{c}$ & 41 &  0.47 &  $-$82.0 &   3.35 & $   22\pm  6$ & $ 0.9\pm 0.2$ & $ -77.6\pm 8.9$ & $  4.4\pm  8.9$ & $   -1\pm   6$ & $   -2\pm   4$ & $ -0.14\pm  0.46$ & $ -0.22\pm  0.38$ \\
1908$-$201 &  1       & 11 &  5.26 &     54.2 &  43.60 & $  388\pm119$ & $22.3\pm 6.8$ & $  29.4\pm11.7$ & $ 24.8\pm 12.0$ & $ -130\pm 141$ & $ -155\pm 125$ & $ -0.71\pm  0.80$ & $ -0.85\pm  0.73$ \\
           &  2$^{b}$ & 38 &  2.84 &     39.1 &  23.53 & $  200\pm 18$ & $11.5\pm 1.1$ & $  55.1\pm 5.1$ & $ 16.0\pm  5.1$ & $  -45\pm  21$ & $  -20\pm  19$ & $ -0.48\pm  0.23$ & $ -0.22\pm  0.20$ \\
           &  3       &  7 &  1.12 &     31.4 &   9.27 & $  293\pm208$ & $16.8\pm11.9$ & $  66.5\pm19.1$ & $ 35.1\pm 20.7$ & $  864\pm 789$ & $  315\pm 433$ & $  6.24\pm  7.22$ & $  2.28\pm  3.52$ \\
           &  4$^{a}$ & 19 &  0.82 &   $-$3.8 &   6.80 & $  187\pm 26$ & $10.7\pm 1.5$ & $   3.0\pm 3.3$ & $  6.8\pm  3.4$ & $  -23\pm  48$ & $    1\pm  19$ & $ -0.26\pm  0.55$ & $  0.01\pm  0.22$ \\
\end{tabular}}
\end{center}
\end{sidewaystable*}

\begin{sidewaystable*}
\begin{center}
Table 6 (Continued) \\
{\scriptsize \begin{tabular}{l c c r r r r r r r r r r r} \tableline \tableline \\[5pt]
\multicolumn{1}{c}{Source} & Comp. & N & \multicolumn{1}{c}{$\langle{r}\rangle$} & \multicolumn{1}{c}{PA} & 
\multicolumn{1}{c}{$d_{proj}$} & \multicolumn{1}{c}{$\mu$} & \multicolumn{1}{c}{$\beta_\mathrm{app}$} & 
\multicolumn{1}{c}{$\phi$} & \multicolumn{1}{c}{$|PA-\phi|$} & \multicolumn{1}{c}{$\dot{\mu}_{\parallel}$} &
\multicolumn{1}{c}{$\dot{\mu}_{\perp}$} & \multicolumn{1}{c}{$\dot{\eta}_{\parallel}$} &
\multicolumn{1}{c}{$\dot{\eta}_{\perp}$} \\
 & & & \multicolumn{1}{c}{(mas)} & \multicolumn{1}{c}{(deg)} & \multicolumn{1}{c}{(pc)} & 
\multicolumn{1}{c}{($\mu as$ yr$^{-1}$)} & & \multicolumn{1}{c}{(deg)} & \multicolumn{1}{c}{(deg)} & 
\multicolumn{1}{c}{($\mu as$ yr$^{-2}$)} & \multicolumn{1}{c}{($\mu as$ yr$^{-2}$)} & \multicolumn{1}{c}{(yr$^{-1}$)} & \multicolumn{1}{c}{(yr$^{-1}$)} \\
\multicolumn{1}{c}{(1)} & \multicolumn{1}{c}{(2)} & \multicolumn{1}{c}{(3)} & \multicolumn{1}{c}{(4)} & \multicolumn{1}{c}{(5)} &
\multicolumn{1}{c}{(6)} & \multicolumn{1}{c}{(7)} & \multicolumn{1}{c}{(8)} & \multicolumn{1}{c}{(9)} & \multicolumn{1}{c}{(10)} &
\multicolumn{1}{c}{(11)} & \multicolumn{1}{c}{(12)} & \multicolumn{1}{c}{(13)} & \multicolumn{1}{c}{(14)} \\ \tableline \\[5pt]
1921$-$293 &  1       & 39 &  6.20 &     25.7 &  30.55 & $  191\pm 27$ & $ 4.2\pm 0.6$ & $   2.8\pm10.1$ & $ 23.0\pm 10.2$ & $    0\pm  21$ & $    0\pm  21$ & $  0.01\pm  0.15$ & $  0.01\pm  0.15$ \\
           &  2       & 10 &  2.99 &     21.3 &  14.73 & $  142\pm 73$ & $ 3.1\pm 1.6$ & $  17.1\pm25.0$ & $  4.2\pm 25.2$ & $   83\pm  62$ & $   -5\pm  63$ & $  0.79\pm  0.71$ & $ -0.05\pm  0.60$ \\
           &  3       & 11 &  1.27 &     13.2 &   6.28 & $  268\pm108$ & $ 5.8\pm 2.3$ & $  -1.3\pm21.7$ & $ 14.6\pm 22.2$ & $  -27\pm 299$ & $   70\pm 288$ & $ -0.14\pm  1.50$ & $  0.36\pm  1.46$ \\
1954$-$388 &  1       & 14 &  2.42 & $-$117.5 &  16.62 & $  258\pm 81$ & $ 9.4\pm 3.0$ & $ -51.6\pm21.0$ & $ 65.9\pm 21.5$ & $ -177\pm 113$ & $  -49\pm 113$ & $ -1.12\pm  0.79$ & $ -0.31\pm  0.72$ \\
           &  2       & 16 &  0.83 &  $-$81.6 &   5.71 & $  137\pm 31$ & $ 5.0\pm 1.2$ & $-121.2\pm14.2$ & $ 39.6\pm 15.0$ & $  -19\pm  39$ & $  -29\pm  37$ & $ -0.23\pm  0.47$ & $ -0.35\pm  0.45$ \\
2145+067   &  1$^{c}$ & 28 &  5.39 &    131.7 &  43.52 & $    1\pm 34$ & $ 0.1\pm 1.8$ & ...             & ...             & ...            & ...            & ...               & ...               \\
           &  2       & 13 &  2.51 &    130.5 &  20.24 & $   65\pm 32$ & $ 3.5\pm 1.7$ & $ -73.3\pm36.6$ & $156.2\pm 36.7$ & $  -13\pm  30$ & $  -33\pm  28$ & $ -0.41\pm  0.95$ & $ -1.02\pm  1.00$ \\
           &  3       & 24 &  1.15 &    124.5 &   9.25 & $  126\pm 20$ & $ 6.7\pm 1.1$ & $ 122.8\pm 8.2$ & $  1.8\pm  8.4$ & $  -13\pm  14$ & $   20\pm  15$ & $ -0.21\pm  0.24$ & $  0.32\pm  0.26$ \\
           &  4$^{a}$ & 35 &  0.81 &    128.8 &   6.55 & $   86\pm  5$ & $ 4.6\pm 0.3$ & $ 130.5\pm 3.7$ & $  1.7\pm  3.7$ & $  -12\pm   6$ & $   -4\pm   6$ & $ -0.30\pm  0.15$ & $ -0.09\pm  0.15$ \\
           &  5$^{c}$ & 18 &  0.53 &    123.0 &   4.30 & $   28\pm 15$ & $ 1.5\pm 0.8$ & $  84.5\pm23.9$ & $ 38.5\pm 24.0$ & $   37\pm  43$ & $   77\pm  31$ & $  2.62\pm  3.38$ & $  5.38\pm  3.69$ \\
2200+420   &  1       &  6 &  7.74 &    168.1 &  10.26 & $ 1075\pm 82$ & $ 5.0\pm 0.4$ & $ 146.3\pm 3.2$ & $ 21.8\pm  3.2$ & $  287\pm 143$ & $  268\pm 136$ & $  0.29\pm  0.14$ & $  0.27\pm  0.14$ \\
           &  2       & 21 &  7.48 &    166.2 &   9.92 & $  593\pm 87$ & $ 2.7\pm 0.4$ & $ 149.8\pm 7.1$ & $ 16.4\pm  7.2$ & $  -75\pm 149$ & $   50\pm 162$ & $ -0.14\pm  0.27$ & $  0.09\pm  0.29$ \\
           &  3$^{a}$ & 13 &  3.93 & $-$179.6 &   5.21 & $  667\pm129$ & $ 3.1\pm 0.6$ & $ 174.8\pm 5.0$ & $  5.6\pm  5.0$ & $  -53\pm 402$ & $ -699\pm 188$ & $ -0.09\pm  0.65$ & $ -1.12\pm  0.37$ \\
           &  4$^{a}$ & 10 &  3.06 & $-$178.6 &   4.05 & $  916\pm 26$ & $ 4.2\pm 0.1$ & $ 157.1\pm 1.6$ & $ 24.3\pm  1.7$ & $  354\pm  42$ & $ -180\pm  43$ & $  0.41\pm  0.05$ & $ -0.21\pm  0.05$ \\
           &  5$^{a}$ & 34 &  4.00 &    177.5 &   5.30 & $  861\pm 28$ & $ 4.0\pm 0.1$ & $ 166.0\pm 2.5$ & $ 11.5\pm  2.7$ & $   81\pm  30$ & $   -6\pm  38$ & $  0.10\pm  0.04$ & $ -0.01\pm  0.05$ \\
           &  6$^{a}$ & 37 &  2.80 & $-$168.9 &   3.72 & $  566\pm 20$ & $ 2.6\pm 0.1$ & $-176.4\pm 1.3$ & $  7.6\pm  1.4$ & $   31\pm  21$ & $  -80\pm  17$ & $  0.06\pm  0.04$ & $ -0.15\pm  0.03$ \\
           &  7$^{a}$ & 12 &  2.10 & $-$159.5 &   2.79 & $  596\pm 36$ & $ 2.8\pm 0.2$ & $-167.2\pm 1.7$ & $  7.7\pm  1.8$ & $  140\pm  85$ & $ -299\pm  76$ & $  0.25\pm  0.15$ & $ -0.54\pm  0.14$ \\
           &  8$^{a}$ & 27 &  2.20 & $-$165.1 &   2.92 & $  640\pm 26$ & $ 3.0\pm 0.1$ & $-173.0\pm 1.4$ & $  8.0\pm  1.5$ & $  235\pm  41$ & $   12\pm  25$ & $  0.39\pm  0.07$ & $  0.02\pm  0.04$ \\
           &  9$^{a}$ & 17 &  1.96 & $-$161.3 &   2.60 & $  662\pm 47$ & $ 3.1\pm 0.2$ & $-164.2\pm 2.2$ & $  2.9\pm  2.3$ & $ -153\pm  97$ & $ -111\pm 102$ & $ -0.25\pm  0.16$ & $ -0.18\pm  0.17$ \\
           & 10       & 13 &  1.27 & $-$162.1 &   1.69 & $  208\pm 44$ & $ 1.0\pm 0.2$ & $-166.2\pm 6.6$ & $  4.1\pm  6.7$ & $ -417\pm 131$ & $   99\pm  85$ & $ -2.14\pm  0.82$ & $  0.51\pm  0.45$ \\
           & 11       & 41 &  0.34 & $-$163.6 &   0.45 & $   23\pm  5$ & $ 0.1\pm 0.0$ & $   7.3\pm 8.6$ & $170.9\pm  8.7$ & $  -13\pm   4$ & $    6\pm   3$ & $ -0.58\pm  0.23$ & $  0.28\pm  0.18$ \\
2223$-$052 &  1$^{a}$ & 14 &  5.76 &     97.7 &  49.20 & $  282\pm 55$ & $18.9\pm 3.7$ & $  90.6\pm 3.9$ & $  7.1\pm  3.9$ & $  -95\pm  47$ & $  -17\pm  21$ & $ -0.81\pm  0.44$ & $ -0.15\pm  0.19$ \\
           &  2       & 22 &  3.15 &    102.7 &  26.91 & $  105\pm 28$ & $ 7.1\pm 1.9$ & $  81.6\pm11.7$ & $ 21.1\pm 11.7$ & $   -8\pm  22$ & $   15\pm  19$ & $ -0.18\pm  0.51$ & $  0.35\pm  0.46$ \\
           &  3       &  9 &  1.30 &    105.1 &  11.10 & $   51\pm 32$ & $ 3.4\pm 2.2$ & $ 112.5\pm61.8$ & $  7.4\pm 62.0$ & $   45\pm  85$ & $  -65\pm  65$ & $  2.12\pm  4.24$ & $ -3.07\pm  3.64$ \\
           &  4       & 22 &  0.47 &     73.8 &   4.05 & $   92\pm 13$ & $ 6.2\pm 0.9$ & $  90.0\pm 9.1$ & $ 16.2\pm  9.4$ & $   49\pm  17$ & $   38\pm  18$ & $  1.28\pm  0.49$ & $  1.00\pm  0.49$ \\
2234+282   &  1$^{c}$ & 11 &  0.84 & $-$122.0 &   6.32 & $   20\pm 25$ & $ 0.9\pm 1.1$ & $-140.2\pm77.5$ & $ 18.2\pm 77.5$ & $  -28\pm  89$ & $   63\pm  44$ & $ -2.53\pm  8.46$ & $  5.55\pm  8.00$ \\
           &  2       &  9 &  0.50 & $-$133.2 &   3.78 & $   44\pm 19$ & $ 2.0\pm 0.9$ & $ -89.9\pm14.2$ & $ 43.3\pm 14.3$ & $   38\pm  22$ & $   22\pm  21$ & $  1.56\pm  1.14$ & $  0.90\pm  0.96$ \\
           &  3$^{a}$ & 26 &  0.51 & $-$136.4 &   3.87 & $   71\pm  4$ & $ 3.2\pm 0.2$ & $-135.3\pm 3.9$ & $  1.1\pm  4.0$ & $   14\pm   8$ & $   17\pm   8$ & $  0.37\pm  0.20$ & $  0.44\pm  0.22$ \\
2243$-$123 &  1       & 39 & 10.82 &     30.9 &  74.18 & $   66\pm 36$ & $ 2.4\pm 1.3$ & $ 179.6\pm22.2$ & $148.7\pm 22.2$ & $  -21\pm  35$ & $  -28\pm  34$ & $ -0.53\pm  0.92$ & $ -0.69\pm  0.92$ \\
           &  2       & 41 &  3.33 &     17.2 &  22.81 & $   99\pm 12$ & $ 3.6\pm 0.4$ & $  30.2\pm 6.1$ & $ 13.0\pm  6.1$ & $  -13\pm  12$ & $   11\pm  11$ & $ -0.21\pm  0.20$ & $  0.18\pm  0.19$ \\
           &  3$^{a}$ & 40 &  1.40 &   $-$2.7 &   9.62 & $   89\pm  6$ & $ 3.3\pm 0.2$ & $   9.0\pm 1.1$ & $ 11.7\pm  1.2$ & $   16\pm   6$ & $    4\pm   3$ & $  0.29\pm  0.12$ & $  0.08\pm  0.07$ \\
\tableline
\end{tabular}}
\end{center}
{\bf Notes.} (1) Source name; (2) Component ID; (3) Number of epochs; (4) Weighted mean radial separation
from core; (5) Weighted mean position angle; (6) Weighted mean projected radial distance in parsecs; (7) Proper
motion; (8) Apparent speed in units of the speed of light; (9) Proper motion position angle.
No entry is given in this or subsequent columns if the fitted error in this quantity exceeds $90\arcdeg$;
(10) Absolute difference between weighted
mean position angle and proper motion position angle; (11) Angular acceleration parallel to the proper motion position angle;
(12) Angular acceleration perpendicular to the proper motion position angle; (13) Relative parallel acceleration;
(14) Relative perpendicular acceleration.\\
$^{a}$ Component is a member of the 48 and 64-component subsamples used in studying apparent accelerations
(see $\S$~\ref{acc}).\\
$^{b}$ Component is a member of the 64-component subsample
used in studying apparent accelerations
(see $\S$~\ref{acc}).\\
$^{c}$ Component satisfies the criteria for an LPS component (see $\S$~\ref{speedwithin}).
\end{sidewaystable*}

\subsection{Speed Variations Within Sources}
\label{speedwithin}
Figure~5 shows a histogram of the measured apparent 
velocity magnitude $\beta_\mathrm{app}$ from Table~\ref{acctab}, for all components
from Table~\ref{acctab} ($N=224$ for the 65 sources with redshifts; 1657$-$261 does not have a measured redshift).
The mean apparent component speed from Figure~5 is 7.2$c$, and the median apparent speed is 4.5$c$.
We discuss the variation in apparent speed from component to component within individual sources here,
and then discuss the apparent speed variations among different sources in the next subsection.

Overall, we confirm the general trend seen in other studies of apparent speed distributions:
that while the vast majority of component motion is outward, there also exist a small but non-negligible number
of apparently inwardly moving components and nearly stationary components (L09; Britzen et al. 2008).
In the RDV survey, 185 of the 218 components in Table~\ref{acctab} with a measured value for
$|PA-\phi|$ are moving `outward' ($|PA-\phi|\le 90\arcdeg$), while only 33 are moving
`inward' ($|PA-\phi|> 90\arcdeg$) --- and most of those 33 measurements are not statistically significant.
Only 10 of these 33 components are moving inward with a significance $>3\sigma$, these all
also have a negative measured value for their radial apparent speed in Table~\ref{speedtab}, as expected. 
L09 discuss in detail five different geometrical effects that can lead to the `illusion'
of apparent inward motion\footnote{Briefly, these are: blending of the core with a new jet component,
a jet component misidentified as the core, a jet that curves back across the line-of-sight,
a backwards moving pattern in the flow, or changes in the internal brightness distribution of
a component.}; none of these represent the real bulk inward motion of jet material, and it is
likely that some combination of these five effects is acting on this small subset of components here.

\begin{figure*}[!t]
\begin{center}
\includegraphics[angle=90,scale=0.45]{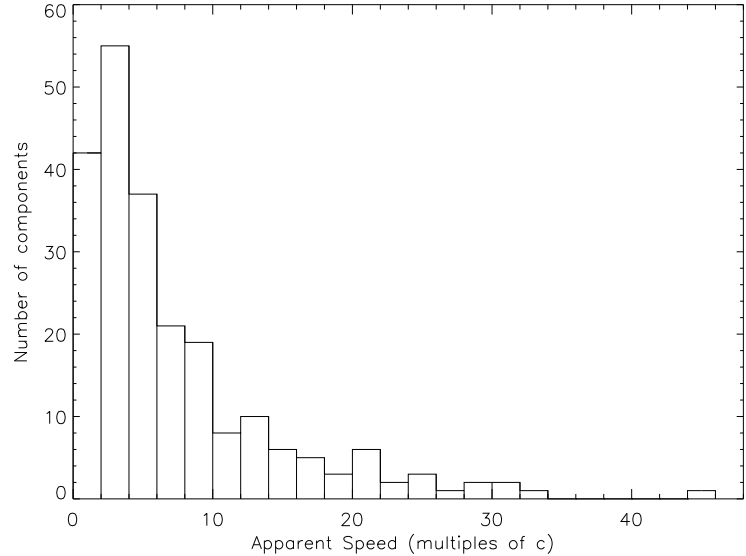}
\end{center}
\caption{
Distribution of apparent speed for the 224 components in the RDV survey
with measured redshifts.}
\end{figure*}

We also confirm the existence of a subset of slowly moving or
nearly stationary (Low Pattern Speed, or LPS) components at
similar numbers and core distances as found by the MOJAVE survey (L09).
There are a total of 43 components in Table~\ref{acctab} with a proper motion of
less than 50 $\mu$as yr$^{-1}$, which is the limiting proper motion for an LPS component
in L09. Of these 43, 21 also meet the other two criteria for an LPS component used
by L09: no significant acceleration, and speed significantly
slower than other components in the same jet. Here we quantify these two criteria
specifically by the significance of both $\dot{\eta}_{\parallel}$ and $\dot{\eta}_{\perp}$
being less than $2\sigma$, and the value of $\beta_\mathrm{app}$ being less than the weighted mean
value of $\beta_\mathrm{app}$ for the other components in the jet by at least $2\sigma$.
The 21 components in 19 sources that meet all three of these criteria are noted
as LPS components in Table~\ref{acctab}.

With these criteria, LPS components make up about 9\% of the
jet components in the RDV survey, and we find that
these LPS components occur closer to the core than the
general population of jet components. Over half
(12 out of 21) of the LPS components are clustered within projected distances of $\sim4$~pc from the core
(and the remaining 9 scatter out to projected distances of $\sim40$~pc), while
the median projected distance from the core for the non-LPS components is 9~pc.
In 8 of the 19 sources with LPS components, the LPS component is the closest one to the core.
For comparison, L09 find an overall occurrence rate of about 6\% (31 out of 526 components),
and typical projected core distances of $<$6~pc for LPS components.
Such apparently stationary features could be due to projection effects if the jet
passes very close to the line of sight, or they could be intrinsically stationary features
such as stable recollimation shocks
that are expected from jet simulations (e.g., Gomez et al. 1995),
and that may tend to occur at similar distances from the core.
In the RDV survey then, about 1/6 of the sources (12 out of 66)
have what could be interpreted as a stationary feature such as a recollimation shock within $\sim4$~pc
projected of the core (several tens of parsecs de-projected for the expected small viewing angles). 
Of the 21 LPS components, 20 are in quasars, 1 is in a galaxy, and none are
in the BL~Lac objects, based on the optical identifications in Table~\ref{sources}.
While L09 find the occurrence rate of LPS features to be higher in their BL~Lac objects,
we note that there are only 7 BL~Lac objects in the RDV sample.

If jets in the RDV survey are on average accelerating or decelerating, then we might
expect there to be a consistent sense of variation in the measured apparent speeds from component
to component in a source, as
components farther from the core would be either systematically faster or 
systematically slower than components closer to the core.
We investigate this relation between apparent component speed and average distance from the core here, and 
we then investigate accelerated motion of individual components in $\S$~\ref{acc}.
For each of the 56 sources in Table~\ref{acctab} that have at least two non-LPS components 
with $\ge1\sigma$ significance apparent velocity measurements,
we performed a fit to $\ln\beta_\mathrm{app}$ versus $\ln\langle{r}\rangle$ using measured values
from Table~\ref{acctab} (excluding all apparent speeds of $<1\sigma$ significance and all LPS components, see above). 
A constant positive apparent acceleration along the length of the jet
in a source would yield a slope of 0.5 for such a fit. 
Figure~6 shows an example of one of these fits (for the source 1514$-$241, with a slope of 0.5, close to
the mean slope for all of the fits), and Figure~7 shows a histogram of all 56 fitted slopes.
For the 56 individual fits, 43 fits yield a positive slope and 13 fits yield a negative slope, with the
mean slope being 0.55 (close to the value of 0.5 expected for constant acceleration within a source),
and the median slope being 0.34.
The binomial probability of measuring 43 positive slopes if they were randomly distributed
between positive and negative values is only $P=4\times10^{-5}$.
If we count only fitted slopes of at least $2\sigma$ significance, so that we may be sure of the sign,
then we find 22 positive slopes and 7 negative, with a binomial probability of $P=4\times10^{-3}$.
We therefore conclude that, on average in our sample, components farther from the core
have larger apparent speeds than components closer to the core 
in a given source, with high statistical significance.
We discuss the relation of this result to other results in the literature in $\S$~\ref{discussion}.
If components farther from the core are moving faster than the closer components, then individual components
must on average undergo positive apparent parallel accelerations. 
We discuss the apparent parallel accelerations of individual
components in $\S$~\ref{paracc}.

\begin{figure*}
\begin{center}
\includegraphics[angle=90,scale=0.45]{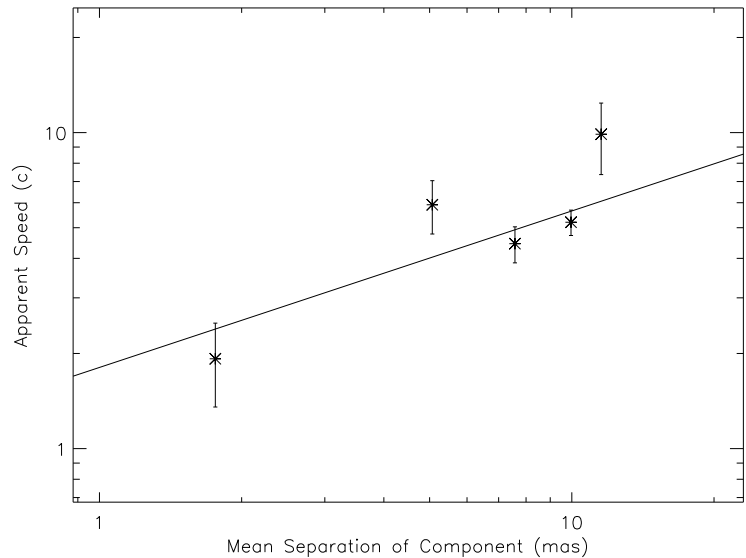}
\end{center}
\caption{An example of a linear fit to $\ln\beta_\mathrm{app}$ versus $\ln\langle{r}\rangle$ for 
a sample individual source, where $\langle{r}\rangle$ is the weighted mean separation of a component,
and $\beta_\mathrm{app}$ is its apparent speed, with values taken from Table~6.
This example is for the source 1514$-$241, which has a fitted slope of 0.5,
close to the mean slope for all of the fits. See $\S$~\ref{speedwithin} for 
further discussion.}
\end{figure*}

\begin{figure*}
\begin{center}
\includegraphics[angle=90,scale=0.45]{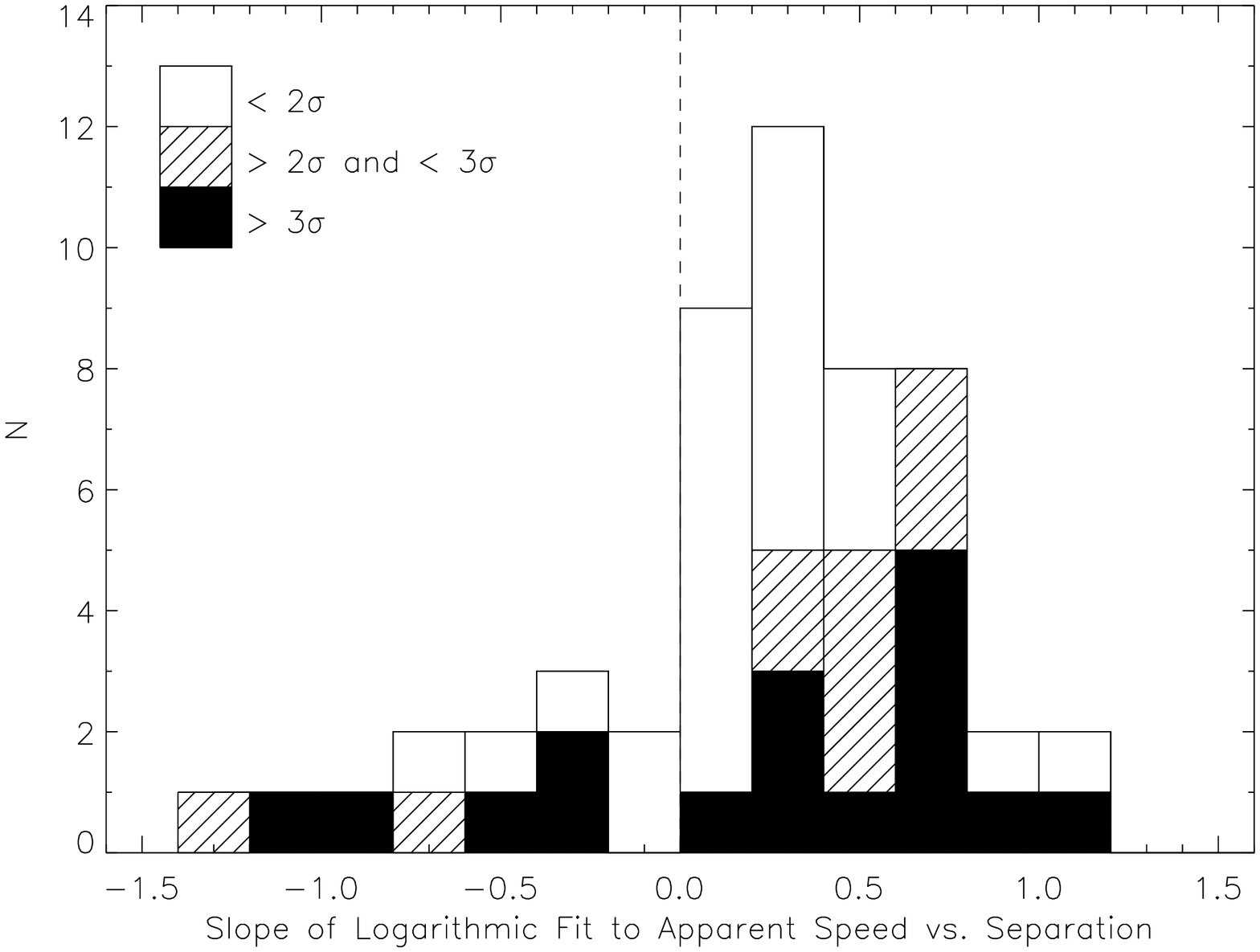}
\end{center}
\caption{
Histogram of the slopes of linear fits to $\ln\beta_\mathrm{app}$ versus $\ln\langle{r}\rangle$
for 56 individual sources, where $\langle{r}\rangle$ is the weighted mean separation of a component,
and $\beta_\mathrm{app}$ is its apparent speed, with values taken from Table~6.
Constant positive acceleration along a jet would yield a slope of 0.5 for these fits.
Hashed and solid fill styles indicate slopes significant at the $2\sigma-3\sigma$ and
$\ge 3\sigma$ levels, respectively. Three of the 56 sources are outside
the plotting window. See $\S$~\ref{speedwithin} for further discussion.}
\end{figure*}

We also confirm an important result that was also found in a number of
previous studies (e.g., L09; Kellermann et al. 2004; Paper I):
that the variation in apparent speeds from component to component within a source is significantly
less than the variation in apparent speeds from source to source within the sample.
As in L09, we quantify this by
computing the standard deviation in the measured apparent speeds for each multi-component
source. The median of these standard deviations is 3.1$c$; this represents a typical
variation in the measured apparent speeds within a single source.
We also compute the median apparent speed for each source in the sample, and find the
standard deviation of this set of speeds to be 7.6$c$; this represents the typical
variation in apparent speed from source to source within the sample.
This shows that there is a characteristic physical speed associated with each source
in the sample, which is plausibly the bulk speed of the jet flow.
Individual components within a source are then measured to have a relatively small range of speeds about this
characteristic speed.
If the flow is accelerating, then there may actually be a physical range of bulk Lorentz factors within a source that
is nevertheless smaller than the range of bulk Lorentz factors from source to source.
Components may also move within a range of pattern Lorentz factors from zero for a standing
shock, to higher speeds for `trailing features' (e.g., Kadler et al. 2008; Perucho et al. 2008), 
up to the peak bulk Lorentz factor of the flow.
Since we are interested in the peak bulk Lorentz factor attained
by each jet over the ten year monitoring time,
we henceforth use the fastest observed apparent component speed in each
source as the measured apparent speed associated with that source.
This also allows for direct comparison with the MOJAVE results, as they also use the
fastest measured apparent speed in each source to characterize that source (L09).

\subsection{Speed Variations Between Sources}
\label{speedbetween}
Figure~8 shows the histogram of the fastest measured apparent speed in each source
from Table~\ref{acctab}, for all sources in Table~\ref{acctab} with a measured redshift ($N=65$).
This distribution has a peak at an apparent speed of about 5$c$, a long tail extending out
to a maximum apparent speed of 44$c$ (for component 1~in 1313$-$333), 
a mean apparent speed of 11.5$c$, and a median apparent speed of 8.3$c$.
The shape of this distribution is similar to the equivalent distribution
measured by the MOJAVE survey (L09), and a Kolmogorov-Smirnov (KS) test confirms that there is
no significant difference between these two distributions.
However, when compared with the equivalent distribution of fastest apparent speeds from Paper I, 
the mean fastest apparent speed has increased from 5.9$c$ in Paper I to 11.5$c$ here.
L09 have also discussed this phenomenon that
apparent speed measurements have tended to increase as survey temporal coverage has increased
from older VLBI surveys (e.g., Britzen et al. 2008; Paper I) to newer surveys, showing that 
high angular resolution and excellent temporal coverage may facilitate the identification
of fast-moving components that would otherwise be missed.
Alternatively, the detected increase of the fastest speeds might be
due to the longer period of observations that provide more opportunity to see fast-moving components.
Whatever the cause, both of these two recent surveys of blazar apparent speeds at 
lower frequencies (L09 and this paper) have measured typical apparent speeds 
that are similar to those measured in surveys done at higher frequencies 
like 43~GHz ($\sim10c$, e.g., Jorstad et al. 2005).

\begin{figure*}
\begin{center}
\includegraphics[angle=90,scale=0.45]{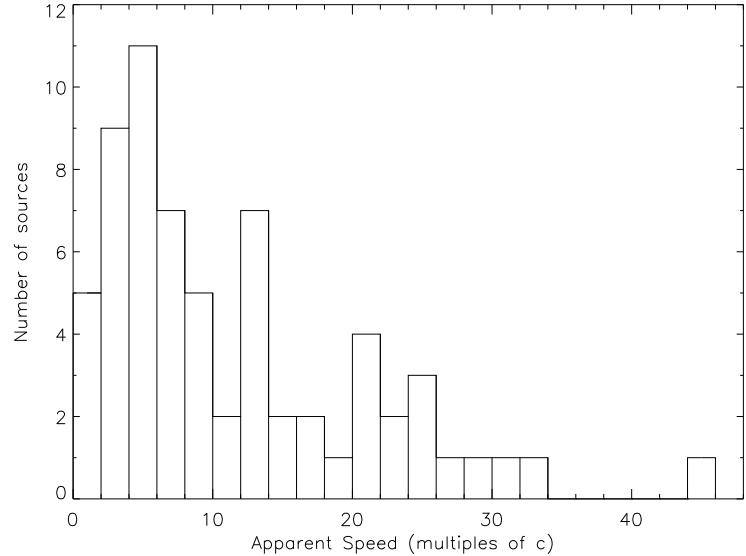}
\end{center}
\caption{
Distribution of fastest measured apparent speed for the 
65 sources in the RDV sample with measured apparent speeds.}
\end{figure*}

\begin{figure*}
\begin{center}
\includegraphics[angle=90,scale=0.45]{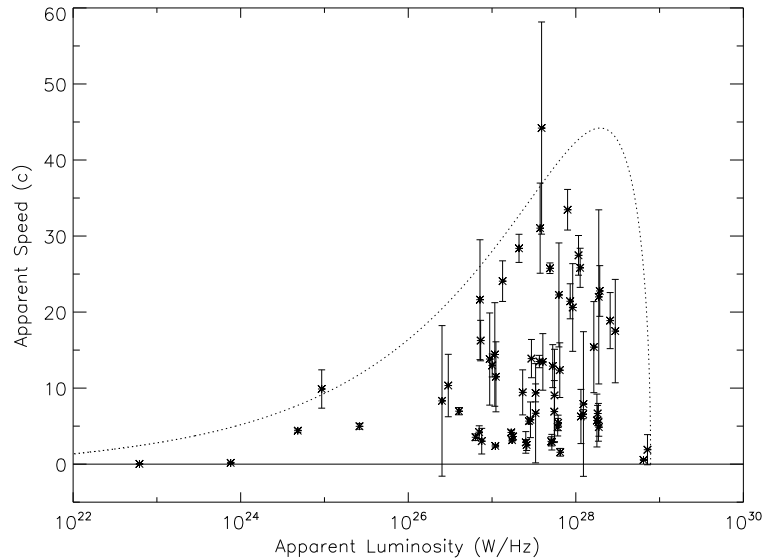}
\end{center}
\caption{Fastest measured apparent speed versus median apparent luminosity
for the 65 sources in the RDV sample with measured apparent speeds.
Some error bars are smaller than the plotting symbols.
The dotted curve corresponds to a jet with a bulk Lorentz factor of 44
and an intrinsic luminosity of $1\times10^{25}$~W~Hz$^{-1}$, as the viewing
angle $\theta$ varies, assuming Doppler boosting by a factor of $\delta^{2}$,
where $\delta$ is the Doppler factor.} 
\end{figure*}

Because the bulk Lorentz factor $\Gamma\ge (\beta_\mathrm{app}^{2}+1)^{1/2}$, the peak observed apparent
speed in Figure~8 of about 44$c$ indicates that bulk Lorentz factors 
in the parent population reach values of at least $\Gamma\sim44$.
This maximum value also agrees well with the peak apparent speeds found by both Jorstad et al. (2005) of 46$c$
and L09 of 51$c$.
The tapering off of apparent speed distributions at higher speed values observed
in Figure~8 and in other surveys can be reproduced in Monte Carlo simulations 
(e.g., Lister \& Marscher 1997; L09) by 
assuming an intrinsic power-law Lorentz factor distribution 
of slope $\sim-1.5$ in the blazar parent population.

Figure~9 shows the fastest apparent speed in each source in the RDV sample from Table~\ref{acctab}
versus its median 8~GHz apparent VLBI luminosity over all of the epochs used in this paper.
The luminosities are calculated according to $L=4\pi D_{l}^{2}S_{8}(1+z)^{-1}$
(using a $k$-correction with an assumed spectral index $\alpha=0$), where
$D_{l}$ is the luminosity distance and $S_{8}$ is the median total 8~GHz VLBI flux density.
There is an upper envelope to the distribution similar to that seen
in the CJF survey (Vermeulen 1995), the 2~cm Survey (Kellermann et al. 2004), and 
the MOJAVE survey (L09). As described by Cohen et al. (2007), the upper envelope 
of this distribution appears to be well-matched by an `aspect curve' that traces
out a single source of given bulk Lorentz factor and intrinsic luminosity in the
$(L,\beta_\mathrm{app})$ plane as the viewing angle $\theta$ changes. Such an aspect
curve is plotted on Figure~9 for a jet with a bulk Lorentz factor of 44
and an intrinsic luminosity of $1\times10^{25}$~W~Hz$^{-1}$, 
assuming Doppler boosting by a factor of $\delta^{2}$, where $\delta=1/(\Gamma(1-\beta\cos\theta))$
is the Doppler factor, and the exponent is for a smooth flat-spectrum jet
and should be appropriate for the core region (Cohen et al. 2007).
While it has been shown that this upper envelope is not due to selection effects (Cohen et al. 2007),
its precise physical origin is unclear; L09 speculate
that such an envelope may arise because of an intrinsic relation
between jet speed and luminosity in the parent population, although the statistics of
current samples cannot fully address this.

\begin{figure*}
\begin{center}
\includegraphics[angle=0,scale=0.53]{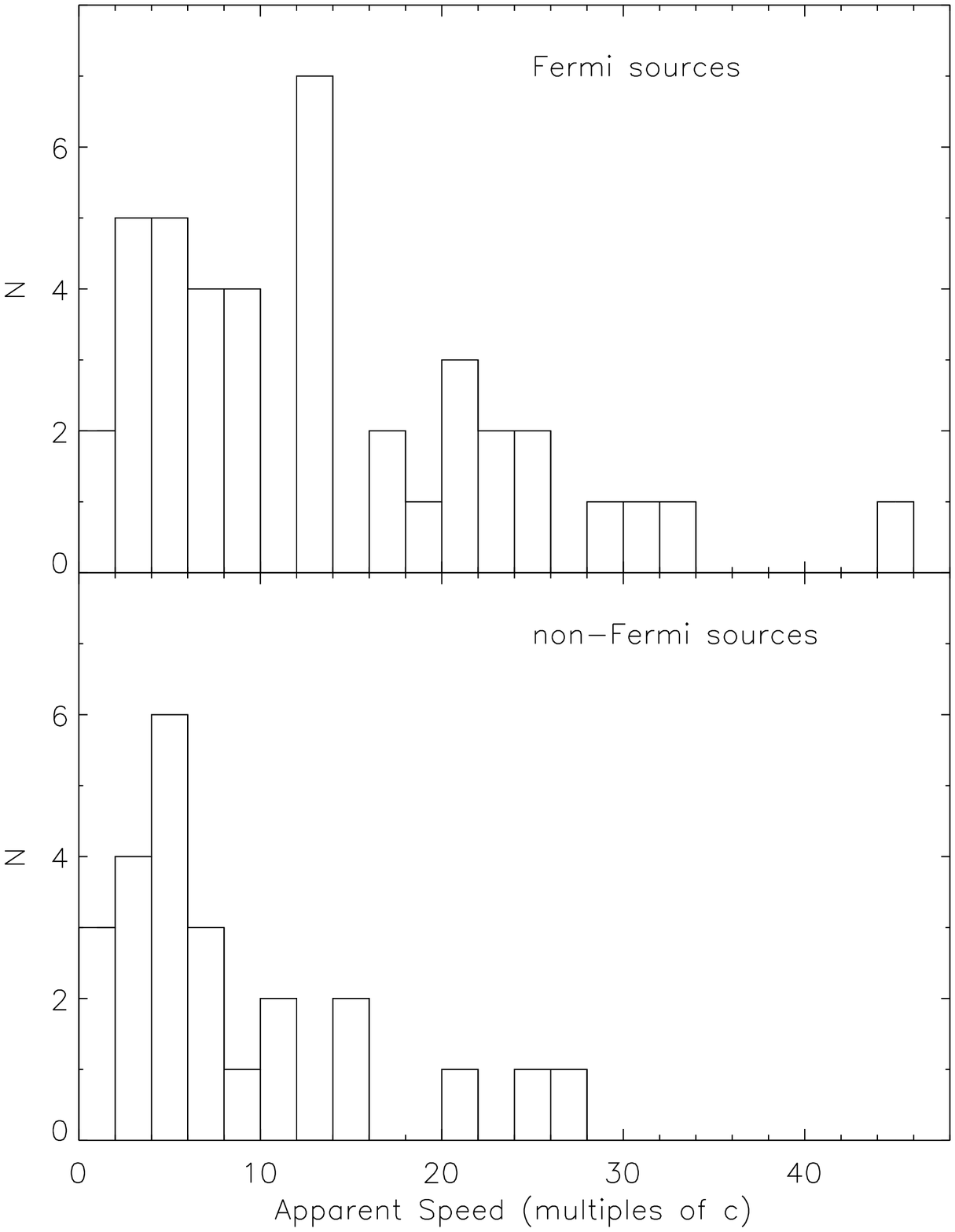}
\end{center}
\caption{
Distribution of fastest measured apparent speed for the 
65 sources in Figure~8,
separated into {\em Fermi} LAT-detected sources ($N=41$, top) and non-detected
sources ($N=24$, bottom).}
\end{figure*}

\subsection{Gamma-Ray Bright Sources}
\label{gamma}
It has been noted since the time of the EGRET gamma-ray telescope that those sources that were
detected in GeV gamma-rays tended to have faster apparent speeds than sources that were not
detected (e.g., Jorstad et al. 2001). This is explained if
the gamma-ray emission is boosted by a higher power of the Doppler factor than the radio emission,
so that the gamma-ray sources tend to have higher Lorentz factors and smaller viewing angles (Pushkarev et al. 2009),
leading to faster apparent speeds (e.g., Lister 1999).
This trend for faster speeds has continued to be noted with the blazars detected by the {\em Fermi} LAT gamma-ray
telescope. Lister et al. (2009c) found a significant difference in the speeds of
the LAT-detected and non-detected sources in the MOJAVE survey, with the LAT-detected sources being faster.
Figure~10 shows the distribution of fastest apparent speeds in the RDV sample from Table~\ref{acctab},
separated into LAT detections and non-detections,
from the list of LAT detections in Table~\ref{sources}.
The median apparent speed of the LAT-detected sources is $12.4c$, while the median apparent speed
of the non-detected sources is only $5.7c$. This difference in the medians of the two distributions is
significant at the 98.5\% confidence level, according to a Wilcoxon rank-sum test.
(For comparison, a students' T-test on the difference in the means, $13.4c$ and
$8.3c$, gives a significance of 97.7\%, and an unbinned KS test gives a significance of 90.7\%.)
We thus confirm that the 2FGL {\em Fermi} LAT-detected
blazars display faster apparent speeds than the non-detected sources in the RDV sample as well.
Note that the situation with the powerful blazars typical of both the RDV sample
and the MOJAVE sample that are detected at GeV energies by {\em Fermi}
contrasts with the lower-luminosity TeV gamma-ray blazars, which tend to have slower apparent speeds at parsec-scales
compared to radio-selected samples (Piner et al. 2010). This is discussed more fully in $\S$~\ref{discussion}.

Other radio properties measured from the RDV survey besides the apparent speed are also related
to Fermi LAT detection status, or to the measured Fermi gamma-ray flux.
A correlation between the non-simultaneous 8~GHz VLBI flux density, including RDV series data,
and Fermi gamma-ray flux is shown by Kovalev (2009).
Kovalev (2009) also show that the LAT-detected sources have higher 8~GHz VLBI flux densities,
when compared to the non-detected sources.
Pushkarev \& Kovalev (2012) use a subsample of 370 sources from 19 of the RDV experiments in Table~\ref{obstab}
between 1998 and 2003 to show that the LAT-detected sources have higher VLBI core flux densities
and brightness temperatures at 8 and 2~GHz, and flatter spectral indices in their VLBI jets between 8 and 2~GHz,
when compared to the non-detected sources.
Despite these significant flux correlations from the RDV data, such comparisons are best done with quasi-simultaneous
flux data because of the variable nature of the sources.
Such studies have been done for quasi-simultaneous VLBA and Fermi data from the MOJAVE survey by e.g.,
Kovalev et al. (2009), Pushkarev et al. (2010), and Lister et al. (2011).

\begin{figure*}
\begin{center}
\includegraphics[angle=0,scale=0.50]{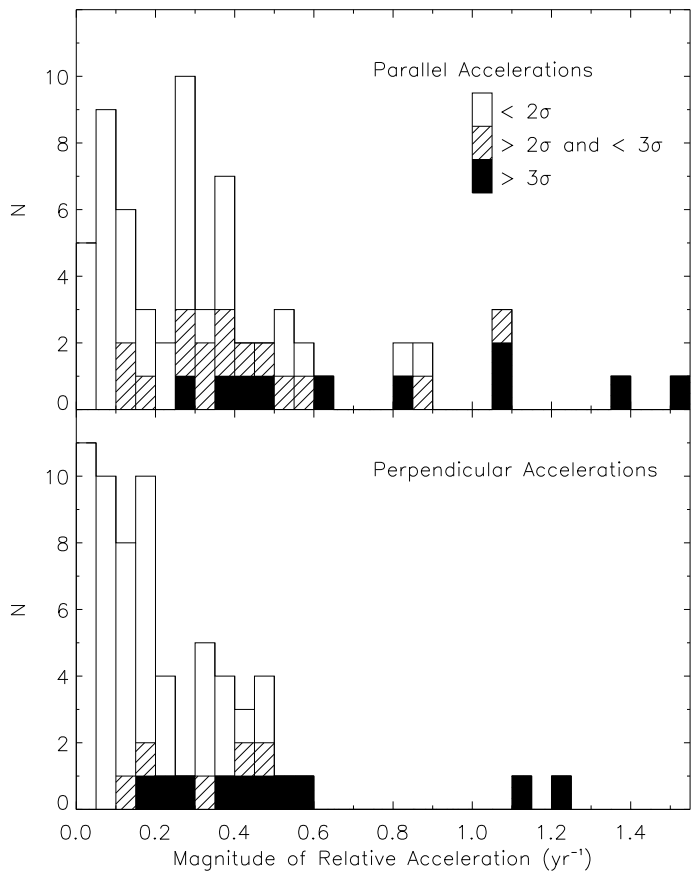}
\end{center}
\caption{
Histograms of magnitudes of relative parallel (top) and perpendicular (bottom)
accelerations for the 64-component subsample. Hashed and solid fill styles
indicate relative accelerations significant at the $2\sigma-3\sigma$ and
$\ge 3\sigma$ levels, respectively.}
\end{figure*}

\section{Apparent Accelerations}
\label{acc}
\subsection{Parallel accelerations}
\label{paracc}
The fitting method used to construct
the apparent acceleration analysis in Table~\ref{acctab} is described in $\S$~\ref{fit},
and the relative parallel and perpendicular accelerations that are used throughout
this section are defined by Equations~(\ref{relpar}) and (\ref{relperp}).
We apply two sets of cuts to the nonlinear fits in Table~\ref{acctab} to yield subsamples of the
highest quality fits for the acceleration analysis. The first set of cuts is identical to the
cuts used by H09 for their acceleration analysis
(the component is observed at at least 10 epochs, with a proper motion significance
of at least $3\sigma$, and an uncertainty in the direction of
motion relative to the weighted mean component position angle, $(PA-\phi)$, of 5$\arcdeg$ or less).
These cuts yield a high-quality subsample of 48 components in 26 sources.
To produce a somewhat larger statistical sample for distributions where we wish to apply
the KS test (since this study contains only about half the total number of components
as the MOJAVE survey studied by H09), we also extend the cuts to allow errors in the direction of
motion relative to the weighted mean component position angle, $(PA-\phi)$, of up to 6$\arcdeg$; this yields a
second and somewhat larger subsample of 64 components in 34 sources.
Components belonging to these subsamples are indicated in Table~\ref{acctab}.
Fitted acceleration values for components that do not make these quality cuts are included in
Table~\ref{acctab} for completeness and to show the current state of the processed RDV data for
each component, but we caution against using those acceleration values
unless the fits can be supplemented with additional data.
A relative acceleration is defined to be ``high'' by H09
if its magnitude is at least $2\sigma$ above 0.1~yr$^{-1}$.
For either of these subsamples, we find that about 1/4 of the components have a
high relative parallel acceleration and 1/7 have a high relative perpendicular acceleration
with magnitudes that are at least $2\sigma$ above 0.1~yr$^{-1}$,
the same occurrence rates of ``high'' parallel and perpendicular accelerations found by H09.

\begin{figure*}
\begin{center}
\includegraphics[angle=90,scale=0.45]{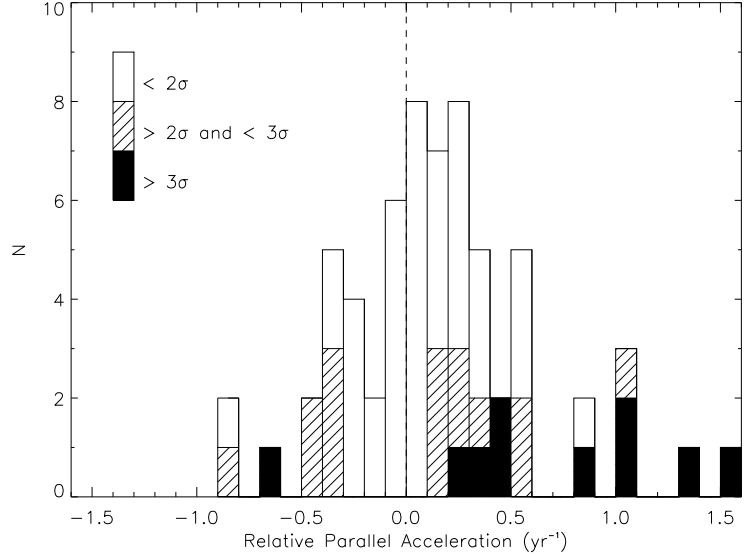}
\end{center}
\caption{
Histogram of relative parallel accelerations for the 64-component subsample,
taking into account the sign of the acceleration. Hashed and solid fill styles
indicate relative parallel accelerations significant at the $2\sigma-3\sigma$ and
$\ge 3\sigma$ levels, respectively.}
\end{figure*}

As in H09, in order to determine whether the observed jet accelerations might be due only to jet bending,
or if they are also due to changes in the component Lorentz factor, we compare the
distributions of relative parallel and perpendicular accelerations.
In a typical beamed jet sample, if the observed accelerations are due only to jet
bending, then the magnitudes of the observed parallel accelerations are expected to be about
60\% of the magnitudes of the observed perpendicular accelerations;
see the discussion following Equation~(6) in H09
\footnote{Also note that some fundamental equations for the kinematics of accelerating
jets are derived in Appendices 1 and 2 of H09.}.
(For a viewing angle of $\sin\theta=1/(n\Gamma)$ with $n\ge1$, the ratio of parallel to perpendicular
acceleration magnitudes due only to jet bending is given approximately by
$(n^2-1)/(n^2+1)$, for $\Gamma>>1$. H09 consider the case $n=2$ which is
typical of beamed jet samples. For larger $n$ the ratio is larger but is always less than one.)
Figure~11 shows histograms of the magnitudes of the relative parallel and perpendicular
accelerations for the 64-component subsample described above.
A KS test shows a significant difference between the two distributions
at the 99\% confidence level, with the
weighted mean magnitude of the relative parallel accelerations being
larger, at $0.20\pm0.01$~yr$^{-1}$, compared
to a weighted mean magnitude of only $0.12\pm0.01$~yr$^{-1}$ for the relative perpendicular accelerations.
From the discussion above, based on the mean magnitude of the relative perpendicular accelerations,
we would expect a mean relative parallel acceleration magnitude of only
about 0.07~yr$^{-1}$ based on jet bending alone, the actual
magnitude is about 3 times larger than this.
This confirms a result found also by H09: the distributions of relative
parallel and perpendicular accelerations are statistically distinct, with the
parallel accelerations having a larger average magnitude by a factor of about 1.7.
This implies that there are intrinsic changes in component Lorentz factors at the
parsec scales that dominate over jet bending in producing the observed parallel accelerations.
Note though that these changes may be in either the bulk or the pattern Lorentz factor, if the component is moving
at a pattern speed that is different from the bulk flow speed.

To determine if components are predominantly accelerating (increasing Lorentz factor) or
decelerating (decreasing Lorentz factor) at these scales, we investigate the signs of the
relative parallel accelerations. We find that positive parallel accelerations
statistically dominate the RDV survey compared to negative parallel accelerations.
For the 64-component subsample, 41 of the 64 components have
a positive parallel acceleration, while only 22 have a negative parallel acceleration
(and one is zero within the round off of the values).
The binomial probability of obtaining 41 or more positive accelerations from a sample
of 63, if they were randomly distributed, is only $P=0.01$.
The weighted mean of these relative parallel accelerations (now taking into account the sign) is
$0.133\pm0.014$~yr$^{-1}$, statistically distinct from zero with high significance.
This relative parallel acceleration distribution is shown in Figure~12.
For comparison, the relative perpendicular acceleration distribution for the
64-component subsample contains 34 positive and 30 negative accelerations, with a weighted mean of
$-0.016\pm0.012$~yr$^{-1}$, statistically consistent with the relative perpendicular
accelerations being randomly distributed
between positive and negative values with a mean of zero, as expected.

\begin{figure*}
\begin{center}
\includegraphics[angle=90,scale=0.45]{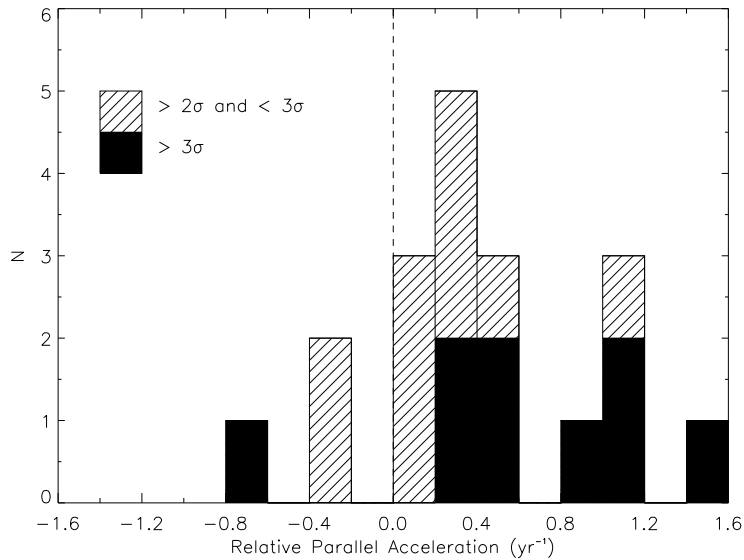}
\end{center}
\caption{
Histogram of relative parallel accelerations, taking into account the sign of the acceleration,
for the 19 components in the 48-component subsample
that have relative parallel accelerations significant at the $\ge 2\sigma$ level.
Hashed and solid fill styles
indicate relative parallel accelerations significant at the $2\sigma-3\sigma$ and
$\ge 3\sigma$ levels, respectively.}
\end{figure*}

If we increase the quality cuts on the fits, then the bias toward positive
parallel accelerations becomes more significant. If we restrict the analysis to
the 48-component subsample described above, and of those look at only components with a relative
parallel acceleration of $\ge 2\sigma$ significance (so that we may be sure of
the sign of the acceleration) then we find a total
of 19 components: 16 with positive parallel acceleration and 3 with negative.
The binomial probability of this many positive accelerations is only
$P=0.002$. The weighted mean of these relative parallel accelerations is
$0.227\pm0.020$~yr$^{-1}$, which is again statistically distinct from zero with high
significance. This relative parallel acceleration distribution is shown in Figure~13.
The equivalent distribution for the relative perpendicular accelerations
contains a total of 13 components, 7 with negative accelerations
and 6 with positive, again consistent with a random distribution.
We also note that a bias toward positive
parallel accelerations remains even if no cuts at all on the data in Table~\ref{acctab} are made: the
weighted mean relative parallel acceleration for all components in Table~\ref{acctab} is
$0.113\pm0.013$~yr$^{-1}$ (compared with $-0.016\pm0.011$~yr$^{-1}$ for the relative perpendicular accelerations).

These parallel acceleration results differ somewhat from those found in the
MOJAVE survey. H09 found approximately equal numbers of positive and negative parallel
accelerations, with positive accelerations tending to
occur within about 15 parsecs of the core (projected), and negative accelerations
tending to occur at distances beyond about 15 parsecs from the core
(projected). We confirm this distance dependence of the parallel accelerations
at marginal significance: for the 64-component subsample
a plot of the 25 relative parallel accelerations (in 18 sources)
that are significant at or above the 2$\sigma$ level versus the
weighted mean projected distance of the component from the core
is shown in Figure~14.
There is a negative correlation between these two quantities significant at the 94\% confidence level, as measured
by the Kendall rank correlation coefficient.
However, as can be seen in Figure~14, the RDV survey simply does not have very many components
in this subsample with significant parallel accelerations that lie beyond 15~parsecs (projected) from the core
(6 out of 25 components in Figure~14).
The results from the RDV survey and the MOJAVE survey therefore seem to be mutually consistent
for the regions of the jet for which both have substantial numbers of components:
because the RDV survey has predominantly measured the accelerations of components
that fall within the positively accelerating part of the jet as measured by MOJAVE,
we expect to find an excess of positive accelerations.
These results may also imply that high-quality subsample components tend to be
closer to the core in the RDV survey than the MOJAVE survey, and indeed the median distance
of a component from the core in the 48-component subsample is slightly less than
in the equivalent sample from H09, although this is significant at only 86\% confidence
according to a Wilcoxon rank-sum test. This is discussed further in $\S$~\ref{discussion}.

\begin{figure*}
\begin{center}
\includegraphics[angle=90,scale=0.45]{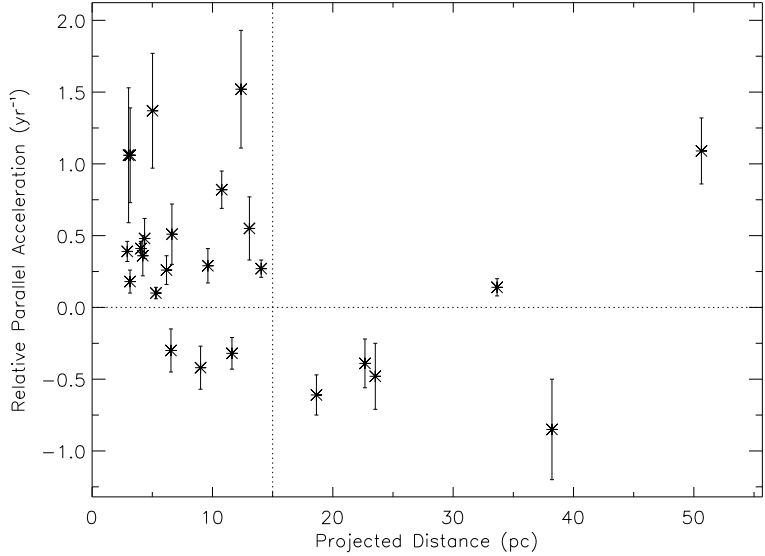}
\end{center}
\caption{
Relative parallel accelerations from the 64-component subsample
that are significant at or above the 2$\sigma$ level (25 components in 18 sources) versus the
weighted mean projected distance of the component from the core.
The horizontal dotted line shows the boundary between positive and negative acceleration.
The vertical dotted line at 15~parsecs (projected)
from the core shows the nominal location at which H09 found the acceleration
to switch from positive to negative.}
\end{figure*}

\subsection{Non-radial Motion}
\label{perpacc}
Parsec-scale jets that appear strongly bent are a common feature of VLBI images of blazars.
There are two general ways that such bent jets could be produced:
through ballistic (radial) motion of components that are ejected at different position angles,
or through a common bent (non-radial) path that is followed by all components.
In the nonlinear fits described in $\S$~\ref{fit} and tabulated in Table~\ref{acctab},
a component whose velocity vector is pointing directly out from the core will have
$\phi=PA$, or $|PA-\phi|=0$, and will be moving radially outward.
The distribution of $|PA-\phi|$ thus indicates the amount of non-radial
motion that is present in the sample. A histogram of $|PA-\phi|$
for the 64-component subsample is shown in Figure~15.
Non-radial motion is common in the RDV survey: 34 out of the 64 components in
Figure~15 have $|PA-\phi|>0$ at $\ge 2\sigma$ significance, and  
25 have $|PA-\phi|>0$ at $\ge 3\sigma$ significance.

\begin{figure*}
\begin{center}
\includegraphics[angle=90,scale=0.45]{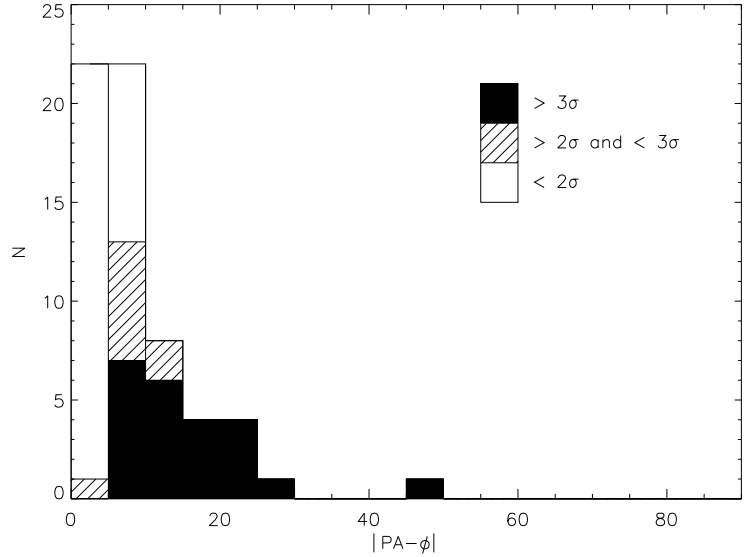}
\end{center}
\caption{
Histogram of velocity vector misalignment $|PA-\phi|$ for the 64-component subsample
from 34 sources. Hashed and solid fill styles
indicate non-radial motion significant at the $2\sigma-3\sigma$ and
$\ge 3\sigma$ levels, respectively. Two components lie beyond the right edge of the plot.}
\end{figure*}

If jet components are following a common bent channel, then the observed non-radial
motions should tend to align components with features that are farther out in the jet. 
For the 23 components in Figure~15 that have $|PA-\phi|>0$ at $\ge 3\sigma$ significance
and that are not moving apparently inward ($|PA-\phi|\le 90$), we have checked to see if
the sign of $(PA-\phi)$ is in the correct direction to move the component
toward the position angle of the downstream jet structure.
The position angle of the downstream jet structure is determined from the position
angle of the next component out in Table~\ref{acctab}, or from the 2~GHz images
produced from the experiments in Table~\ref{obstab} for the two sources where the component is
already the outermost component in the 8~GHz images. 
For 17 of these 23 components, $(PA-\phi)$ has the correct sign to move the component
toward the downstream structure; the chance probability of this is $P=0.02$.
Similarly, there is a correlation between $(PA-\phi)$ and $PA_{N}-PA_{N-1}$, where
$PA_{N}$ is the weighted mean position angle of the component with significant non-radial motion
and $PA_{N-1}$ is the weighted mean position angle of the next component out. The significance
of the Spearman rank correlation coefficient between these two quantities has a chance
probability of $P=0.04$.
These results show that jet components tend to follow a common flow channel pre-determined 
by the downstream structure, although what controls the exact shape of this bent jet
path is not determined by these observations.
We also note that the intrinsic jet bends will be significantly smaller than
these apparent bends, which are amplified by projection effects.
Similarly high percentages of non-radial motion, and a similar connection
between non-radial motion and downstream jet structure are also found in the MOJAVE survey by H09,
and were also previously found in the 2~cm Survey by Kellermann et al. (2004).

H09 also established a link in the MOJAVE survey data between the relative perpendicular acceleration
$\dot{\eta}_{\perp}$ and the velocity vector misalignment $(PA-\phi)$;
in other words, components moving in a non-radial direction tended to have a significant
acceleration in that direction, as might be expected.
In H09, 37 components out of their 203 component subsample
had significant values for both $\dot{\eta}_{\perp}$ and $(PA-\phi)$, and in 30 of those (or about 3/4)
the sign of $\dot{\eta}_{\perp}$ was in the proper direction to produce the
observed $(PA-\phi)$, which was a statistically significant result.
In the 64-component RDV subsample, 8 out of 64 components have highly significant values for
both $\dot{\eta}_{\perp}$ and $(PA-\phi)$, of these 6 out of 8 (or 3/4) have consistent directions.
While this is the same fraction found by the MOJAVE survey, the fewer total number of components
in the RDV survey (see Table~\ref{comptab}) means that this fraction is not statistically significant.

\section{Discussion}
\label{discussion}
In the previous sections of this paper we presented the kinematic analysis of a
large sample of VLBI data from the RDV experiment series. We concluded that this
analysis shows statistically significant evidence for positive parsec-scale jet accelerations in 
this particular jet sample
based on two mutually consistent lines of evidence. These are the observation that components farther
from the core tend to appear faster than components closer to the core in most jets in the sample ($\S$~\ref{speedwithin}),
and that individual components tend to have accelerations that increase their apparent speeds
($\S$~\ref{paracc}). Here we relate these observations to other results in the literature and to
the intrinsic physical properties of the sources. 

Other multi-jet studies have also reported a tendency for more distant jet components
to be faster than components closer to the core in the same object
(e.g., Homan et al. 2001; Piner et al. 2006; Britzen et al. 2008). 
Homan et al. (2001) observed this effect in five out of six sources (out of their sample of 12)
that had multiple components with measurable proper motion; it was also noted in all
three of the high-speed blazars studied by Piner et al. (2006).
Most significantly, Britzen et al. (2008)
performed fits to $\beta_\mathrm{app}$ versus $r$ for 105 sources out of the 293~in the CJF survey;
those fits are similar to our fits to the same two quantities described in $\S$~\ref{speedwithin}.
They concluded based on these fits that there was a slight trend toward positive acceleration; however,
their results were somewhat hindered by having only 3 epochs per source and lower angular resolution.

There is an apparent contradiction to these observations that has also been noted:
that in older VLBI surveys done at lower frequencies (and therefore lower resolutions, and
so observing components farther from the core)
the observed speeds have tended to be lower compared to surveys done at higher frequencies
(and therefore at higher resolutions, and observing components closer to the core). 
For example, apparent speeds measured in the CJF survey
by Britzen et al. (2008) at 5~GHz were on average slower than in the 2~cm Survey at 15~GHz 
(Kellermann et al. 2004), which in turn were slower than in the 22 and 43~GHz survey of EGRET
blazars by Jorstad et al. (2001). This seemed to imply that components were instead faster closer to the core.
However, as discussed in $\S$~\ref{speedbetween},
in newer VLBI surveys that are much better sampled in time at both lower and higher
frequencies, much of this apparent difference has disappeared.
The better temporal coverage in the newer surveys may aid in the identification
of fast-moving components missed in previous surveys.
As we noted in $\S$~\ref{speedbetween}, the surveys of blazar apparent speeds at
15~GHz and at 8~GHz by L09 and in this paper have now measured typical apparent speeds
that are similar to those measured at higher frequencies ($\sim10c$, e.g., Jorstad et al. 2005).
In any event, since these different samples were chosen based on different physical attributes, 
some for their radio emission and some for their gamma-ray
emission, some differences in jet speeds between the samples are not surprising,
and need not be due to component distance from the core.

In $\S$~\ref{paracc} we reported results on the measured apparent parallel accelerations
of individual jet components. We found that parallel accelerations were considerably larger
than perpendicular accelerations, showing that the parallel accelerations are dominated by
changes in the component Lorentz factor rather than jet bending.
Among the parallel accelerations, positive accelerations dominated over negative accelerations.
Mean relative parallel accelerations were in the range of 0.1-0.2~yr$^{-1}$, depending on the subsample.
To relate these observed apparent accelerations to changes in intrinsic source properties, note that
if the apparent parallel accelerations are due entirely to changes in the Lorentz factor then
the observed relative parallel acceleration is given by
\begin{equation}
\dot{\eta}_{\parallel}\equiv\frac{\dot{\beta}_\mathrm{\parallel app}}{\beta_\mathrm{app}}\approx \frac{\dot{\Gamma}}{\Gamma}\delta^{2}.
\end{equation}
This result can be obtained from differentiating Equation (\ref{speedeqn}) with respect to time in
the observer's reference frame with $\theta$ constant, see also the discussion of this equation in H09.
So, $\dot{\Gamma}/\Gamma\approx\dot{\eta}_{\parallel}/\delta^{2}$,
and with typical observed apparent speeds of 10$c$, we have in terms of orders of magnitude
$\delta\sim 10$ and $\dot{\eta}_{\parallel}\sim 0.1$~yr$^{-1}$, for 
$\dot{\Gamma}/\Gamma\sim 10^{-3}$~yr$^{-1}$ in the reference frame of
the host galaxy. The typical distance of a component 
from the core in the RDV survey is about 10 parsecs (projected), or about 100 parsecs de-projected.
Thus, a typical component in the RDV survey has $\dot{\Gamma}/\Gamma\sim 10^{-3}$~yr$^{-1}$ in the reference frame of
the host galaxy, at $\sim 100$
parsecs from the core. This is the same order of magnitude of
intrinsic accelerations found in the MOJAVE sample by H09. The intrinsic changes in the
Lorentz factor corresponding to the observed accelerations are relatively modest.
Such a level of intrinsic acceleration, if constant, is not sufficient to produce the
high Lorentz factors that are observed at these distances from the core. The intrinsic acceleration must be at
least an order of magnitude larger closer to the core to produce the high bulk Lorentz factors
that are observed at these distances.
Therefore, a typical component must get accelerated to $\Gamma\sim10$ by the time it
reaches distances of $\sim10$ parsecs from the core
(de-projected). Beyond these distances,
there are typical accelerations $\dot{\Gamma}/\Gamma\sim 10^{-3}$~yr$^{-1}$
that would correspond to about another 30\% increase in the Lorentz factor by $\sim 100$
parsecs from the core (de-projected). Beyond that, the acceleration must decrease (or even become negative
as observed by H09), as the jet transitions to the kiloparsec scale.

These observations of positive parsec-scale accelerations are consistent with the existence of an extended magnetic
acceleration region like that proposed by Vlahakis \& K\"{o}nigl (2004).
Those authors argue that magnetic acceleration should still be
active on parsec scales, and they also argue that 
some previously observed extended accelerations are unlikely to have had a purely hydrodynamic (non-magnetic) origin.
However, other magnetic acceleration models (e.g., Granot et al. 2011; McKinney 2006) predict that the acceleration
should be nearly complete by about 0.1~pc from the central engine.
There are also several other arguments that the magnetic to kinetic energy conversion should be nearly complete
by parsec scales. Flaring activity observed in blazar cores shows
that the jet is already matter-dominated at that point, if the flares are due to internal shocks.
However, if such flaring activity is instead related to magnetic 
energy dissipation rather than to internal shocks, then a matter-dominated jet
may not be required (e.g., Giannios 2011; Sikora 2005).
Following a different argument, Celotti \& Ghisellini (2008) conclude that the
jets of powerful blazars are matter-dominated on parsec scales through SED modeling of the emitting regions.
While the observations in this paper do show that there are modest accelerations 
on the parsec-scale, they cannot by themselves 
differentiate between a magnetic or hydrodynamic cause of these accelerations.
Also, since we are observing the motions of brightness centroids in the flow, we
cannot discount the possibility of pattern rather than bulk accelerations for any individual component.
Note that bulk accelerations should also yield changes in component flux density
through changing Doppler boosting, but since such changes would be coupled with intrinsic changes in flux density,
it would be difficult to use this as a diagnostic in practice.
However, the dominance of positive over negative parallel accelerations 
does suggest a link with a physical property of
the jet, such as the bulk flow speed, rather than the motions of random patterns.

A difference between the acceleration results presented here and those presented
for the MOJAVE survey by H09 is that while the RDV survey
is dominated by positive parallel accelerations, H09 found
approximately equal numbers of positive and negative parallel accelerations, with a transition
from positive to negative acceleration occurring at about 15~parsecs from the core (projected).
As can be seen from Figure~14, because we have studied fewer total jet components than
MOJAVE (see Table~\ref{comptab}), after the various quality cuts are applied we are not
left with enough components with significant accelerations at $>15$~parsecs (projected) from the core to make a conclusive         
statement about this region of the jet. However, interior to 15~parsecs (projected)
from the core we have demonstrated conclusively that jet components tend to have
a positive parallel acceleration: 16 of the 19 components in Figure~14 within 15~parsecs
(projected) of the core have a positive acceleration. This agrees with the results
of H09 for this region of the jet.
We also note that Jorstad et al. (2005) find a similar bias for positive parallel accelerations in a study
that favors components within 15~parsecs (projected) of the core because of its
high observing frequency. We thus interpret the results of the RDV survey, H09, and
Jorstad et al. (2005) as all being mutually consistent with the Lorentz factors of
jet components in powerful blazars tending to increase throughout the region of the jet
interior to 15~parsecs (projected) from the core.

Finally, we also wish to stress the difference between the sample of high-power blazars studied in this
paper and in Paper I
and the sample of less-powerful TeV blazars studied by, e.g.,
Piner et al. (2010). 
Significant evidence has been assembled indicating the substantial deceleration of the
flow before the parsec scales observed with VLBI in the low-power, nearby TeV sources, in contrast to the 
slight overall positive acceleration that has been measured in this paper for a sample of high-power sources.
Such a significant difference between the acceleration and deceleration length scales
for the high and low-power sources is likely to be related to 
fundamental differences between the central engine and/or the environment in these two source classes.

\section{Conclusions}
\label{conclusions}
We studied the parsec-scale kinematics of a sample of 68 extragalactic
jets using global VLBI observations at 8~GHz from the RDV experiment series, significantly
expanding upon our previous such study from Paper I. We included in this study all sources
observed at 20 or more epochs during a series of 50 VLBI experiments from 1994 to 2003.
We produced and analyzed 2753 VLBI images from these experiments, with a median of 43 epochs of observation per source.
In terms of angular resolution and temporal coverage, this RDV survey is similar to the
MOJAVE survey (L09; H09).
We fit Gaussian models to the visibilities associated with each image, and
identified a total of 225 jet components in 66 sources that could be followed from epoch to epoch.
Second-order polynomials were fit to $x(t)$ and $y(t)$ for each component to
study its velocity and acceleration.
Observational results related to the measured apparent speeds can be summarized as follows:
\begin{enumerate}
\item{When multiple moving components are present in a jet, components farther from the core
tend (about 75\% of the time) to have larger apparent speeds than components closer to the core,
with high statistical significance.}
\item{The variation in apparent speeds from component to component within a source is significantly
less than the variation in apparent speeds from source to source within the sample,
showing the existence of a characteristic speed associated with each source.}
\item{The distribution of the fastest measured apparent speed in each source
shows a maximum of 44$c$ and a median of 8.3$c$.}
\item{Sources detected by the {\em Fermi} LAT gamma-ray telescope display higher apparent speeds,
with a median of $12.4c$, than those that have not been detected,
which have a median of $5.7c$.}
\item{Apparently stationary or slowly moving Low Pattern Speed (LPS) components
are found in 19 sources. These LPS components
are clustered within $\sim4$~pc projected from the core, and may represent
truly stationary features such as recollimation shocks.}
\end{enumerate}

We identified high-quality subsamples of the full set of 225 components
for acceleration analysis, and for each of these components we analyzed the relative acceleration
both parallel and perpendicular to the direction of the average velocity vector, as well as the difference
between the direction of the average velocity vector and the weighted mean position angle.
Observational results related to the measured accelerations and non-radial motions can be summarized as follows:
\begin{enumerate}
\item{Significant non-radial motion is common, occurring in about
half of the components at $\ge 2\sigma$ significance. When non-radial motion occurs, it tends to
align the component with the downstream jet structure.}
\item{`High' relative accelerations (magnitudes that are at least
$2\sigma$ above 0.1~yr$^{-1}$) are fairly common, and comprise
about 1/4 of the parallel accelerations and 1/7 of the perpendicular accelerations, the
same rates of high accelerations found by the MOJAVE survey (H09).}
\item{The distributions of relative
parallel and perpendicular accelerations are statistically distinct, with the
parallel accelerations having a larger average magnitude by a factor of about 1.7.
This difference implies that there are intrinsic changes in component Lorentz factors
that dominate over jet bending in producing the observed parallel accelerations.}
\item{Positive parallel accelerations statistically dominate over negative parallel accelerations.
The weighted mean relative parallel acceleration for the 64-component subsample is 0.133$\pm$0.014~yr$^{-1}$.
A typical observed relative parallel acceleration of 0.1~yr$^{-1}$
corresponds to an increase in the bulk or pattern
Lorentz factor in the reference frame of
the host galaxy of order $\dot{\Gamma}/\Gamma\sim10^{-3}$~yr$^{-1}$ at
a distance of order 100 parsecs (de-projected) from the core.}
\end{enumerate}

In summary, blazar jets each have a characteristic speed within about 100 parsecs (de-projected) of the
supermassive black hole, with any sideways motions tending to
move components down a channel in the direction of the previous component. An
average component increases its apparent speed at an observed rate of about 10\% per year
at distances of about 100 parsecs (de-projected) from the core.
This apparent acceleration corresponds to an increase in the Lorentz factor
at a rate of about one part in 10$^{3}$ per year in the reference frame of
the host galaxy. A minority of components have an apparent deceleration at these distances.

All of the above conclusions are statistical in nature, and will not necessarily
apply to any particular individual source.
When taken together with the similar kinematic results from the MOJAVE survey by H09, the
acceleration results reported here show that modest changes in bulk or pattern Lorentz factors on parsec scales
are a relatively common feature of relativistic jets.
This observational result has now been confirmed in a mutually consistent manner
with high statistical significance by two large VLBI surveys
(although note that these two surveys are not completely statistically
independent, because they have 37 sources in common).
These observations are consistent with modest increases in the 
bulk kinetic energy on parsec-scales, although the source of this
energy is not determined from these observations. 

This paper and Paper I represent the tip of the iceberg of the astrophysics that can be done
with the RDV data. With $\sim100$ experiments observed to date, the total number of potential
images is approximately 10,000 each at 8 and 2~GHz. For example, adding the $\sim50$ experiments
that have been observed since 2003 to this study could double the size of the kinematic survey presented here.
Many things can also be studied other than kinematics; including
flux variability and multiwavelength correlations, spectral index and
core opacity (e.g., Kovalev et al. 2008; Pushkarev \& Kovalev 2012),
jet ridgelines and bending, and transverse structures in jets.

\vspace{-0.1in}
\acknowledgments
We thank both Dan Homan and
the anonymous referee for helpful comments that improved the paper.
The National Radio Astronomy Observatory is a facility of the National
Science Foundation operated under cooperative agreement by Associated Universities, Inc.
We acknowledge the International VLBI Service for Geodesy and Astrometry (IVS),
which organizes the non-VLBA stations for the RDV sessions.
This research has made use of the NASA/IPAC Extragalactic Database (NED)  
which is operated by the Jet Propulsion Laboratory, California Institute of Technology,
under contract with the National Aeronautics and Space Administration.
YYK and PAV were supported in part by the Russian Foundation for
Basic Research grant 11-02-00368, and by the basic research program
``Active processes in galactic and extragalactic objects'' of the Physical Sciences Division
of the Russian Academy of Sciences. YYK also thanks the
Dynasty Foundation for support.
This work was supported by the National Science Foundation under Grant 0707523
(PI Glenn Piner).

{\it Facilities:} \facility{VLBA ()}, \facility{IVS ()}


\begin{references}

\reference{}Ackermann, M., 
Ajello, M., Allafort, A., et al. 2011, ApJ, 743, 171

\reference{}Britzen, S., Vermeulen, R.~C., Campbell, R.~M., et al. 2008, A\&A, 484, 119

\reference{}Celotti, A., \& Ghisellini, G. 2008, MNRAS, 385, 283 

\reference{}Cohen, M.~H., Lister, M.~L., Homan, D.~C., et al.,
2007, ApJ, 658, 232

\reference{}Daly, R.~A., \& Marscher, A.~P. 1988, ApJ, 334, 539

\reference{}Dondi, L., \& Ghisellini, G. 1995, MNRAS, 273, 583

\reference{}Fey, A.~L., \& Charlot, P. 1997, ApJS, 111, 95

\reference{}Fey, A.~L., \& Charlot, P. 2000, ApJS, 128, 17

\reference{}Fey, A.~L., Clegg, A.~W., \& Fomalont, E.~B. 1996, ApJS, 105, 299

\reference{}Fey, A.~L., Ma, C., Arias, 
E.~F., et al. 2004, AJ, 127, 3587 

\reference{}Giannios, D. 2011, Journal of Physics Conference Series, 283, 012015

\reference{}Gomez, J.~L., Marti, J.~M.~A., Marscher, A.~P., Ibanez, J.~M.~A., 
\& Marcaide, J.~M. 1995, ApJ, 449, L19

\reference{}Granot, J., Komissarov, S.~S., \& Spitkovsky, A. 2011, MNRAS, 411, 1323

\reference{}Hartman, R.~C., 
B{\"o}ttcher, M., Aldering, G., et al. 2001, ApJ, 553, 683 

\reference{}Homan, D.~C., Kadler, M., 
Kellermann, K.~I., et al. 2009, ApJ, 706, 1253 (H09)

\reference{}Homan, D.~C., Lister, 
M.~L., Kellermann, K.~I., et al. 2003, ApJ, 589, L9

\reference{}Homan, D.~C., Ojha, R., 
Wardle, J.~F.~C., et al. 2001, ApJ, 549, 840 

\reference{}Jorstad, S.~G., 
Marscher, A.~P., Lister, M.~L., et al. 2005, AJ, 130, 1418

\reference{}Jorstad, S.~G., 
Marscher, A.~P., Mattox, J.~R., et al. 2001, ApJS, 134, 181

\reference{}Kadler, M., Ros, E., 
Perucho, M., et al. 2008, ApJ, 680, 867 

\reference{}Kellermann, K.~I., 
Lister, M.~L., Homan, D.~C., et al. 2004, ApJ, 609, 539

\reference{}Komissarov, S.~S.\ 2011, \memsai, 82, 95

\reference{}K\"{o}nigl, A. 2010, International Journal of Modern Physics D, 19, 635

\reference{}Kovalev, Y.~Y. 2009, ApJ, 707, L56

\reference{}Kovalev, Y.~Y., Aller, 
H.~D., Aller, M.~F., et al. 2009, ApJ, 696, L17 

\reference{}Kovalev, Y.~Y., 
Kellermann, K.~I., Lister, M.~L., et al. 2005, AJ, 130, 2473

\reference{}Kovalev, Y.~Y., Lobanov, A.~P., Pushkarev, A.~B., \& Zensus, J.~A. 2008, A\&A, 483, 759

\reference{}Lister, M.~L. 1999, Ph.D.~Thesis 

\reference{}Lister, M.~L., Aller, 
H.~D., Aller, M.~F., et al. 2009a, AJ, 137, 3718

\reference{}Lister, M.~L., Aller,
M., Aller, H., et al. 2011, ApJ, 742, 27

\reference{}Lister, M.~L., Cohen, 
M.~H., Homan, D.~C., et al. 2009b, AJ, 138, 1874 (L09)

\reference{}Lister, M.~L., Homan, 
D.~C., Kadler, M., et al. 2009c, ApJ, 696, L22 

\reference{}Lister, M.~L., \& Marscher, A.~P. 1997, ApJ, 476, 572

\reference{}Lyutikov, M., \& Lister, M. 2010, ApJ, 722, 197 

\reference{}McKinney, J.~C. 2006, MNRAS, 368, 1561 

\reference{}Ojha, R., Kadler, M., B{\"o}ck, M., et al. 2010, A\&A, 519, A45

\reference{}Perucho, M., Agudo, I., G{\'o}mez, J.~L., et al. 2008, A\&A, 489, L29

\reference{}Petrov, L., Gordon, D., 
Gipson, J., et al. 2009, Journal of Geodesy, 83, 859

\reference{}Petrov, L., \& Ma, C. 2003, JGRB, 108, 2190

\reference{}Piner, B.~G., Bhattarai, D., Edwards, P.~G., \& Jones, D.~L. 2006, ApJ, 640, 196

\reference{}Piner, B.~G., Mahmud, M., Fey, A.~L., \& Gospodinova, K. 2007, AJ, 133, 2357 (Paper I)

\reference{}Piner, B.~G., Pant, N., \& Edwards, P.~G. 2010, ApJ, 723, 1150

\reference{}Pushkarev, A.~B., \& Kovalev, Y.~Y. 2012, A\&A, 544, A34

\reference{}Pushkarev, A.~B., Kovalev, Y.~Y., Lister, M.~L. 2010, ApJ, 722, L7

\reference{}Pushkarev, A.~B., Kovalev, Y.~Y., Lister, M.~L., \& Savolainen, T. 2009, A\&A, 507, L33 

\reference{}Sikora, M., Begelman, M.~C., Madejski, G.~M., \& Lasota, J.-P. 2005, ApJ, 625, 72

\reference{}Tingay, S.~J., Preston, 
R.~A., Lister, M.~L., et al. 2001, ApJ, 549, L55

\reference{}Unwin, S.~C., Wehrle, 
A.~E., Lobanov, A.~P., et al. 1997, ApJ, 480, 596

\reference{}Vermeulen, R.~C. 1995, Proceedings of the National Academy of Science, 921, 11385

\reference{}V{\'e}ron-Cetty, M.~P., \& V{\'e}ron, P. 2010, A\&A, 518, A10 

\reference{}Vlahakis, N., \& K\"{o}nigl, A. 2004, ApJ, 605, 656

\end{references}
\end{document}